\documentclass[12pt]{article}
\usepackage{amsmath, amsfonts, amsthm, amssymb, color, graphicx, bm, url, xr, zref, comment, setspace}

\RequirePackage[colorlinks, linkcolor=red, citecolor=blue, urlcolor=blue]{hyperref}

\usepackage[top=2.5cm, bottom=2.5cm, left=2.6cm, right=2.6cm]{geometry}

\newtheorem{theorem}{Theorem}
\newtheorem{lemma}{Lemma}
\newtheorem{proposition}{Proposition}
\newtheorem{corollary}{Corollary}

\newcommand{\by}{y}

\newcommand{\yza}{\by_{0,a}}

\usepackage[square,numbers]{natbib}
\doublespace
\title{Approximating the Operating Characteristics of Bayesian Uncertainty Directed Trial Designs}

\begin{document}

\author{
Marta Bonsaglio $^{1}$, 
Sandra Fortini  $^{1}$, 
Steffen Ventz  $^{2,3}$, 
Lorenzo Trippa $^{2,3}$\\
$^{1}${\small Department of Decision Sciences, Universit\`a Bocconi, Italy},
$^{2}${\small Dana-Farber Cancer Institute, US},\\
$^{3}${\small Department of Biostatistics, Harvard T.H. Chan School of Public Health, US}\\
}

\maketitle

\begin{abstract}
\noindent 
\small
Bayesian response adaptive clinical trials   are currently evaluating experimental therapies for several diseases.  
Adaptive decisions, such as pre-planned  variations  of the   randomization probabilities,  attempt  to  accelerate  the  development of  new  treatments.  The  design of response adaptive  trials,   in  most cases,  requires time consuming  simulation studies to describe  operating characteristics, such as type I/II error rates, across plausible scenarios. We investigate large sample approximations of pivotal operating characteristics in Bayesian Uncertainty directed trial Designs (BUDs).  A BUD trial  utilizes an   explicit   metric  $u$ to quantify the   information  accrued  during  the  study  on   parameters of  interest, for  example  the treatment effects.  The randomization  probabilities  vary  during  time  to  minimize  the  uncertainty  summary  $u$  at  completion of the study.       

We  provide  an asymptotic analysis  (i)  of the  allocation of patients to treatment arms and  (ii)  of the randomization probabilities.   
For  BUDs with outcome distributions belonging to the natural exponential family with quadratic variance function, we  illustrate the asymptotic normality of    the  number of  patients  assigned to  each  arm  and of the randomization probabilities. 
We use these   results   to  approximate  relevant  operating  characteristics such as  the power of the BUD. 
We evaluate  the accuracy of     the   approximations through simulations under several scenarios for  
binary, time-to-event and continuous outcome models. 

\end{abstract}

\newpage
\section{Introduction}  
Randomized clinical trials (RCTs) are essential to demonstrate  the efficacy of novel experimental therapies \citep{council1948}.  
The  landscape of clinical studies has  changed during the last decades, 
 with an increasing     number  of  trials   that  utilize adaptive designs, 
   in some cases to evaluate several experimental treatments  in  biomarker-defined subpopulations \citep{ barker2009spy, ventz2017bayesian}.
Adaptive   designs are attractive to 
reduce  the duration of the study and to allocate  efficiently limited resources  \cite{berry2004bayesian}.    
Most  adaptive designs use  data  generated  during the clinical trial for interim  decisions \citep{berry2004bayesian}, 
for example to vary   the randomization probabilities  during the study \citep{barker2009spy, berry1995adaptive, ventz2017bayesian, zhou2008bayesian} or to discontinue  the  evaluation of an experimental treatment  \cite{ventz2017bayesian}.
In multi-arm studies adaptive randomization algorithms unbalance  the  randomization probabilities, in most cases, towards the most  promising   treatments.
This  can increase power compared to balanced randomization,   
and   it can reduce the overall   sample size  necessary to test  experimental treatments \cite{wason2014comparison}.
 Adaptive  randomization procedures have been developed for  several  designs, 
 including multi-arm studies  \citep{berry1995adaptive, berry2010bayesian},  platform  and basket studies  \citep{ barker2009spy, ventz2017bayesian, zhou2008bayesian}. 

The decision theoretic  paradigm has  been  used   to  develop  trial designs 
\citep{berry1978modified, berry1985bandit, ding2008bayesian}. The  study
 aims  and costs  are  represented by  a utility 
 function  $u(\cdot)$
  of the data $\Sigma$  generated during the trial and the study design $d$.  
Using a Bayesian joint  model  for   patient   profiles, outcomes  and  other  key  variables,   
candidate  designs $d$ can be compared by computing    their expected utility $E[u(\Sigma,d)].$ 
The   optimal design   maximizes  $E[u(\Sigma,d)]$ among all candidate designs. 
    %
    Several approximations of  the  described   optimization    have  been proposed.
 For  example,   \cite{ventz2018} discussed  {\it Bayesian Uncertainty directed trial Designs} (BUDs), 
a  class of approximate decision theoretic designs.
The utility function $u$ in BUDs   coincides with  an  information metric. 
In different words, the goal is to  minimize  uncertainty at completion of the study.
     BUDs  for    dose-finding   and basket trials  have been discussed in \cite{domenicano2019bayesian} and \cite{trippa2016bayesian}. 
Previous  work,  related to  BUDs, 
proposed information-based sampling schemes \citep{berry2002adaptive, muller2006bayesian, russo2017learning}.

There is  a rich literature on large sample analyses of  adaptive  designs. 
For instance,  \citep{  bai1999asymptotic, wei1979generalized}  
studied the  behavior of sequential urn schemes. 
See also \cite{bai2005,ghiglietti2017central,rosenberger2002randomized,zhang2016} for a recent summary on large sample results for urn schemes.
The limiting behavior of adaptive  biased coin designs have been investigated, among others,  
by  \cite{eisele1995central, hu2004asymptotic} and  \cite{hu2009efficient}.
Relevant work connecting stochastic approximation with  response-adaptive clinical trials include  \cite{bartroff2010approximate} and \cite{laruelle2013randomized}. 

In this  manuscript  we  focus on the asymptotic characteristics of  BUDs.
The design of adaptive clinical  trials requires  the  estimation of pivotal operating characteristics, 
such as type I and II  error rates and the distribution of patients randomized  to each  treatment arm. 
In most cases  these  estimates  are  based on   time consuming 
 Monte Carlo simulations, conducted for  different candidate  designs and varying key parameters,  including   sample sizes, enrollment  rates, and outcome distributions.
Approximations  of the operating characteristics, beyond    simulations,  using asymptotic results, are  crucial to compare  designs across plausible scenarios. 
%

 The  need for  computationally  efficient  approximations of design-specific operating characteristics motivates  our study.  
We show the almost sure convergence and asymptotic normality  of the relative allocation of patients to treatment arms in  BUDs.  
We first derive  analytic  results   assuming  that the treatment-specific outcome distributions belong to  natural exponential family \citep{diaconis1979}, and  later  relax this assumption.
In our analysis, we  represent BUD randomization procedures  as stochastic approximations (SAs).   
We study the ordinary differential equations associated with the resulting SAs and 
the stability of the stationary points, following the framework developed in \cite{benveniste2012adaptive} and  using results of   
 \cite{kushner2003stochastic, laruelle2013randomized}.  
%
We  illustrate  through examples   the accuracy  of  the asymptotic   approximations  
by  comparing    asymptotic     and
 Monte-Carlo estimates of  operating characteristics  of BUDs. \\
 Our  asymptotic  results  allow  investigators
  to quickly  approximate the  distribution of the  number  of  patients
  that will be assigned to each  arm and the power of  BUDs.

\section{Trial Design}

We consider a  clinical study
that assigns  $n$ patients sequentially to  $K$  arms.
We use  $A_t \in \mathcal A = \{0, \cdots, K-1\}$  to indicate  the assignment of   individual $t=1, \cdots, n$ to treatment arm $A_t$  and  
$Y_t \in \mathbb  R$  is  the response of individual $t$. We summarize
the accumulated data up to  enrollment $t$  
by $\Sigma_t= \{ \left( A_\ell, Y_\ell  \right); \ell \leq t\}$.

The BUD
 is defined by  first specifying a Bayesian model.  
Outcomes are  conditionally independent  $Y_t \sim f_{\theta}(Y_t|A_t)$ and
$\theta \sim \pi(\theta)$  indicates  the prior for the unknown parameter    $\theta$.
%
The  function $u(\cdot)$    
translates  the posterior distribution $\pi(\theta| \Sigma_t)$  into  utilities,
and we use it  to   quantify the information $u(\Sigma_t)$ generated by the experiment up to stage $t$.
Large values of $u(\Sigma_t)$ correspond to  low  uncertainty levels.
The utility  $u(\cdot)$ in a BUD is a convex functional of the posterior distribution of $\theta$, for example
$u(\Sigma_t)= -\text{Var}(\theta |\Sigma_t)$.
By Jensen's inequality,   the information, on average,  increases with each enrollment,
\begin{small}
\begin{align}\label{Increment}
\Delta_t(a) & := E \left[ u(\Sigma_{t+1}) | A_{t+1}=a, \Sigma_t \right] - u(\Sigma_t) \geq 0,
\end{align}\end{small}
for  every $a \in \mathcal A$.
The myopic and non randomized policy  
$A_{t+1}  = \mathop{\arg \max}_{a \in \mathcal A }  \Delta_{t}( a ),$
which is  often inappropriate for clinical experiments \citep{berry1995adaptive},
is  relaxed in BUDs by   a  randomized  version,  with randomization  probabilities
\begin{align}\label{BUD::General}
p_{t,a} := p( A_{t+1} = a \mid  \Sigma_t ) \propto \Delta_{t}(a)^{h},
\end{align}
where $h\geq 0$ is a  tuning  parameter.  The  randomization probabilities  coincide  with  the  myopic policy   when $h\rightarrow \infty$,
while with   $h=0$ the  randomization  probabilities  become   identical across arms.

\subsection{Outcome distributions within the  natural exponential family  }
We will  focus on outcome distributions $ f_{\theta}$    
in the natural exponential family (NEF) \citep{bernardo2009},
\begin{equation}
f_{\theta}(y | A_t=a) = f_{\psi_a}(y)
\propto \exp\{y\psi_a-b(\psi_a)\}, \label{exponentialparametrization}
\end{equation}
where $\psi_a \in \mathbb R$ is the canonical parameter and
$b(\cdot)$ is the cumulant transform.
We indicate the  mean  with
$
\theta_a=E_{\psi} [ Y_t |A_t=a ] = b'(\psi_a)
$
and we  use  the  equivalent  parametrization $ f_{\psi_a}$ and $ f_{\theta_a}$ interchangeably. We use independent conjugate prior distributions \citep{diaconis1979} for $\psi_a$,
\begin{small}
\begin{equation}
\pi( \psi_a\mid n_{0,a},\yza)
\propto  \exp\{ n_{0,a} \tilde{y}_{0,a} \psi_a-n_{0,a} b(\psi_a)\}, \label{conjugateexp}
\end{equation}
\end{small}
with hyper-parameters $n_{0,a}>0$ and $\tilde y_{0,a} \in \mathbb R$.
The posterior distribution for $\psi=(\psi_0, \cdots, \psi_{K-1})$
is $\pi(\psi \mid \Sigma_t)=\prod_{a=0}^{K-1} \pi(\psi_a \mid \Sigma_t)$,
where  $\pi(\psi_a \mid \Sigma_t)$ has the  same form as  (\ref{conjugateexp})
with updated parameters $n_{t,a}=n_{0,a}+t \hat p_{t,a}$ and
$\tilde y_{t,a} =\big( n_{0,a}  \tilde y_{0,a}+\sum_{s=1}^t Y_{s} 1{(A_s=a)} \big) \big / n_{t,a}$.
Here
$\hat p_{t,a}$ is the proportion of patients assigned to treatment $a$ by time $t$ and
$1(A_s=a)=1$ if  patient $s$  received  treatment $a$ and zero otherwise.    Let $\sigma_a^2=\int y^2 f_{\psi_a}(y) dy -\left(\int y f_{\psi_a}(y) dy \right)^2.$\\
We  consider  
\begin{equation}
u(\Sigma_t) = - \sum_{a=0}^{K-1} \text{Var}(\theta_a | \Sigma_t),  \label{u:va}
\end{equation}
and the expected information increment is
\begin{align}\label{Increment}
\Delta_t(a)= \text{Var}(\theta_a\mid \Sigma_t)
- E[\text{Var}(\theta_a\mid \Sigma_{t+1})\mid A_{t+1}=a,\Sigma_t].
\end{align}
We recall a  useful result from the  literature  on conjugate  Bayesian models \citep{bernardo2009, diaconis1979},
$\tilde{y}_{t,a}=E(\theta_a\mid \Sigma_t )$.  
Since $A_{t+1}$ and  $\theta_a$ are  conditionally independent,  
given $\Sigma_t$,  
the information gain equals
\begin{align*}
\Delta_t(a)
&= \text{Var}(E(\theta_a\mid \Sigma_{t+1})\mid A_{t+1}=a,\Sigma_t) \label{VarE}\\
&=\text{Var}\left(\frac{n_{0,a}+ \tilde y_{0,a} +\sum_{s=1}^{t+1}Y_{s}1(A_s=a)}{n_{0,a}+t\hat p_{t,a}+1}\mid A_{t+1}=a, \Sigma_t\right) \\
&=\text{Var}\left(\frac{Y_{t+1} }{n_{0,a} +t\hat p_{t,a}+1}\mid A_{t+1}=a, \Sigma_t\right),
\end{align*}
where the first  equality follows from  the law of total variance.
We can  therefore write  
\begin{align*}
\Delta_t(a)=\frac{\sigma_{t,a}^2}{(n_{0,a}+t\hat p_{t,a}+1)^2},
\end{align*}
where   $\sigma_{t,a}^2=\text{Var}(Y_{t+1}\mid A_{t+1}=a,\Sigma_t)$.

\section{Asymptotic properties}\label{Ref::Asymptotic:properties} 

We discuss  asymptotic properties of BUDs  with 
 sum of  the (negative) posterior variances of  $\theta_a, a=0, \dots, K-1$, as information measure $u(\Sigma_t)$.
In \cite{ventz2018} a criterion is given for the allocation proportions to have a limit.  Based on this result, we first prove convergence of allocation proportions and randomization probabilities under the assumption that the outcome distributions belong to the natural exponential family.  We then investigate the rate of convergence of these quantities in the case $K=2$.
 
\begin{proposition}\label{Prop:1}
Consider a two arm BUD, $K=2$,
with outcome distribution belonging to the NEF (\ref{exponentialparametrization}),
 conjugates prior (\ref{conjugateexp})
and
 information metric $u(\Sigma_t)$ in (\ref{u:va}).
Then, as {\small$ t \rightarrow \infty,$}\\
(i)  the allocation of patients to treatments $a=0, 1$  converges almost surely (a.s.),
\label{propa}
\begin{equation}
\widehat p_{t,a}  \longrightarrow \rho_a:= \frac{\sigma_a^{\frac{2h}{2h+1}}}{\sigma_0^{\frac{2h}{2h+1}}+\sigma_1^{\frac{2h}{2h+1}}}
 \, \, \text{a.s. as }  \small t \rightarrow \infty \label{limitexp}
 \end{equation}
(ii)  the randomization probability converges a.s. to the same limit,  
\begin{equation} p_{t,a} \longrightarrow \rho_a  \, \, \text{a.s.  as } \small t \rightarrow \infty. \label{limitexp2}
\end{equation}
\end{proposition}

\noindent The proofs of the proposition and of all subsequent results are  included in the Supplementary material. 
The following corollary states the extension of   Proposition 1 for multi-arm settings $K>2$.
\begin{corollary}
Under the same assumptions  of Proposition \ref{propa}, if  $K>2$, then, as {\small$ t \rightarrow \infty,$} the allocation of patients to treatments $(\widehat{p}_{t,0}, \dots, \widehat{p}_{t,K-1})$ and the randomization probabilities $({p}_{t,0}, \dots, {p}_{t,K-1})$ converge a.s. to $(\rho_0, \dots, \rho_{K-1})$, where for $a \in \{0, \dots, K-1\}$
\begin{equation}
\rho_a=\dfrac{\sigma_{a}^{\frac{2h}{2h+1}}}{\sum_{i=0}^{K-1}\sigma_{i}^{\frac{2h}{2h+1}}}.
\end{equation}
 \label{Coro}
\end{corollary}

 We recall that the NEFs  with quadratic variance function consist of all NEFs
  such that $\sigma^2_a= v_0 +v_1 \theta_a+v_2 \theta_a^2$ for  some constants $v_0, v_1, v_2$. In different words, the variance  is a polynomial function of order  $\leq 2$ of the mean  \cite{morris1982}.
This class  contains    models, such as
 the normal, Poisson, gamma, negative binomial and binomial distributions.  
We refer to Morris \cite{morris1982, morris1983} for a detailed study of   this class of distributions.

We  derive the rate of convergence and
show the asymptotic normality of the randomization probabilities $p_{t,a} $ and
of the allocation proportions $\widehat p_{t,a}$ in BUDs  with utility  $u(\Sigma_t) = - \sum_{a=0}^{K-1} \text{Var}(\theta_a | \Sigma_t)$
when $K$ equals $2$ and  the  model $f_{\theta_a}$ belongs to the  NEF  with quadratic variance. 
Lemma \ref{lemma1} 
 approximates, for $a \in \{0,1\}$, the variables $\sigma_{t,a}$ and $ \tilde{y}_{t+1,a} $ with    functions  of     $(\tilde{y}_{t,a},p_{t,1})$ and  $(Y_{t+1}, A_{t+1})$.
For $a=0,1,$ we  use  the  following  notation:
$$\Delta \sigma^2_{t,a}=\text{Var}(Y_{t+2}\mid A_{t+2}=a,\Sigma_{t+1})-\text{Var}(Y_{t+1}\mid A_{t+1}=a,\Sigma_t),$$
$v(\tilde{y}_{t,a})=\left(v_0+v_1\tilde{y}_{t,a}+v_2\tilde{y}_{t,a}^2\right)^{\frac{1}{2}} $,
$W_t = [p_{t,1}, \tilde{y}_{t,1}, \tilde{y}_{t,0}]' \,$
and $\,k_a(W_t) = \left[1+  \left(\dfrac{p_{t,a}}{ p_{t,1-a} }\right)^{\frac{1}{2h}} \dfrac{v(\tilde{y}_{t,1-a})}{v(\tilde{y}_{t,a})} \right].$ \\%
We also write  $X(t)= \mathcal{O}_P(t^{-\alpha})$, 
for $\alpha>0$, if  for all $\epsilon>0$ there exist finite $T,M>0$  such that $P( |X(t)| > M t^{-\alpha})< \epsilon$ for all $t>T.$

\begin{lemma}\label{lemma1}
 If the outcome distributions $f_{\psi_a}, a=0,1$, of the two-arm BUD
belong to the NEF with quadratic variance function,
then 
\begin{itemize}
\item[(i)]$\sigma_{t,a}= v(\tilde{y}_{t,a}) +\mathcal{O}_P(t^{-1})$
\item[(ii)] $\tilde{y}_{t+1,a}= \tilde{y}_{t,a}+(Y_{t+1}-\tilde{y}_{t,a}) \dfrac{1(A_{t+1}=a) }{t} k_a(W_t) +\mathcal{O}_P(t^{-2})$
\item[(iii)]$\Delta \sigma_{t,a}^2=(v_1+2v_2\tilde{y}_{t,a}) (Y_{t+1}-\tilde{y}_{t,a})
 \dfrac{1(A_{t+1}=a)}{t} k_a(W_t) +\mathcal{O}_P(t^{-2}).$
\end{itemize}
\end{lemma}

In spirit  to the previous result, the next Lemma illustrates  that $p_{t+1,1}$ can be  approximated
by  a  function of  $W_t, Y_{t+1}, A_{t+1}$   with an  $ \mathcal{O}_P(t^{-2})$ error  term.

\begin{lemma}\label{lemma2}
 Let the outcome distributions $f_{\psi_a}, a=0,1$ of the two-arm BUD belong to the NEF with quadratic variance function. 
 For the  randomization probabilities $p_{t,1}, t\geq 1$ it holds that
{\small
\begin{align}
p_{t+1,1}= p_{t,1}+
 h p_{t,1} (1-p_{t,1})
  \Bigg \{ & \left[  
\dfrac{(v_1+2v_2\tilde{y}_{t,1})(Y_{t+1}-\tilde{y}_{t,1}) }
{v(\tilde{y}_{t,1})^2} -2\right] \dfrac{A_{t+1}}{t} k_1(W_t)  \nonumber \\
+\! &  \left[ 2- \dfrac{(v_1+2v_2\tilde{y}_{t,0})(Y_{t+1}-\tilde{y}_{t,0})}{v(\tilde{y}_{t,0})^2} \right] \dfrac{(1-A_{t+1})}{t} k_0(W_t)
 \Bigg \} 
+  \mathcal{O}_P(t^{-2}). &
\label{eqlemma1}
\end{align}
}
\end{lemma}
The two Lemmas above suggest how to approximate $\tilde{y}_{t+1,a}- \tilde{y}_{t,a}$ for $a\in\{0,1\}$ and  $p_{t+1,1}-p_{t,1}$.  
For $t\geq 1$, we define  the random vector $\tilde{G}_{t+1}=[G_{t+1}, G_{t+1,1}, G_{t+1,0}]'$, whose components are approximations of  $t(p_{t+1,1}-p_{t,1})$ and $t(\tilde{y}_{t+1,a}- \tilde{y}_{t,a})$, respectively, where
\begin{align*}
G_{t+1} :=
h p_{t,1} (1-p_{t,1})
  \Bigg \{ & \left[  
\dfrac{(v_1+2v_2\tilde{y}_{t,1})(Y_{t+1}-\tilde{y}_{t,1}) }
{v(\tilde{y}_{t,1})^2} -2\right] A_{t+1} k_1(W_t) + \nonumber \\
+\! &  \left[ 2- \dfrac{(v_1+2v_2\tilde{y}_{t,0})(Y_{t+1}-\tilde{y}_{t,0})}{v(\tilde{y}_{t,0})^2} \right]  (1-A_{t+1}) k_0(W_t)
 \Bigg \}, 
 \end{align*}
and
$G_{t+1,a} := 1(A_{t+1}=a) (Y_{t+1}-\tilde{y}_{t,a}) k_a(W_t)$ for  $a=0,1$.
By computing  the conditional expectations $\tilde{g}(W_t)= -E_\psi(\tilde{G}_{t+1}\mid \Sigma_t)$  we  define the map $\tilde{g}(\cdot)= [g(\cdot), g_1(\cdot), g_0(\cdot)]'$, whose components are 
\begin{align*}
g(W_t) & :=
-{2h} \dfrac{v(\tilde{y}_{t,1})}{v(\tilde{y}_{t,0})} \dfrac{(1-p_{t,1})^{\frac{2h+1}{2h}}}{ p_{t,1}^{\frac{1-2h}{2h}}} k_1(W_t)^2 \left( \dfrac{1}{ k_1(W_t)  }- p_{t,1} \right)-h p_{t,1}(1-p_{t,1})\times \\ &\left[  p_{t,1} \dfrac{   (v_1+2v_2\tilde{y}_{t,1})(b'(\psi_1)-\tilde{y}_{t,1})   }{v(\tilde{y}_{t,1})^2} k_1(W_t)  -  
(1-p_{t,1}) \dfrac{ (v_1+2v_2\tilde{y}_{t,0})(b'(\psi_0)-\tilde{y}_{t,0})}{v(\tilde{y}_{t,0})^2} k_0(W_t)  \right] ,\\
g_a(W_t)&:= - p_{t,a}(b'(\psi_a)-\tilde{y}_{t,a}) k_a(W_t)\quad \text{for } a \in \{0,1\}.
\end{align*}
In Proposition \ref{propb} we rewrite $t(W_{t+1}-W_t)$ as the sum of (i) a function of $W_t$, (ii) a $\Sigma_t$ -martingale-difference sequence $\Delta \tilde{M}_{t+1}$  and (iii) a $\Sigma_{t+1}$-measurable sequence of remainder terms. In particular, $\Delta \tilde{M}_{t+1}= [\Delta M_{t+1},\Delta {M}_{t+1,1},\Delta {M}_{t+1,0}]' $ is defined by $\tilde{G}_{t+1}+ \tilde{g}(W_t)$.
 \begin{proposition}
\label{propb}
 Let the outcome distributions $f_{\psi_a}, a=0,1$, of the two-arm BUD, with 
  information metric $u(\Sigma_t)$ in (\ref{u:va}),
 belong to the NEF with quadratic variance function.
Then, 
\begin{equation}
W_{t+1}= W_t - \dfrac{1}{t}\tilde{g}(W_t)+ \dfrac{1}{t}(\Delta \tilde{M}_{t+1}+\tilde{r}_{t+1}), \label{SAexp}
\end{equation}
where the reminder terms $\tilde{r}_{t+1}:=[r_{t+1}, r_{t+1,1}, r_{t+1,0}]$ are three $\mathcal{O}_P(t^{-1})$ sequences. \\
\end{proposition}
Using Proposition \ref{propb} we  use  the theory of stochastic  approximation  
 \cite{ benveniste2012adaptive, kushner2003stochastic} to  derive the asymptotic distribution 
 of $W_t$. 
Following the stochastic  approximation  framework, 
we consider equation \eqref{SAexp}  together with the ordinary differential equation (ODE) 
\begin{equation}
\frac{d W_t}{d t} = - \tilde{g}(W_t), \label{oode}
\end{equation} 
 where  $t \in (0, +\infty)$ denotes continuous time.  The ODE has  arbitrary initial conditions. 
 Note  that  if we  ignore  the  residual term $\tilde{r}_{t+1}$,  the  difference $W_{t+1}-W_t$ in \eqref{SAexp}  is  equal  to  $-\frac{1}{t}\tilde{g}(W_t)$  plus  a $\Sigma_t$-martingale-difference sequence. 
We  describe the distribution of $W_t$, for  large  $t$,  using on an 
asymptotic analysis of the ODE \eqref{oode}. 
By 
identifying the stationary point $[\rho_1, b'(\psi_1), b'(\psi_0)]$ of the ODE, 
assessing its stability, and  using some regularity conditions on $\Delta \tilde{M}_{t+1}$ and $\tilde{r}_{t+1}$, 
we prove a central limit type result for $W_t$.  
In particular,  Theorem \ref{clt} indicates the asymptotic normality of the randomization probability $p_{t,1}$.

\begin{theorem}\label{clt}
Under the assumptions of Proposition \ref{propb}, 
$t^{1/2}(p_{t,1} - \rho_1) \rightarrow \mathcal{N}\Big(0, \dfrac{\Gamma}{1+4h} \Big),$
as $t\rightarrow +\infty$,
where
\begin{small}
\begin{equation}
\Gamma = h^2 \rho_1^2(1-\rho_1)^2\left[\dfrac{(v_1+2v_2b'(\psi_1))^2}{\rho_1\sigma_1^2}+  \dfrac{(v_1+2v_2b'(\psi_0))^2}{(1-\rho_1)\sigma_0^2} +\dfrac{4}{\rho_1}+\dfrac{4}{1-\rho_1}\right].\label{Gamma}
\end{equation}
\end{small}
\label{theo}
\end{theorem}
The following corollary 
verifies the asymptotic normality
of  the relative allocation $\hat{p}_{t,1}$ by applying the Delta method and Slutsky's Theorem.

\begin{corollary}\label{clt2}
Under the assumptions of Theorem \ref{theo}, as $t\rightarrow +\infty$,
 $$t^{1/2}(\hat{p}_{t,1} - \rho_1) \longrightarrow \mathcal{N}\Big(0, \frac{\Gamma}{4h^2(1+4h)} + \dfrac{\rho_1(1-\rho_1)^2}{4\sigma_1^2}(v_1+2 v_2 b'(\psi_1))^2+ \dfrac{\rho_1^2(1-\rho_1)}{4\sigma_0^2}(v_1+2 v_2 b'(\psi_0))^2 \Big).$$
 \label{prop4}
\end{corollary}


\section{Applications  and   examples}
We apply the results in the previous section  to the design of clinical trials. 
We consider three common  outcomes,  binary, time to event  and  continuous  outcomes. \\
\medskip \\
{\it Binary outcomes.} 
For   $Y_t \in \{0,1\}$, we  use the  Bernoulli model  
 $f_{\psi_a}(1) = 1-f_{\psi_a}(0) = \theta_a$, $\theta_a=1/(1+e^{-\psi_a})$,
and conjugated  prior
$\theta_a  \sim \mbox{Beta}(  \alpha,  \beta).$
The outcome variance  $\sigma^2_a $ in expression (\ref{limitexp}) is  $  \theta_a  (1-\theta_a)$, and
the parameters of the quadratic variance function in \eqref{Gamma}  are $v_1=1$ and  $v_2=-1$.
Therefore, $t^{1/2}(\hat{p}_{t,1} - \rho_1)$ converges in distribution to  a  mean zero Gaussian variable  
with variance
$$
\dfrac{\rho_1^2(1-\rho_1)^2}{4}
\left[ \left(  \dfrac{(1-2\theta_1)^2}{\rho_1\sigma_1^2} + \dfrac{(1-2\theta_0)^2}{(1-\rho_1)\sigma_0^2}\right)\left(1+\dfrac{1}{1+4h} \right)  +\dfrac{4}{\rho_1(1+4h)}+\dfrac{4}{(1-\rho_1)(1+4h)} \right].
$$

\begin{figure}[htbp]
\includegraphics[scale=.8]{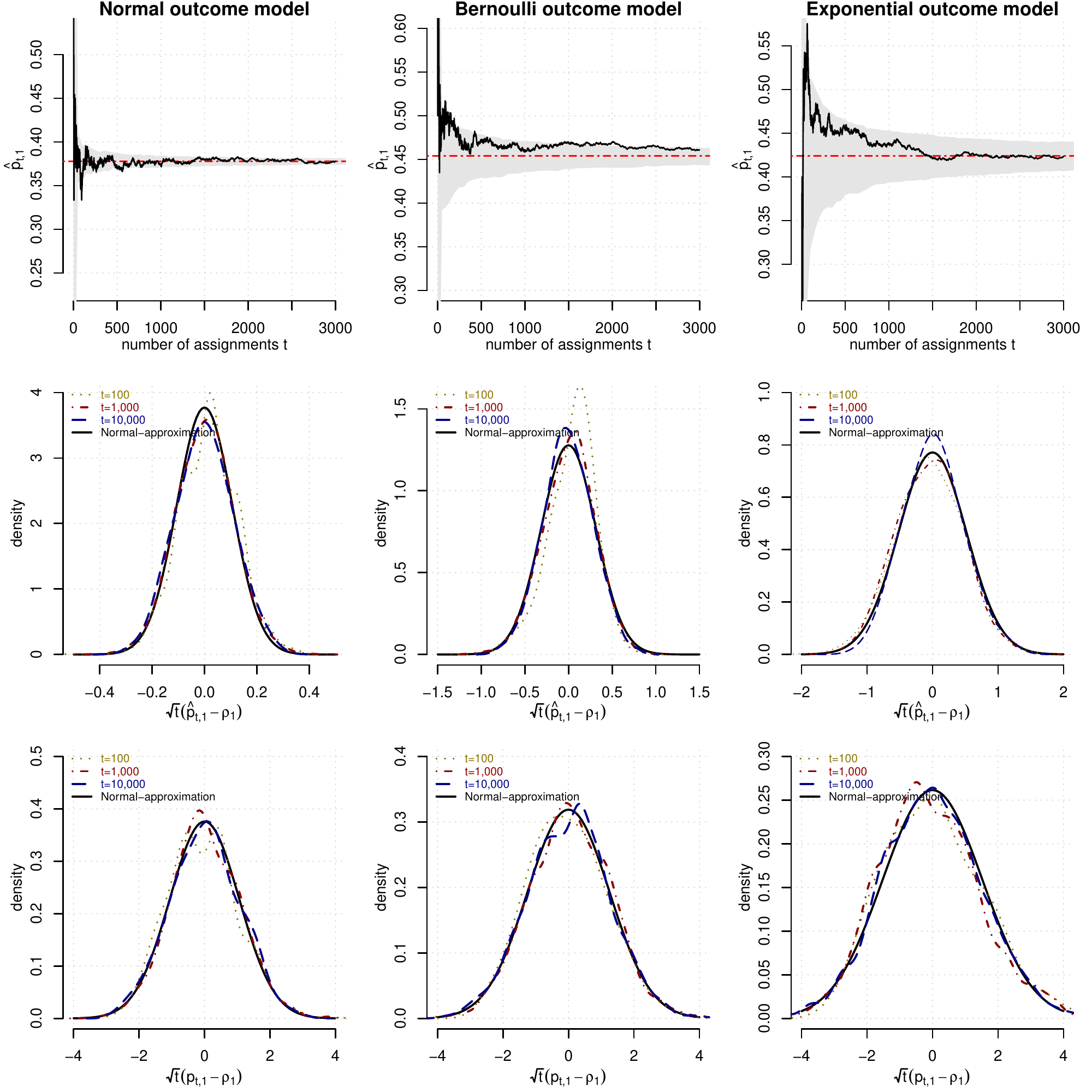}
\caption{The panels in the first row compare $\hat p_{t,1}$ with the limit $\rho_1$ (red line) in  each  of the three  examples (binary, continuous and time to event  outcomes).  The other panels compare  asymptotic  and  empirical distributions  of   randomization probabilities 
$p_{t,1}$  and  allocation proportions  $\hat{p}_{t,1}$. The  empirical  distributions  in  each  of the three  examples (binary, continuous and time to event  outcomes)  are  based  on 1000 simulations  of the two-arm BUD  trial.}\label{Fig-1}
\end{figure}

The top panel of the second column of  Figure  \ref{Fig-1} shows a trajectory $\hat{p}_{t,1}, t=1, \cdots, 10,000$ 
for a single simulated two-arm BUD trial (black curve).
The  response  probabilities $(\theta_0, \theta_1) $  are  set  equal to $0.2$ and $ 0.4$. 
We used  $\alpha =  \beta=2$ and $h=5$.
The shaded area shows (point-wise at each $t$) upper and lower 2.5\% quantiles of the distribution of  $\hat{p}_{t,1}$ across 1,000 simulations. 
The second row illustrates the distribution of $t^{1/2}(\hat{p}_{t,1} - \rho_1)$ across 1000 simulations of the two-arm BUD trial.  The  empirical  
     distribution of   $t^{1/2}(\hat{p}_{t,1} - \rho_1)$ has been smoothed with a kernel density estimator.  
      The panel  compares  the  $\mathcal{N}(0, 0.097)$  density (asymptotic approximation)  to the empirical  distribution of $t^{1/2}(\hat{p}_{t,1} - \rho_1)$ across simulations, when $t=100, 1000$ and $10,000$.
The last row compares the  empirical  distribution distribution of   $t^{1/2}(p_{t,1} - \rho_1)$   
to the $\mathcal{N}\Big(0, \dfrac{\Gamma}{1+4h} \Big)$ density.\\

\medskip
{\it Time-to-event outcomes.} We consider 
an exponential model  $f_{\psi_a}(y)=\exp\{-y \psi_a\} \psi_a, y\ge 0$ with mean $\theta_0 =1/\psi_a$, 
and we use the conjugated  gamma prior $\psi_a \sim \mbox{Gamma}( \alpha, \beta )$.
The outcome variance  $\sigma^2_a  $ in  expression (\ref{limitexp}) is $ 1/ \psi_a^2$, the parameters of  the quadratic variance function in \eqref{Gamma}  are $(v_1,v_2)=(0,1)$ and  $h=5$.
Therefore, the asymptotic  variance of $t^{1/2}(\hat{p}_{t,1} - \rho_1)$  is
$$
\rho_1^2(1-\rho_1)^2 \left(   \dfrac{1}{\rho_1}+\dfrac{1}{1-\rho_1}  \right)\left( \dfrac{2}{1+4h}+1 \right).
$$
The third column of Figure \ref{Fig-1} compares 
  the asymptotic   and empirical distributions of $t^{1/2}(\hat{p}_{t,1} - \rho_1)$ and $t^{1/2}(p_{t,1} - \rho_1)$, based on 
 1000 simulations of the BUD trial.  In this example 
$(\theta_0, \theta_1)=(5, 7)$ and $( \alpha, \beta )=(3,3).$\\

\medskip
{\it Continuous outcomes.}  We consider a normal outcome model $\mathcal{N}(\theta_a, \sigma^2_a)$ 
 with known variance $ \sigma^2_a$.
We use a conjugated prior  $\theta_a  {\sim}  \mathcal{N}(0, v^2_{0,a}).$ 
In this case  $v_1=v_2=0$, $h=5$, and 
\begin{equation}
t^{1/2}(\hat{p}_{t,1} - \rho_1) \underset{{\small t \rightarrow \infty}}{\longrightarrow}\mathcal{N}(0, \dfrac{\rho_1(1-\rho_1)}{1+4h} ). \label{convnorm}
\end{equation}
Column 1 of  Figure \ref{Fig-1}  illustrates  the empirical distribution of  $t^{1/2}(\hat{p}_{t,1} - \rho_1)$,
$t= 100, 1000$ or $10 000$,
and the  normal approximation.\\
\medskip

{\it Power analysis and sample size selection.} 
We use the  results  in Section \ref{Ref::Asymptotic:properties}   to select the sample size of  BUD studies accordingly to the targeted type I and II error rates. 
We approximate  the  power function of  
the BUD  under fixed scenarios using  Corollary \ref{clt2}.

We assume  that  the  primary aim of the clinical trial is to test   
the one-sided null hypothesis  $H_0: \theta_{0,1} =\theta_{0,0}$ against the  alternative   $H_1: \theta_{0,1} > \theta_{0,0}$. 
We verified  (Supplementary material) that, under the sequential BUD design, 
the maximum-likelihood estimates (MLE) 
$\widehat \theta_{t,a} $ 
of the unknown true mean response $\theta_{0,a}$ to treatment 
$a=0,1$ within the NEF of outcome models
 have the same limiting distribution as the MLE of a study design  with fixed  arm-specific  sample sizes, i.e.
\begin{equation}
t^{1/2} \begin{bmatrix} \hat{\theta}_{t,0} - \theta_{0,0} \vspace{0.15 cm} \\ \hat{\theta}_{t,1} - \theta_{0,1}
\end{bmatrix}  \underset{{\small t \rightarrow \infty}}{\longrightarrow}\mathcal{N} \left( \bm{0}, Diag(\eta_{0,0}, \eta_{0,1}) \right), \label{mle}
\end{equation}
where 
 $\eta_{0,a}:=\left(\rho_a I_{\theta_{0,a}}\right)^{-1} $ and  $I_{\theta_{0,a}}$ is the Fisher information of $f_{\theta_{0,a}}$. 

We use a standard Wald-statistics,  
$ Z_a  = \dfrac{ \sqrt{t}  \times ( \hat{\theta}_{t,1}-\hat{\theta}_{t,0}) }{  \sqrt{ \widehat \eta_{t,a}+\widehat \eta_{t,1} } },$
 where  $ \widehat \eta_{t,a} = 1/ (\widehat \rho_{a} \times  I_{ \widehat  \theta_{t,a}})$,
and  the MLE  $\widehat \sigma^2_a$ for $\widehat \rho_a =  \rho_a( \widehat \sigma^2_a)$ in (\ref{limitexp}) 
to test $H_{0}$. 
The power function of the BUD design is approximated by  
$\Phi \Big(  z_{1-\alpha}  -  \dfrac{ \sqrt{t} (\theta_{0,1} -\theta_{0,0}) }{ \sqrt{\eta_{0,1} + \eta_{0,0}} }  \Big)$
where $\Phi (\cdot)$ is the cumulative distribution function of a 
standard normal random distribution  and  $\Phi(z_{1-\alpha}) = 1-\alpha.$
Therefore $\widehat t _{1-\alpha, 1-\beta} =  \dfrac{ ( z_{1-\alpha} - z_{1-\beta}   )^2 (\eta_{0,0} + \eta_{0,1})}{ (\theta_{0,1} -\theta_{0,0})^2 }$ approximates the sample size of the BUD study  to  achieve   
a   power  equal to 1-$\beta$ and type I error rate $\alpha$.

\begin{figure}[htbp]
\includegraphics[scale=.8]{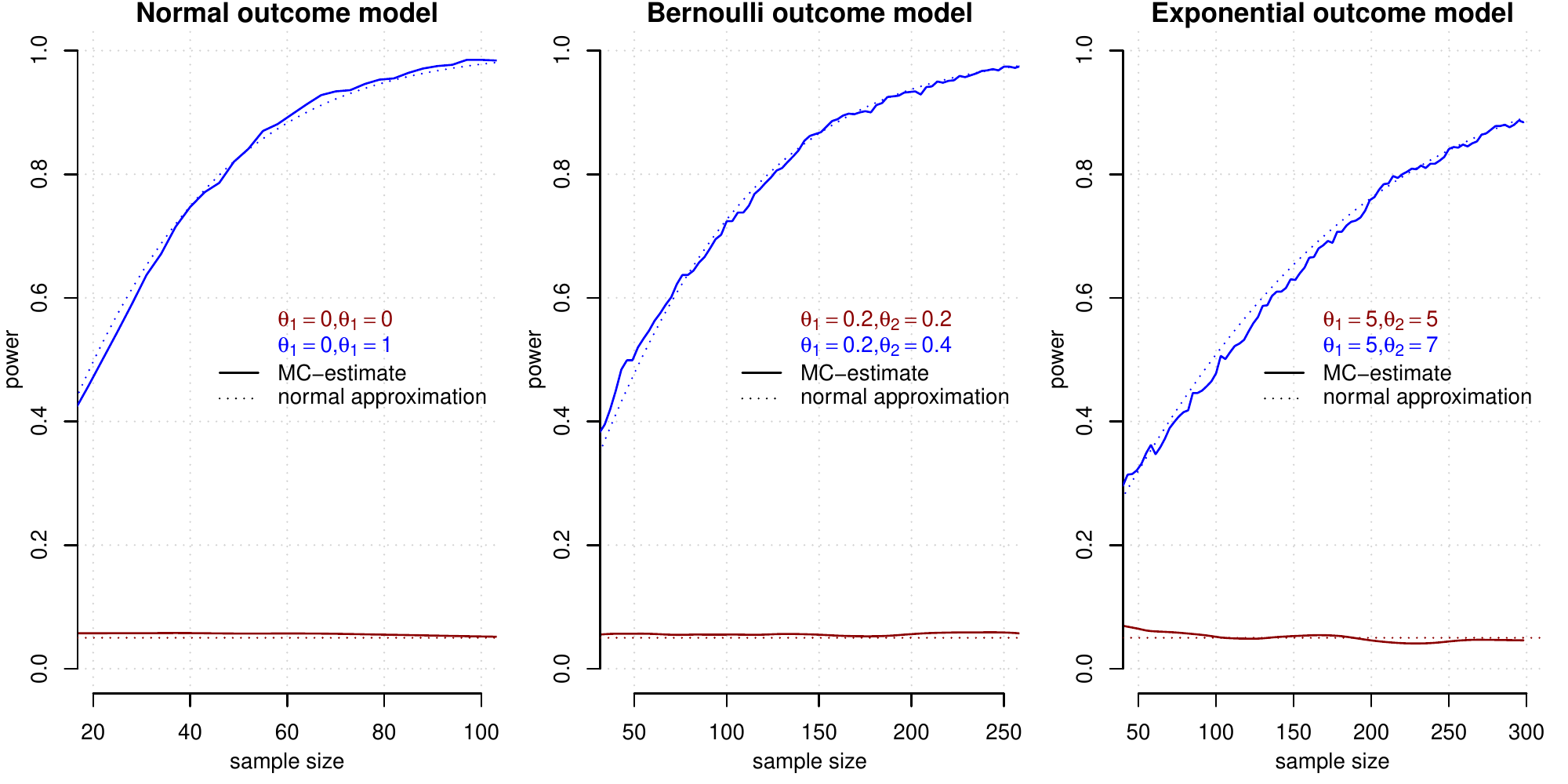}
\caption{
Power  (blue lines) and type I error (red lines): comparison of estimates based on asymptotic approximations (dotted lines)
and  standard  Monte Carlo  simulations (1000 simulated trials, solid lines), for binary, continuous and time-to-event outcomes.
}
\label{Fig-2}
\end{figure}

 Figure \ref{Fig-2}  compares, for  three BUD  designs (binary,  time-to-event and continuous outcomes), 
   power  estimates    based   on  asymptotic approximations 
(blue dotted lines)   and  on    Monte Carlo  simulations  (1000 simulated trials, blue solid lines).   
The computational time for  the  simulation-based  calculations   is  orders  of  magnitude larger  than the normal approximation.
We   also  show the  empirical estimates  of the  type I error rates (brown solid lines) 
for   the  outlined  asymptotic testing  procedure with target  type I error rate of $\alpha=0.05$ (brown dotted lines).
%
For the normal outcome model, $\sigma^2_0=1, \sigma_1^2=3$, and $\theta_0 = \theta_1 = 0$  (null  scenario,  brown  lines)
 or  $(\theta_1,\theta_2)=(0,1)$  (positive  treatment   effect,  blue  lines).  
 Similarly,  for  the  Bernoulli and Exponential  models  the  parameter  values  $\theta$
   that  defined    null    (brown  lines)  and  alternative scenarios  (blue  lines)
   are  indicated in the  panels  of  Fig \ref{Fig-2}.

\section{Convergence results beyond the NEF}
We extend  the almost sure convergence of the allocation proportion and randomization probability of a BUD  (Proposition \ref{propa}) to outcome  distributions beyond  the NEF. 
The following Lemma introduces approximations of the information increment (\ref{Increment}) of BUDs  with utility 
$u(\Sigma_t) = - \sum_{a=0}^{1} \text{Var}(\theta_a | \Sigma_t),$
where  $\theta_a$ is not  required to be the mean of the outcomes as in Section 3.
We use   $X(t)= o_P(a(t))$ to indicate that $X(t)/a(t)$ converges to zero in probability.

\begin{lemma}\label{lemma4}
 Consider two-arm BUDs with 
 information metric $u(\Sigma_t)$ in (\ref{u:va}). 
  The parameter space $\Theta\subset \mathbb R$ is a bounded open interval,  the    parameter $\theta_{0,a}$ is an interior point of $\Theta$ for $a \in \{0,1\}$ and  the prior is the uniform distribution on $\Theta$. 
If
(i) $\inf_{y, \theta_a} f_{\theta_{a}}(y) >0$,
(ii) $ \sup_{y, \theta_a} f_{\theta_{a}}(y) < \infty  $, and
(iii) $ \sup_{y, \theta_a}  \bigg \vert \dfrac{\partial^k f_{\theta_a}(y)}{\partial \theta_a^k} \bigg \vert < \infty $ for $k=1,2,3,$
then 
\begin{equation}
\Delta_t(a)= I_{\theta_{0,a}}^{-1} \times (t \hat{p}_{t,a})^{-2}+o_P((t\hat{p}_{t,a})^{-2}). \label{approxus1}
\end{equation}
\end{lemma}

The following proposition states that, under the assumptions of Lemma \ref{lemma4}, 
the asymptotic convergence (\ref{limitexp}) and  (\ref{limitexp2}) also hold outside the NEF.

\begin{proposition}
\label{propc}
Under the assumptions of Lemma \ref{lemma4}, it holds that\\
\begin{equation}
\hat p_{t,a}\longrightarrow  \rho_a:=
\frac{
	I_{\theta_{0,a}}^{-\frac{h}{2h+1}}
}{I_{\theta_{0,0}}^{-\frac{h}{2h+1}}+I_{\theta_{0,1}}^{-\frac{h}{2h+1}}}
\, \, \text{a.s. as }  \small t \rightarrow \infty.
\end{equation}
and
\begin{equation}
p_{t,a}\longrightarrow \rho_a \, \, \text{a.s. as }  \small t \rightarrow \infty.
\end{equation}
for $a \in \{0,1\}$.
\end{proposition}

\noindent Note that assumption (i) of Lemma \ref{lemma4}   implies that the support of the outcome distribution $f_{\theta_a}$ is bounded. 
The regularity conditions of Lemma \ref{lemma4}  can  be  modified, for example 
to cover  settings  where the  outcome support is  unbounded.  
A  list of alternative assumptions is specified in the Supplementary material.

\begin{figure}[htbp]
\includegraphics[scale=.75]{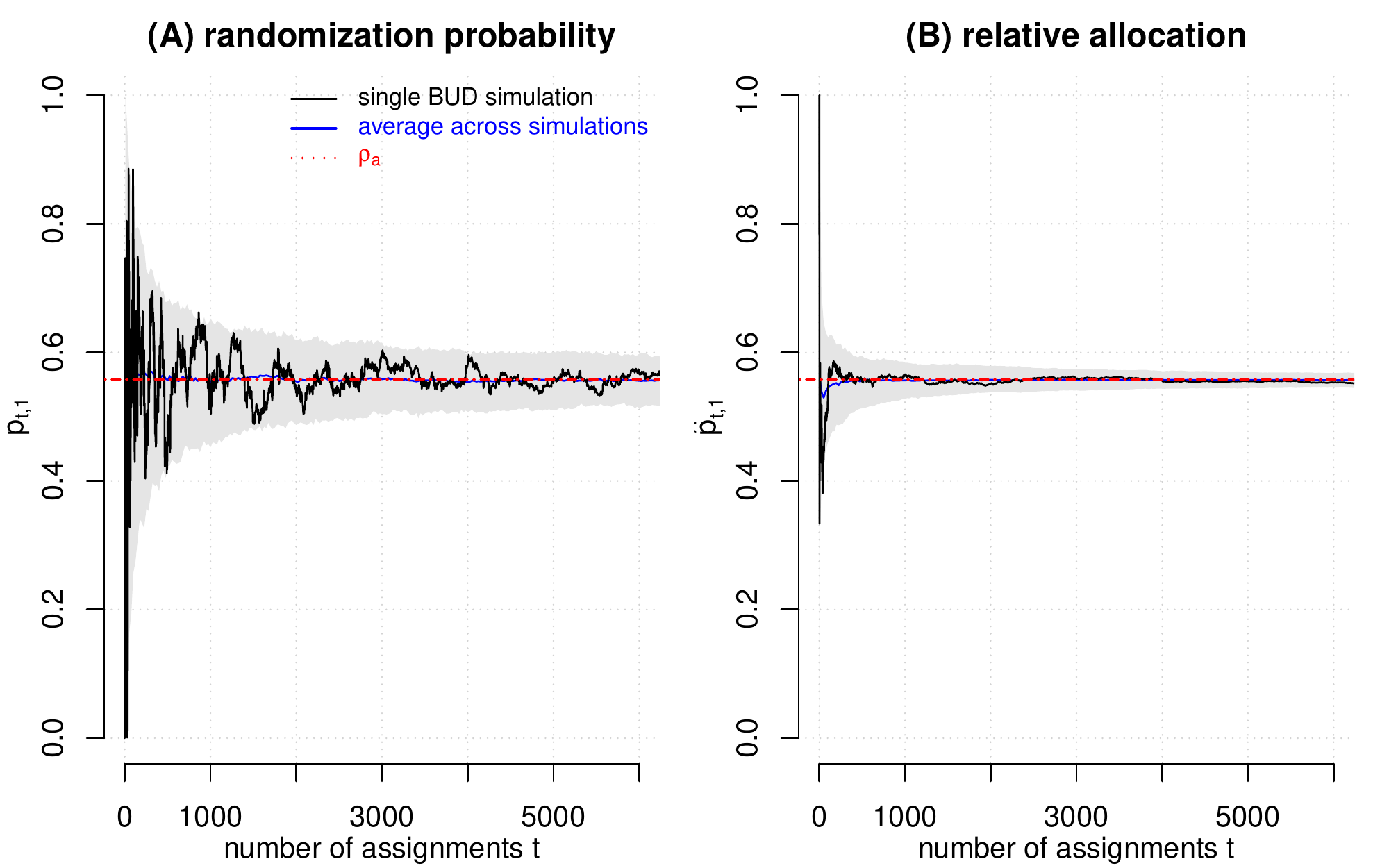}
\caption{Allocation proportions and  randomization probabilities  of a two-arm BUD  design.
 The primary outcomes are modeled with a truncated  Weibull distribution. 
 The average allocation proportion and  randomization probability  across  1000 simulations (blue  lines)  are  close  to their
   limit ($t\rightarrow \infty$, red lines). }\label{Fig-3}
\end{figure}

To illustrate the result
we consider as an outcome  model   $f_{\theta_a}$ beyond  the NEF 
the  truncated Weibull  model
$f_{\theta_a}(y) = 
\dfrac{ e^{ - ( y r )^{\theta_a} } (r y)^{\theta_a - 1} r \theta_a  }{1- e^{ - ( t_0 r )^{\theta_a} } }$,  $y \in (0, t_0)$, 
with unknown shape parameter $\theta_a$ and known rate $r$ parameter.
Panels A and B of Figure  \ref{Fig-3} show, 
similar to Figure  \ref{Fig-1}, 
a trajectory of $p_{t,1}$  (Panel A) 
and $\hat{p}_{t,1}$ (Panel B) $t=1,2 \cdots, 6000$
for a single simulated two-arm BUD trial with $r=1, \theta_0=1, \theta_1=1.5$  (black curve).
The shaded area shows (point-wise at each $t$) upper and lower 2.5\% quantiles of the empirical distribution of 
 $p_{t,1}$ and  $\hat{p}_{t,1}$ across 1,000 simulations.
   The  blue  lines indicate  the means  of $p_{t,1}$ and  $\hat{p}_{t,1}$ across these simulations, while the  red horizontal lines indicate  their  limit $ \rho_a$.

\newpage
\section{Discussion}
Asymptotic analyses of Bayesian adaptive procedures simplify the design of clinical trials 
and reduce the need for time-consuming simulations to evaluate  operating characteristics across potential trial scenarios.
We derived asymptotic results for the randomization probabilities and the allocation proportions of BUDs using stochastic approximation techniques. 
BUD's randomization procedure was expressed as a sequence of recursive equations which  allowed the application of techniques from classical stochastic approximation theory. 
This allowed us to derive a central limit theorem for the relative allocation of patients to treatments  and for the randomization probabilities. 
Potential applications of stochastic approximation theory in the analysis of trial designs have been previously discussed by \citep{laruelle2013randomized}. 
In our work we showed that they allow to  evaluate  major operating characteristics of BUDs. 
We considered for example the variability of the allocation proportions during the trials and the power of the BUD design with a fixed sample size under a parameter  $\theta$ of interest. 
The stochastic approximation framework, as we showed in our examples, enables useful approximations of the patients' assignment variability and  other characteristics.    

\bibliography{Bibliography.bib}

\begin{thebibliography}{10}

\bibitem{bai1999asymptotic}
Zhi-Dong Bai and Feifang Hu.
\newblock Asymptotic theorems for urn models with nonhomogeneous generating
  matrices.
\newblock {\em Stochastic Processes and Their Applications}, 80(1):87--101,
  1999.

\bibitem{bai2005}
Zhi-Dong Bai and Feifang Hu.
\newblock Asymptotics in randomized urn models.
\newblock {\em The Annals of Applied Probability}, 15(1B):914--940, 2005.

\bibitem{barker2009spy}
Anna Barker, Caroline Sigman, G~Kelloff, N~Hylton, DA~Berry, and L~Esserman.
\newblock I-spy 2: an adaptive breast cancer trial design in the setting of
  neoadjuvant chemotherapy.
\newblock {\em Clinical Pharmacology \& Therapeutics}, 86(1):97--100, 2009.

\bibitem{bartroff2010approximate}
Jay Bartroff, Tze~Leung Lai, et~al.
\newblock Approximate dynamic programming and its applications to the design of
  phase i cancer trials.
\newblock {\em Statistical Science}, 25(2):245--257, 2010.

\bibitem{benveniste2012adaptive}
Albert Benveniste, Michel M{\'e}tivier, and Pierre Priouret.
\newblock {\em Adaptive algorithms and stochastic approximations}, volume~22.
\newblock Springer Science \& Business Media, 2012.

\bibitem{bernardo2009}
Jos{\'e}~M Bernardo and Adrian~FM Smith.
\newblock {\em Bayesian theory}, volume 405.
\newblock John Wiley \& Sons, 2009.

\bibitem{berry1978modified}
Donald~A Berry.
\newblock Modified two-armed bandit strategies for certain clinical trials.
\newblock {\em Journal of the American Statistical Association},
  73(362):339--345, 1978.

\bibitem{berry2004bayesian}
Donald~A Berry.
\newblock Bayesian statistics and the efficiency and ethics of clinical trials.
\newblock {\em Statistical Science}, 19(1):175--187, 2004.

\bibitem{berry1995adaptive}
Donald~A Berry and Stephen~G Eick.
\newblock Adaptive assignment versus balanced randomization in clinical trials:
  a decision analysis.
\newblock {\em Statistics in medicine}, 14(3):231--246, 1995.

\bibitem{berry1985bandit}
Donald~A Berry and Bert Fristedt.
\newblock Bandit problems: sequential allocation of experiments (monographs on
  statistics and applied probability).
\newblock {\em London: Chapman and Hall}, 5:71--87, 1985.

\bibitem{berry2002adaptive}
Donald~A Berry, Peter Mueller, Andy~P Grieve, Michael Smith, Tom Parke, Richard
  Blazek, Neil Mitchard, and Michael Krams.
\newblock Adaptive bayesian designs for dose-ranging drug trials.
\newblock In {\em Case studies in Bayesian statistics}, pages 99--181.
  Springer, 2002.

\bibitem{berry2010bayesian}
Scott~M Berry, Bradley~P Carlin, J~Jack Lee, and Peter Muller.
\newblock {\em Bayesian adaptive methods for clinical trials}.
\newblock CRC press, 2010.

\bibitem{casella2002statistical}
George Casella and Roger~L Berger.
\newblock {\em Statistical inference}, volume~2.
\newblock Duxbury Pacific Grove, CA, 2002.

\bibitem{council1948}
Medical~Research Council et~al.
\newblock Streptomycin treatment of pulmonary tuberculosis.
\newblock {\em British Medical Journal}, 2:769--782, 1948.

\bibitem{diaconis1979}
Persi Diaconis and Donald Ylvisaker.
\newblock Conjugate priors for exponential families.
\newblock {\em The Annals of Statistics}, 7(2):269--281, 1979.

\bibitem{ding2008bayesian}
Meichun Ding, Gary~L Rosner, and Peter M{\"u}ller.
\newblock Bayesian optimal design for phase ii screening trials.
\newblock {\em Biometrics}, 64(3):886--894, 2008.

\bibitem{domenicano2019bayesian}
Ilaria Domenicano, Steffen Ventz, Matteo Cellamare, R.~Mak, and Lorenzo Trippa.
\newblock Bayesian uncertainty-directed dose finding designs.
\newblock {\em Journal of the Royal Statistical Society: Series C (Applied
  Statistics)}, 68(5):1393--1410, 2019.

\bibitem{eisele1995central}
Jeffrey~R Eisele and Michael~B Woodroofe.
\newblock Central limit theorems for doubly adaptive biased coin designs.
\newblock {\em The Annals of Statistics}, 23(1):234--254, 1995.

\bibitem{ghiglietti2017central}
Andrea Ghiglietti, Anand~N Vidyashankar, William~F Rosenberger, et~al.
\newblock Central limit theorem for an adaptive randomly reinforced urn model.
\newblock {\em The Annals of Applied Probability}, 27(5):2956--3003, 2017.

\bibitem{hu2006theory}
Feifang Hu and William~F Rosenberger.
\newblock {\em The theory of response-adaptive randomization in clinical
  trials}, volume 525.
\newblock John Wiley \& Sons, 2006.

\bibitem{hu2004asymptotic}
Feifang Hu and Li-Xin Zhang.
\newblock Asymptotic properties of doubly adaptive biased coin designs for
  multitreatment clinical trials.
\newblock {\em The Annals of Statistics}, 32(1):268--301, 2004.

\bibitem{hu2009efficient}
Feifang Hu, Li-Xin Zhang, and Xuming He.
\newblock Efficient randomized-adaptive designs.
\newblock {\em The Annals of Statistics}, 37(5A):2543--2560, 2009.

\bibitem{johnson1970}
Richard~A Johnson.
\newblock Asymptotic expansions associated with posterior distributions.
\newblock {\em The Annals of Mathematical Statistics}, 41(3):851--864, 1970.

\bibitem{kushner2003stochastic}
Harold Kushner and G~George Yin.
\newblock {\em Stochastic approximation and recursive algorithms and
  applications}, volume~35.
\newblock Springer Science \& Business Media, 2003.

\bibitem{laruelle2013randomized}
Sophie Laruelle and Gilles Pag{\`e}s.
\newblock Randomized urn models revisited using stochastic approximation.
\newblock {\em The Annals of Applied Probability}, 23(4):1409--1436, 2013.

\bibitem{morris1982}
Carl~N Morris.
\newblock Natural exponential families with quadratic variance functions.
\newblock {\em The Annals of Statistics}, 10(1):65--80, 1982.

\bibitem{morris1983}
Carl~N Morris.
\newblock Natural exponential families with quadratic variance functions:
  Statistical theory.
\newblock {\em The Annals of Statistics}, 11(2):515--529, 1983.

\bibitem{muller2006bayesian}
Peter M{\"u}ller, Don~A Berry, Andrew~P Grieve, and Michael Krams.
\newblock A bayesian decision-theoretic dose-finding trial.
\newblock {\em Decision analysis}, 3(4):197--207, 2006.

\bibitem{rosenberger2002randomized}
William~F Rosenberger.
\newblock Randomized urn models and sequential design.
\newblock {\em Sequential Analysis}, 21(1-2):1--28, 2002.

\bibitem{russo2017learning}
Daniel Russo and Benjamin Van~Roy.
\newblock Learning to optimize via information-directed sampling.
\newblock {\em Operations Research}, 66(1):230--252, 2017.

\bibitem{trippa2016bayesian}
Lorenzo Trippa and Brian~Michael Alexander.
\newblock Bayesian baskets: a novel design for biomarker-based clinical trials.
\newblock {\em Journal of Clinical Oncology}, 2016.

\bibitem{van2000asymptotic}
A~W Van~der Vaart.
\newblock {\em Asymptotic statistics}, volume~3.
\newblock Cambridge University Press, 2000.

\bibitem{ventz2017bayesian}
Steffen Ventz, William~T Barry, Giovanni Parmigiani, and Lorenzo Trippa.
\newblock Bayesian response-adaptive designs for basket trials.
\newblock {\em Biometrics}, 73(3):905--915, 2017.

\bibitem{ventz2018}
Steffen Ventz, Matteo Cellamare, Sergio Bacallado, and Lorenzo Trippa.
\newblock Bayesian uncertainty directed trial designs.
\newblock {\em Journal of the American Statistical Association},
  114(527):962--974, 2018.

\bibitem{wason2014comparison}
James~MS Wason and Lorenzo Trippa.
\newblock A comparison of bayesian adaptive randomization and multi-stage
  designs for multi-arm clinical trials.
\newblock {\em Statistics in medicine}, 33(13):2206--2221, 2014.

\bibitem{wei1979generalized}
L~J Wei.
\newblock The generalized polya's urn design for sequential medical trials.
\newblock {\em The Annals of Statistics}, 7(2):291--296, 1979.

\bibitem{zhang2016}
Li-Xin Zhang.
\newblock Central limit theorems of a recursive stochastic algorithm with
  applications to adaptive designs.
\newblock {\em The Annals of Applied Probability}, 26(6):3630--3658, 2016.

\bibitem{zhou2008bayesian}
Xian Zhou, Suyu Liu, Edward~S Kim, Roy~S Herbst, and J~Jack Lee.
\newblock Bayesian adaptive design for targeted therapy development in lung
  cancer - a step toward personalized medicine.
\newblock {\em Clinical Trials}, 5(3):181--193, 2008.

\end{thebibliography}

\newpage
\section{Supplementary Material: ``Approximating the Operating Characteristics of Bayesian Uncertainty Directed Trial Designs''}

\begin{proof}
(Proposition~\ref{propa})
It is enough to prove \eqref{limitexp} and \eqref{limitexp2} for $a=1$.
 First, define 
\begin{small}
\[
F_t=-\hat p_{t,1}+
\frac{ 
	\left( \frac{\sigma_{t,1}^2}{(t_0+t\hat p_{t,1}+1)^2}   \right)^h
}{
\left( \frac{\sigma_{t,0}^2}{(t_0+t\hat p_{t,0}+1)^2}   \right)^h
+
\left( \frac{\sigma_{t,1}^2}{(t_0+t\hat p_{t,1}+1)^2}   \right)^h
}
\]
\end{small}
and
\begin{small}
\[
\tilde F_t=-\hat p_{t,1}+\frac{\hat p_{t,1}^{-2h}\sigma_1^{2h}}{\hat p_{t,0}^{-2h}\sigma_0^{2h}+\hat p_{t,1}^{-2h}\sigma_1^{2h}}.
\]
\end{small}
As a function of $\hat p_{t,1}$, $\tilde F_t$ is strictly decreasing. 
The unique root of $\tilde F_t=0$ is
\begin{small}
\begin{equation}
\rho_1:=\frac{\sigma_1^{2h/(2h+1)}}{\sigma_0^{2h/(2h+1)}+\sigma_1^{2h/(2h+1)}}
\end{equation}
\end{small}
Now, we show that $F_t-\tilde F_t$ converges to zero a.s. as $t\rightarrow\infty$.
The proof is based on the following elementary facts:\\
\begin{itemize}
\item[a] If $a_n$, $b_n$, $a_n'$ and $b_n'$ are sequences of positive numbers, then
\begin{small}
\[
\left |
\frac{a_n}{a_n+b_n}
-
\frac{a_n'}{a_n'+b_n'}
\right |\leq 
\min\left(
\left |
\frac{a_n}{b_n}-\frac{a_n'}{b_n'}\right |,
\left | \frac{b_n}{a_n}-\frac{b_n'}{a_n'}\right |
\right)
\]
\end{small}
Indeed
\begin{small}
\[
\left |
\frac{a_n}{a_n+b_n}
-
\frac{a_n'}{a_n'+b_n'}
\right |=
\left |
\frac{1}{1+b_n/a_n}
-
\frac{1}{1+b_n'/a_n'}
\right |
=
\left |
\frac{b_n'/a_n'-b_n/a_n}{(1+b_n/a_n)(1+b_n'/a_n')}
\right |\leq \mid b_n'/a_n'-b_n/a_n\mid
\]
\end{small}
and
\begin{small}
\[
\left |
\frac{a_n}{a_n+b_n}
-
\frac{a_n'}{a_n'+b_n'}
\right |
=
\left |
1-\frac{a_n}{a_n+b_n}
-1+
\frac{a_n'}{a_n'+b_n'}
\right |
=
\left |
\frac{b_n}{a_n+b_n}
-
\frac{b_n'}{a_n'+b_n'}
\right |
\]
\end{small}
\item[b] If $a_n$, $b_n$, $a_n'$ and $b_n'$ are bounded sequences of numbers such that $a_n-a_n'\rightarrow 0$ and $b_n-b_n'\rightarrow 0$, then
$a_nb_n-a_n'b_n'\rightarrow 0$. 
Indeed,
\begin{small}
\[
\mid a_nb_n-a_n'b_n' \mid\leq 
\mid a_nb_n-a_n'b_n+a_n'b_n-a_n'b_n'\mid\leq \mid b_n\mid \mid a_n-a_n'\mid+\mid a_n'\mid \mid b_n-b_n'\mid
\]
\end{small}
\item[c] If $a_n$ and $a_n'$ are bounded sequences such that $a_n-a_n'\rightarrow 0$ and $r$ is a positive real number, then $a_n^r-a_n'^r\rightarrow 0$.
The thesis is obvious if $r=1$. If $r>1$, and $M$ is an upper bound for both sequences, then
\begin{small}
\[
\mid
a_n^r-a_n'^r
\mid\leq 2r M^{r-1}\mid a_n-a_n'\mid
\]
\end{small}
If $r<1$, then
\begin{small}
\[
\mid a_n^r-a_n'^r\mid\leq \mid a_n-a_n'\mid^r
\]
\end{small}
\end{itemize}
\noindent Let us now prove that $F_t-\tilde F_t\rightarrow 0$ a.s.\\
By (a),
\begin{small}
\[
\mid F_t-\tilde F_t\mid \leq \min\left(
\left | \frac{\sigma_{t,0}^{2h}((n_{0,1}+1)/t+\hat p_{t,1})^{2h}}{\sigma_{t,1}^{2h}((n_{0,0}+1)/t+\hat p_{t,0})^{2h}}-\frac{\sigma_{0}^{2h}\hat p_{t,1}^{2h}}{\sigma_{1}^{2h}\hat p_{t,0}^{2h}}
\right |,
\left | \frac{\sigma_{t,1}^{2h}((n_{0,0}+1)/t+\hat p_{t,0})^{2h}}{\sigma_{t,0}^{2h}((n_{0,1}+1)/t+\hat p_{t,1})^{2h}}-\frac{\sigma_{1}^{2h}\hat p_{t,0}^{2h}}{\sigma_{0}^{2h}\hat p_{t,1}^{2h}}
\right |
\right)
\]
\end{small}
Hence
\begin{small}
\[
\begin{aligned}
\mid F_t-\tilde F_t\mid
&\leq 
\left | \frac{\sigma_{t,0}^{2h}((n_{0,1}+1)/t+\hat p_{t,1})^{2h}}{\sigma_{t,1}^{2h}((n_{0,0}+1)/t+\hat p_{t,0})^{2h}}-\frac{\sigma_{0}^{2h}\hat p_{t,1}^{2h}}{\sigma_{1}^{2h}\hat p_{t,0}^{2h}}
\right | 1_{(\hat p_{t,0}>1/2)}+
\left | \frac{\sigma_{t,1}^{2h}((n_{0,0}+1)/t+\hat p_{t,0})^{2h}}{\sigma_{t,0}^{2h}((n_{0,1}+1)/t+\hat p_{t,1})^{2h}}-\frac{\sigma_{1}^{2h}\hat p_{t,0}^{2h}}{\sigma_{0}^{2h}\hat p_{t,1}^{2h}}
\right | 1_{(\hat p_{t,0}\leq 1/2)}\\
& 
\leq 
\left |\frac{\sigma_{t,0}^{2h}\sigma_{1}^{2h}((n_{0,1}+1)/t+\hat p_{t,1})^{2h}\hat p_{t,0}^{2h}
-\sigma_{0}^{2h}\sigma_{t,1}^{2h}((n_{0,0}+1)/t+\hat p_{t,0})^{2h}\hat p_{t,1}^{2h}
}
{\sigma_{t,1}^{2h}\sigma_{1}^{2h}((n_{0,0}+1)/t+\hat p_{t,0})^{2h}\hat p_{t,0}^{2h}}
\right| 
1_{(\hat p_{t,0}>1/2)}\\
&+
\left | \frac{\sigma_{t,1}^{2h}\sigma_{0}^{2h}((n_{0,0}+1)/t+\hat p_{t,0})^{2h}\hat p_{t,1}^{2h}
	-\sigma_{1}^{2h}\sigma_{t,0}^{2h}((n_{0,1}+1)/t+\hat p_{t,1})^{2h}\hat p_{t,0}^{2h}
}
{\sigma_{t,0}^{2h}\sigma_{0}^{2h}((n_{0,1}+1)/t+\hat p_{t,1})^{2h}\hat p_{t,1}^{2h}}
\right |
 1_{(\hat p_{t,1}\geq 1/2)}
\end{aligned}
\]
\end{small}
Thus,
\begin{small}
\[
\begin{aligned}
\mid F_t-\tilde F_t\mid &\leq 
\sigma_{t,1}^{-2h}\sigma_1^{-2h}2^{4h}\mid\sigma_{t,0}^{2h}\sigma_{1}^{2h}((n_{0,1}+1)/t+\hat p_{t,1})^{2h}\hat p_{t,0}^{2h}
-\sigma_{0}^{2h}\sigma_{t,1}^{2h}((n_{0,0}+1)/t+\hat p_{t,0})^{2h}\hat p_{t,1}^{2h}\mid1_{(\hat p_{t,0}> 1/2)}\\
&
+\sigma_{t,0}^{-2h}\sigma_0^{-2h}2^{4h}\mid\sigma_{t,1}^{2h}\sigma_{0}^{2h}((n_{0,0}+1)/t+\hat p_{t,0})^{2h}\hat p_{t,1}^{2h}
-\sigma_{1}^{2h}\sigma_{t,0}^{2h}((n_{0,1}+1)/t+\hat p_{t,1})^{2h}\hat p_{t,0}^{2h}\mid 1_{(\hat p_{t,1}\geq  1/2)}
\end{aligned}
\]
\end{small}
By (c), for every $a=0,1$,
\begin{small}
\[
\left(\hat p_{t,a}+\frac{n_{0,a}+1}{t}\right)^{2h}-\hat p_{t,a}^{2h}\rightarrow 0
\]
\end{small}
Now, if $\omega\in (A_t=0 \;i.o.)\cap (A_t=1\; i.o)$, then $\sigma_{t,0}^2\rightarrow \sigma_0^2$ and $\sigma_{t,1}^2\rightarrow \sigma_1^2$ as $t\rightarrow \infty$. By (b), $F_t-\tilde F_t\rightarrow 0$. \\
On the other hand, if $\omega\in (A_t=0\; ult.)$, then
 for $t$ large enugh, $\sigma_{t,1}\rightarrow \sigma_1$,
$\sigma_{t,0} \rightarrow \sigma_{T,0}$ for a finite stopping time $T$, $\hat p_{t,1}\rightarrow 0$ and
$\hat p_{t,0}\rightarrow 1$. Thus, $ 1_{(\hat p_{t,1}\geq  1/2)}\rightarrow 0$. \noindent Therefore, $F_t-\tilde F_t\rightarrow 0$.
Analogously, if $\omega\in (A_t=1\; ult.)$, then $F_t-\tilde F_t\rightarrow 0$. \\
Now, let $c$ be such that $\tilde F_t<-2c$ if
 $\hat p_{t,1}>\rho_1+\epsilon$ and 
 $\tilde F_t>2c$ if $\hat p_{t,1}<\rho_1-\epsilon$. 
 Since $F_t-\tilde F_t\rightarrow 0$, 
there exists a random time $T$ such that $\mid F_t-\tilde F_t\mid<c$ for all $t\geq T$. 
For every $t\geq T$, $F_t<-c$ if $\hat p_{t,1}>\rho_1+\epsilon$ and $F_t>c$ if $\hat p_{t,1}<\rho_1-\epsilon$.  Based on basics of stochastic approximation, it follows that $\hat p_{t,1}  \rightarrow \rho_1 $ almost surely.
 Additionally, by definition of $p_{t,1}$, we have\\
\begin{small}
\begin{equation}
p_{t,1} = \dfrac{1}{1+ \left(\dfrac{n_{0,1}+t\hat{p}_{t,1}+1}{n_{0,0}+t(1-\hat{p}_{t,1})+1}\right)^{2h}  \dfrac{\sigma_{t,0}^{2h}}{\sigma_{t,1}^{2h}}  }. \label{ppt}
\end{equation}
\end{small}
Hence, applying continuous mapping theorem (Theorem 2.3 of \cite{van2000asymptotic}), 
we have \begin{small}
\begin{equation*}
p_{t,1}  \underset{{\small t \rightarrow \infty}}{\longrightarrow} \rho_1  \, \,{\normalsize  \text{a.s. } } .
\end{equation*}
\end{small}

\end{proof}

\begin{proof}
(Corollary~\ref{Coro})
For any pair of arms $(a_1, a_2)$, the subsequence of samples assigned to these two arms is equivalent to a two arm BUD design. Therefore, Proposition \ref{propa} implies that almost surely $$\dfrac{\widehat p_{t,a_1}}{\widehat p_{t,a_1}+\widehat p_{t,a_2}}\underset{{\small t \rightarrow \infty}}{\longrightarrow} \rho_{a_1,a_2}:= \frac{\sigma_{a_1}^{\frac{2h}{2h+1}}}{\sigma_{a_1}^{\frac{2h}{2h+1}}+\sigma_{a_2}^{\frac{2h}{2h+1}}}.$$
Then, the allocation proportions $(\widehat{p}_{t,0}, \dots, \widehat{p}_{t,K-1})$  converge to a limit $(\rho_0, \dots, \rho_{K-1})$, which is
the unique solution to 
$$\sum_{a=0}^{K-1} \rho_a=1 \; \text{ and } \;  \rho_{a_1}= \rho_{a_1,a_2}(\rho_{a_1}+\rho_{a_2}) \; \text{ for all } \{a_1, a_2\} \subset \{0, \dots, K-1\}. $$
The solution of the above linear system is given by
\begin{equation}
\rho_a=\dfrac{\sigma_{a}^{\frac{2h}{2h+1}}}{\sum_{i=0}^{K-1}\sigma_{i}^{\frac{2h}{2h+1}}} \quad \text{ for } a \in \{0, \dots, K-1\}\label{limitmultiarm}
\end{equation}
Analogously, \eqref{limitmultiarm} defines the limit $(\rho_0, \dots, \rho_{K-1})$ of the randomization probabilities of the BUD in the multi-arm setup.
\end{proof}

\newpage
\begin{proof}
(Lemma~\ref{lemma1})
 We will make use of the following properties of $\mathcal{O}_P(\cdot)$:  {\small \begin{align}\mathcal{O}_P(a)\mathcal{O}_P(b)&= \mathcal{O}_P{(ab)} \nonumber \\ \mathcal{O}_P(a)+\mathcal{O}_P{(a)}&= \mathcal{O}_P{(a)} .\label{Op}
\end{align}}
Moreover, we will invoke the following properties of the distributions in the natural exponential family (see \cite{diaconis1979}):
\begin{small}
\begin{align}
E_{\psi_a}(Y_i)&=b'(\psi_a) \quad \forall i=1,\dots, t+1 \nonumber \\
\text{Var}_{\psi_a}(Y_i )&= b''(\psi_a) \quad \forall i=1,\dots, t+1\nonumber \\
E(b'(\psi_a)\mid \Sigma_t)&= \tilde{y}_{t,a} . \label{properties}
\end{align}
\end{small}
From the law of total variance and the characterization of the distributions in the natural exponential family with quadratic variance function, we have
\begin{small}
 \begin{align}
\sigma^2_{t,a}&= E(\text{Var}_{\psi_a}(Y_{t+1}) \mid \Sigma_t)+ \text{Var}(E_{\psi_a}(Y_{t+1})\mid \Sigma_t) \nonumber\\
&= E(v_0+v_1 b'(\psi_a)+v_2 b'(\psi_a)^2\mid \Sigma_t)+ \text{Var}(b'(\psi_a)\mid \Sigma_t) \nonumber \\
&= v_0 +v_1 \tilde{y}_{t,a} +(v_2+1) E(b'(\psi_a)^2\mid \Sigma_t)- E(b'(\psi_a)\mid \Sigma_t)^2 \nonumber \\
&= v_0 +v_1 \tilde{y}_{t,a} + (v_2+1)( E(b'(\psi_a)^2\mid \Sigma_t)-E(b'(\psi_a)\mid \Sigma_t)^2 )+v_2\tilde{y}_{t,a}^2 \label{quadr1}
\end{align}
\end{small}
for $a \in \{0,1\}.$ Now, from Theorem 5.3 of Morris \cite{morris1983}, 
\begin{small}
\begin{equation}
E(b'(\psi_a)^2\mid \Sigma_t)-E(b'(\psi_a)\mid \Sigma_t)^2= \dfrac{1}{n_{0,a}+t\hat{p}_{t,a}-v_2}(v_0 +v_1 \tilde{y}_{t,a} +v_2\tilde{y}_{t,a}^2)
\end{equation}
\end{small}
and \eqref{quadr1} becomes
\begin{small}
\begin{equation}
\sigma^2_{t,a} =v_0+v_1\tilde{y}_{t,a}+v_2\tilde{y}_{t,a}^2  +\mathcal{O}_P(t^{-1}). \label{quadr2}
\end{equation}
\end{small}
\eqref{quadr2} is a consequence of the convergence of $\hat{p}_{t,a}$, due to \eqref{limitexp}, and of the properties \eqref{Op}. By taking the square root of \eqref{quadr2}, (i) follows.\\
Also, by inverting \eqref{ppt}, we obtain 
\begin{small}
\begin{align}
\hat{p}_{t,1} &= \dfrac{1}{1+ \left(\dfrac{p_{t,1}}{1-p_{t,1}}\right)^{\frac{1}{2h}}  \dfrac{\sigma_{t,0}}{\sigma_{t,1}}  } + \dfrac{1}{t} \dfrac{(n_{0,0}+1)        \left(\dfrac{1-p_{t,1}}{p_{t,1}}\right)^{\frac{1}{2h}}  \dfrac{\sigma_{t,1}}{\sigma_{t,0}}-(n_{0,1}+1)}{1+ \left(\dfrac{1-p_{t,1}}{p_{t,1}}\right)^{\frac{1}{2h}}  \dfrac{\sigma_{t,1}}{\sigma_{t,0}}} \nonumber \\
&= \dfrac{1}{1+ \left(\dfrac{p_{t,1}}{1-p_{t,1}}\right)^{\frac{1}{2h}}  \dfrac{\sigma_{t,0}}{\sigma_{t,1}}  } +  \mathcal{O}_P(t^{-1}).
\label{eqexphat}
\end{align}
\end{small}
and, therefore,
\begin{small}
\begin{equation}
\hat{p}_{t,1}^{-1}=1+ \left(\dfrac{p_{t,1}}{1-p_{t,1}}\right)^{\frac{1}{2h}}  \dfrac{\sigma_{t,0}}{\sigma_{t,1}}   +\mathcal{O}_P(t^{-1}). \label{hp11}
\end{equation}
\end{small}
We have
\begin{small}
 \begin{align}
\tilde{y}_{t+1,1}&=A_{t+1}\left(\tilde{y}_{t,1}+ \dfrac{Y_{t+1}- \tilde{y}_{t,1}}{t \hat{p}_{t,1}+1}\right)+ (1-A_{t+1}) \tilde{y}_{t,1} \nonumber \\
&= \tilde{y}_{t,1}+  A_{t+1}  \dfrac{Y_{t+1}- \tilde{y}_{t,1}}{t \hat{p}_{t,1}} + \mathcal{O}_P(t^{-2}) \label{G10}\\
&=  \tilde{y}_{t,1}+ A_{t+1} \dfrac{(Y_{t+1}-\tilde{y}_{t,1})}{t} \left[1+\left(\dfrac{p_{t,1}}{1-p_{t,1}}\right)^{\frac{1}{2h}}\dfrac{\sigma_{t,0}}{\sigma_{t,1}}\right]  + \mathcal{O}_P(t^{-2}) \label{G12} \\
&= \tilde{y}_{t,1}+ A_{t+1} \dfrac{(Y_{t+1}-\tilde{y}_{t,1})}{t} \left[1+\left(\dfrac{p_{t,1}}{1-p_{t,1}}\right)^{\frac{1}{2h}}\dfrac{v(\tilde{y}_{t,0})}{v(\tilde{y}_{t,1})}\right]  + \mathcal{O}_P(t^{-2}). \label{G11}
\end{align}
\end{small}
Equation \eqref{G12} is obtained by plugging \eqref{hp11} into \eqref{G10}, equation \eqref{G11} is a consequence of (i). So,  we can generalize to arm $a\in\{0,1\}$ as follows

\begin{small}
\begin{equation}
{y}_{t+1,a}= \tilde{y}_{t,a}+ 1(A_{t+1}=a) \dfrac{(Y_{t+1}-\tilde{y}_{t,a})}{t} \left[1+  \left(\dfrac{p_{t,a}}{ p_{t,1-a} }\right)^{\frac{1}{2h}} \dfrac{v(\tilde{y}_{t,1-a})}{v(\tilde{y}_{t,a})} \right] +\mathcal{O}_P(t^{-2}) \label{G122}
\end{equation}
\end{small}
and this proves (ii).
Finally, by using (ii), we get
\begin{small}
\begin{align}
\Delta \sigma^2_{t,1}& = v_0+v_1 \tilde{y}_{t+1,1} + v_2 \tilde{y}_{t+1,1}^2-(v_0+v_1 \tilde{y}_{t,1} + v_2 \tilde{y}_{t,1}^2) \nonumber \\
&= (v_1+2 v_2 \tilde{y}_{t,1} ) A_{t+1} \dfrac{(Y_{t+1}-\tilde{y}_{t,1})}{t} \left[1+\left(\dfrac{p_{t,1}}{1-p_{t,1}}\right)^{\frac{1}{2h}}\dfrac{v(\tilde{y}_{t,0})}{v(\tilde{y}_{t,1})}\right]  + \mathcal{O}_P(t^{-2})
\end{align}
\end{small}
and, analogously,
\begin{small}
\begin{equation}
\Delta \sigma^2_{t,0}= (v_1+2 v_2 \tilde{y}_{t,0} )(1- A_{t+1}) \dfrac{(Y_{t+1}-\tilde{y}_{t,0})}{t} \left[1+\left(\dfrac{1-p_{t,1}}{p_{t,1}}\right)^{\frac{1}{2h}}\dfrac{v(\tilde{y}_{t,1})}{v(\tilde{y}_{t,0})}\right]  + \mathcal{O}_P(t^{-2}).
\end{equation}
\end{small}
This completes the proof of (iii).

\end{proof}

\newpage

\begin{proof}
(Lemma~\ref{lemma2})
Throughout this proof, we consider first-order approximations of $p_{t+1}-p_t$. 
First, by definition of the randomization probabilities of the BUD in terms of the information increments, we have
\begin{small}
\begin{align}
p_{t+1,1} - p_{t,1} 
&= \dfrac{ \left[ \dfrac{\sigma_{t+1,1}^2}{(n_{0,1}+(t+1)\hat p_{t+1,1}+1)^2}   \right]^h}{\left[ \dfrac{\sigma_{t+1,0}^2}{(n_{0,0}+(t+1)\hat p_{t+1,0}+1)^2}   \right]^h+\left[ \dfrac{\sigma_{t+1,1}^2}{(n_{0,1}+(t+1)\hat p_{t+1,1}+1)^2}   \right]^h} -\nonumber \\ & - \frac{ 
	\left[ \dfrac{\sigma_{t,1}^2}{(n_{0,0}+t\hat p_{t,0}+1)^2}   \right]^h}{\left[ \dfrac{\sigma_{t,0}^2}{(n_{0,0}+t\hat p_{t,0}+1)^2}   \right]^h+\left[ \dfrac{\sigma_{t,1}^2}{(n_{0,1}+t\hat p_{t,1}+1)^2}   \right]^h} \nonumber \\
&= \dfrac{1}{ 1+ \dfrac{\sigma_{t+1,0}^{2h}}{\sigma_{t+1,1}^{2h}} \left( \dfrac{n_{0,1} +1 + t\hat{p}_{t,1}+A_{t+1}   }{n_{0,0}+1+t(1-\hat{p}_{t,1}) +1-A_{t+1} }\right)^{2h} }- \dfrac{1}{ 1+ \dfrac{\sigma_{t,0}^{2h}}{\sigma_{t,1}^{2h}} \left( \dfrac{n_{0,1} +1 + t\hat{p}_{t,1}  }{n_{0,0}+1+t(1-\hat{p}_{t,1}) }\right)^{2h} } \label{eqexp00}\\
&= \dfrac{   \dfrac{\sigma_{t,0}^{2h}}{\sigma_{t,1}^{2h}} \left( \dfrac{n_{0,1} +1 + t\hat{p}_{t,1}   }{n_{0,0}+1+t(1-\hat{p}_{t,1})  }\right)^{2h} -  \dfrac{\sigma_{t+1,0}^{2h}}{\sigma_{t+1,1}^{2h}} \left( \dfrac{n_{0,1} +1 + t\hat{p}_{t,1} + A_{t+1}   }{n_{0,0}+1+t(1-\hat{p}_{t,1}) + 1 - A_{t+1}  }\right)^{2h}   }{ \left[ 1+ \dfrac{\sigma_{t,0}^{2h}}{\sigma_{t,1}^{2h}} \left( \dfrac{n_{0,1} +1 + t\hat{p}_{t,1}   }{n_{0,0}+1+t(1-\hat{p}_{t,1})  }\right)^{2h} \right]\left[ 1+ \dfrac{\sigma_{t+1,0}^{2h}}{\sigma_{t+1,1}^{2h}} \left( \dfrac{n_{0,1} +1 + t\hat{p}_{t,1} +A_{t+1}  }{n_{0,0}+1+t(1-\hat{p}_{t,1})+1-A_{t+1}  }\right)^{2h} \right]    } \label{eqexp0a}\\
 & = \dfrac{   \dfrac{\sigma_{t,0}^{2h}}{\sigma_{t,1}^{2h}} \left( \dfrac{n_{0,1} +1 + t\hat{p}_{t,1}   }{n_{0,0}+1+t(1-\hat{p}_{t,1})  }\right)^{2h} -  \dfrac{\sigma_{t+1,0}^{2h}}{\sigma_{t+1,1}^{2h}} \left( \dfrac{n_{0,1} +1 + t\hat{p}_{t,1} + A_{t+1}   }{n_{0,0}+1+t(1-\hat{p}_{t,1}) + 1 - A_{t+1}  }\right)^{2h}   }{ \left[ 1+ \dfrac{\sigma_{t,0}^{2h}}{\sigma_{t,1}^{2h}} \left( \dfrac{n_{0,1} +1 + t\hat{p}_{t,1}   }{n_{0,0}+1+t(1-\hat{p}_{t,1})  }\right)^{2h} \right]^2   } \left(1+ \mathcal{O}_P(t^{-1}) \right) .\label{eqexp0b}
\end{align}
\end{small}
To obtain \eqref{eqexp0b} we have noticed that the two factors of the denominator in the right-hand-side of \eqref{eqexp0a} share the same asymptotic behavior. Thus, we have isolated the principal part of the denominator in \eqref{eqexp0a}  and we have identified a remainder term which appears as $\mathcal{O}_P(t^{-1})$ since, due to Proposition 1, $\hat{p}_{t,1}$ converges almost surely to a limit which is different from 0 and 1 and $\sigma_{t,a}^2$ converges to a finite limit different from 0 almost surely for $a\in\{0,1\}$. 
Now, we split the right-hand-side of \eqref{eqexp0b} into two parts, referring to the possible assignements of treatment $(t+1)$ and using the fact that $A_{t+1}$ takes value 1 when treatment $(t+1)$ is assigned to arm 1 and 0 otherwise. So, when the response $Y_{t+1}$ comes from arm 1, $\sigma_{t+1,0}^2=\sigma_{t,0}^2$ and, instead, when the $(t+1) ^{th}$ treatment is assigned to arm 0, $\sigma_{t+1,1}^2=\sigma_{t,1}^2$. We get
\begin{small}
\begin{align}
&p_{t+1,1} - p_{t,1} = \left\{A_{t+1} \dfrac{  \left[  \dfrac{\sigma_{t,0}^{2h}}{\sigma_{t,1}^{2h}} \left( \dfrac{n_{0,1} +1 + t\hat{p}_{t,1}   }{n_{0,0}+1+t(1-\hat{p}_{t,1})  }\right)^{2h} -  \dfrac{\sigma_{t,0}^{2h}}{\sigma_{t+1,1}^{2h}} \left( \dfrac{n_{0,1} +1 + t\hat{p}_{t,1} + A_{t+1}   }{n_{0,0}+1+t(1-\hat{p}_{t,1})  }\right)^{2h} \right]  }{ \left[ 1+ \dfrac{\sigma_{t,0}^{2h}}{\sigma_{t,1}^{2h}} \left( \dfrac{n_{0,1} +1 + t\hat{p}_{t,1}   }{n_{0,0}+1+t(1-\hat{p}_{t,1})  }\right)^{2h} \right]^2   }+ \right.\nonumber \\
&  \left.+(1-A_{t+1}) \dfrac{ \left[    \dfrac{\sigma_{t,0}^{2h}}{\sigma_{t,1}^{2h}} \left( \dfrac{n_{0,1} +1 + t\hat{p}_{t,1}   }{n_{0,0}+1+t(1-\hat{p}_{t,1})  }\right)^{2h} -  \dfrac{\sigma_{t+1,0}^{2h}}{\sigma_{t,1}^{2h}} \left( \dfrac{n_{0,1} +1 + t\hat{p}_{t,1}  }{n_{0,0}+1+t(1-\hat{p}_{t,1}) +1 - A_{t+1}   }\right)^{2h}   \right]}{ \left[ 1+ \dfrac{\sigma_{t,0}^{2h}}{\sigma_{t,1}^{2h}} \left( \dfrac{n_{0,1} +1 + t\hat{p}_{t,1}   }{n_{0,0}+1+t(1-\hat{p}_{t,1})  }\right)^{2h} \right]^2   } \right\}\left(1+\mathcal{O}_P(t^{-1})\right).  \label{eqexp0}
\end{align}
\end{small}
Second, we invoke the following result, stated as a separate Lemma, whose proof is given subsequently.
\begin{lemma}[Supplementary]\label{lemmaA1}
Under the assumptions of Lemma \ref{lemma2}, we have
\begin{small}
\begin{align}
A_{t+1}   \left[  \dfrac{\sigma_{t,0}^{2h}}{\sigma_{t,1}^{2h}} \right.& \left. \left( \dfrac{n_{0,1} +1 + t\hat{p}_{t,1}   }{n_{0,0}+1+t(1-\hat{p}_{t,1})  }\right)^{2h} -  \dfrac{\sigma_{t,0}^{2h}}{\sigma_{t+1,1}^{2h}} \left( \dfrac{n_{0,1} +1 + t\hat{p}_{t,1} + A_{t+1}   }{n_{0,0}+1+t(1-\hat{p}_{t,1})  }\right)^{2h} \right]  \nonumber \\
& = A_{t+1}h \dfrac{\sigma_{t,0}^{2h} \hat{p}_{t,1}^{2h}}{ \sigma_{t,1}^{2h}(1-\hat{p}_{t,1})^{2h} } \left[ \frac{ \Delta \sigma_{t,1}^2}{\sigma_{t,1}^2} - \frac{2}{t\hat{p}_{t,1}} \right] + \mathcal{O}_P(t^{-2})  \label{eqexp2}
\end{align}
\end{small}\\
and
\begin{small}
\begin{align}
(1-A_{t+1}) &  \left[    \dfrac{\sigma_{t,0}^{2h}}{\sigma_{t,1}^{2h}}  \left( \dfrac{n_{0,1} +1 + t\hat{p}_{t,1}   }{n_{0,0}+1+t(1-\hat{p}_{t,1})  }\right)^{2h} -  \dfrac{\sigma_{t+1,0}^{2h}}{\sigma_{t,1}^{2h}} \left( \dfrac{n_{0,1} +1 + t\hat{p}_{t,1}  }{n_{0,0}+1+t(1-\hat{p}_{t,1}) +1 - A_{t+1}   }\right)^{2h}   \right] \nonumber \\ 
& = (1-A_{t+1})  \dfrac{\hat{p}_{t,1}^{2h}\sigma_{t,0}^{2h}}{\sigma_{t,1}^{2h}( 1-\hat{p}_{t,1})^{2h} }h \left[  - \frac{\Delta \sigma_{t,0}^2 }{\sigma_{t,0}^2} +  \frac{2}{t(1-\hat{p}_{t,1})}   \right] + \mathcal{O}_P(t^{-2}). \label{eqexp3}
\end{align}
\end{small}
\end{lemma}
Thus, we replace the numerators of the two addenda in \eqref{eqexp0} with the right-hand-side of equations \eqref{eqexp2} and \eqref{eqexp3} and we write
\begin{small}
\begin{align}
p_{t+1,1} - p_{t,1}& = \dfrac{h \dfrac{\hat{p}_{t,1}^{2h}\sigma_{t,0}^{2h}}{\sigma_{t,1}^{2h}(1- \hat{p}_{t,1})^{2h}}}{\left[ 1+ \dfrac{\sigma_{t,0}^{2h}}{\sigma_{t,1}^{2h}} \left( \dfrac{n_{0,1} +1 + t\hat{p}_{t,1}   }{n_{0,0}+1+t(1-\hat{p}_{t,1})  }\right)^{2h} \right]^2 } \left[ A_{t+1} \left( \dfrac{ \Delta \sigma_{t,1}^2}{\sigma_{t,1}^2} - \dfrac{2}{t\hat{p}_{t,1}} \right) + \right. \nonumber \\
& \left. +(1-A_{t+1})          \left(  - \frac{\Delta \sigma_{t,0}^2 }{\sigma_{t,0}^2} +  \dfrac{2}{t(1-\hat{p}_{t,1})}   \right) + \mathcal{O}_P(t^{-2}) \right] \left(1+\mathcal{O}_P(t^{-1})\right). \label{eqexp5}
\end{align}
\end{small}
Third, retaining the dominant part of the denominator in equation \eqref{eqexp5}, it follows that
\begin{small}
\begin{align}
p_{t+1,1} - p_{t,1} &= \dfrac{h \dfrac{\hat{p}_{t,1}^{2h}\sigma_{t,0}^{2h}}{\sigma_{t,1}^{2h}(1- \hat{p}_{t,1})^{2h}}}{\left(1+ \dfrac{\sigma_{t,0}^{2h}}{\sigma_{t,1}^{2h}} \left( \dfrac{ \hat{p}_{t,1}   }{1-\hat{p}_{t,1} }\right)^{2h} \right)^2 } \left[ A_{t+1} \left( \dfrac{ \Delta \sigma_{t,1}^2}{\sigma_{t,1}^2} - \dfrac{2}{t\hat{p}_{t,1}} \right) \right.+ \nonumber \\
&\left. + (1-A_{t+1})          \left(  - \frac{\Delta \sigma_{t,0}^2 }{\sigma_{t,0}^2} +  \dfrac{2}{t(1-\hat{p}_{t,1})}   \right) + \mathcal{O}_P(t^{-2}) \right] \left(1+\mathcal{O}_P(t^{-1})\right). \label{eqexp6}
\end{align}
\end{small}
Next, noting that
\begin{small}
\begin{equation}
\left(\dfrac{1-p_{t,1}}{p_{t,1}}\right)^{\frac{1}{2h}}  \dfrac{\sigma_{t,1}}{\sigma_{t,0}}= \dfrac{n_{0,1}+1 +t\hat{p}_{t,1}}{n_{0,0}+1 +t(1-\hat{p}_{t,1})},
\end{equation}
\end{small}
 it holds that \begin{small}
\begin{align}
\left(\dfrac{1-p_{t,1}}{p_{t,1}}\right)^{\frac{1}{2h}}  \dfrac{\sigma_{t,1}}{\sigma_{t,0}}&=\dfrac{\hat{p}_{t,1}}{1-\hat{p}_{t,1}}+ \dfrac{n_{0,1}+1}{n_{0,0}+1+t(1-\hat{p}_{t,1})}- \dfrac{\hat{p}_{t,1}(n_{0,0}+1)}{(t(1-\hat{p}_{t,1})+n_{0,0}+1)(1-\hat{p}_{t,1})}\nonumber \\
&= \dfrac{\hat{p}_{t,1}}{1-\hat{p}_{t,1}} + \mathcal{O}_P(t^{-1})
\end{align}
\end{small}
and, thus,
\begin{small}
\begin{align}
\left( \dfrac{\hat{p}_{t,1}}{1-\hat{p}_{t,1}} \right)^{2h} &=\left[ \left(\dfrac{1-p_{t,1}}{p_{t,1}}\right)^{\frac{1}{2h}}  \dfrac{\sigma_{t,1}}{\sigma_{t,0}}+ \mathcal{O}_P(t^{-1}) \right]^{2h} \nonumber \\
&= \dfrac{1-p_{t,1}}{p_{t,1}}  \dfrac{\sigma_{t,1}^{2h}}{\sigma_{t,0}^{2h}} + \mathcal{O}_P(t^{-1}). \label{opt1}
\end{align}
\end{small}
Plugging \eqref{opt1} and \eqref{eqexphat} into \eqref{eqexp6} yield to\\
\begin{small}
\begin{align}
p_{t+1,1} - p_{t,1} &=\left( h p_{t,1} (1-p_{t,1}) +\mathcal{O}_P(t^{-1})\right)\left\{A_{t+1} \left[  \frac{ \Delta \sigma_{t,1}^2}{\sigma_{t,1}^2} - \dfrac{2}{t}\left(\left(\dfrac{p_{t,1}}{1-p_{t,1}}\right)^{\frac{1}{2h}}  \dfrac{\sigma_{t,0}}{\sigma_{t,1}} +1 \right)  \right] \right. \nonumber \\
& \left. + (1-A_{t+1})\left[ - \frac{ \Delta \sigma_{t,0}^2}{\sigma_{t,0}^2} + \dfrac{2}{t} \left( \left(\dfrac{1-p_{t,1}}{p_{t,1}}\right)^{\frac{1}{2h}}  \dfrac{\sigma_{t,1}}{\sigma_{t,0}}+1 \right)  \right]+\mathcal{O}_P(t^{-2}) \right\} \left(1+\mathcal{O}_P(t^{-1})\right).    \label{eqexp5c}
\end{align}
\end{small}\\
To obtain equation \eqref{eqexp5c} we have noticed that 
\begin{small}
\begin{equation*}
\dfrac{h \dfrac{\hat{p}_{t,1}^{2h}\sigma_{t,0}^{2h}}{\sigma_{t,1}^{2h}(1- \hat{p}_{t,1})^{2h}}}{\left(1+ \dfrac{ \hat{p}_{t,1}^{2h}\sigma_{t,0}^{2h}}{\sigma_{t,1}^{2h}( 1-\hat{p}_{t,1})^{2h}}\right)^2} = h p_{t,1} (1-p_{t,1})+ \mathcal{O}_P(t^{-1}),
\end{equation*}
\end{small}
\begin{small}
\begin{equation}
\hat{p}_{t,1}^{-1}=1+ \left(\dfrac{p_{t,1}}{1-p_{t,1}}\right)^{\frac{1}{2h}}  \dfrac{\sigma_{t,0}}{\sigma_{t,1}}   +\mathcal{O}_P(t^{-1}), \label{hp1}
\end{equation}
\end{small}
\begin{small}
\begin{equation}
(1-\hat{p}_{t,1})^{-1}=1+ \left(\dfrac{1-p_{t,1}}{p_{t,1}}\right)^{\frac{1}{2h}}  \dfrac{\sigma_{t,1}}{\sigma_{t,0}}   +\mathcal{O}_P(t^{-1}).
\end{equation}
\end{small}
Indeed, by properties \eqref{Op}, \eqref{eqexp5c} becomes 
\begin{small}
\begin{align}
p_{t+1,1} - p_{t,1} &= h p_{t,1} (1-p_{t,1}) \left\{A_{t+1} \left[  \frac{ \Delta \sigma_{t,1}^2}{\sigma_{t,1}^2} - \dfrac{2}{t}\left(\left(\dfrac{p_{t,1}}{1-p_{t,1}}\right)^{\frac{1}{2h}}  \dfrac{\sigma_{t,0}}{\sigma_{t,1}} +1 \right)  \right] \right. \nonumber \\
& \left. + (1-A_{t+1})\left[ - \frac{ \Delta \sigma_{t,0}^2}{\sigma_{t,0}^2} + \dfrac{2}{t} \left( \left(\dfrac{1-p_{t,1}}{p_{t,1}}\right)^{\frac{1}{2h}}  \dfrac{\sigma_{t,1}}{\sigma_{t,0}}+1 \right)  \right] \right\} + \mathcal{O}_P(t^{-2}) \nonumber \\
&= h p_{t,1} (1-p_{t,1}) \left\{ \left[  \frac{ \Delta \sigma_{t,1}^2}{\sigma_{t,1}^2} - \dfrac{2A_{t+1}}{t}\left(\left(\dfrac{p_{t,1}}{1-p_{t,1}}\right)^{\frac{1}{2h}}  \dfrac{\sigma_{t,0}}{\sigma_{t,1}} +1 \right)  \right] \right. \nonumber \\
& \left. + \left[  \dfrac{2(1-A_{t+1})}{t} \left( \left(\dfrac{1-p_{t,1}}{p_{t,1}}\right)^{\frac{1}{2h}}  \dfrac{\sigma_{t,1}}{\sigma_{t,0}}+1 \right)  - \frac{ \Delta \sigma_{t,0}^2}{\sigma_{t,0}^2} \right] \right\} + \mathcal{O}_P(t^{-2})\label{eqexp6a} \\
&= h p_{t,1} (1-p_{t,1}) \left\{ \left[  \frac{ \Delta \sigma_{t,1}^2}{\sigma_{t,1}^2} - \dfrac{2A_{t+1}}{t}\left(\left(\dfrac{p_{t,1}}{1-p_{t,1}}\right)^{\frac{1}{2h}}  \dfrac{v(\tilde{y}_{t,0})}{v(\tilde{y}_{t,1})} +1 \right)  \right] \right. \nonumber \\
& \left. +\left[   \dfrac{2(1-A_{t+1})}{t} \left( \left(\dfrac{1-p_{t,1}}{p_{t,1}}\right)^{\frac{1}{2h}}  \dfrac{v(\tilde{y}_{t,1})}{v(\tilde{y}_{t,0})}+1 \right)  - \frac{ \Delta \sigma_{t,0}^2}{\sigma_{t,0}^2} \right] \right\} + \mathcal{O}_P(t^{-2}), \label{eqexp5d}
\end{align}
\end{small}
where \eqref{eqexp6a} follows from the fact that $A_{t+1} \Delta \sigma_{t,1}^2= \Delta \sigma_{t,1}^2$ and $(1-A_{t+1})\Delta \sigma_{t,0}^2=\Delta \sigma_{t,0}^2$ and \eqref{eqexp5d} is a consequence of (i) of Lemma 1.\\
Finally, the statement of Lemma \ref{lemma2} is obtained by plugging the expression for $\sigma^2_{t,1},\Delta \sigma^2_{t,1}, \sigma^2_{t,0} $ and $\Delta \sigma^2_{t,0} $ given in Lemma \ref{lemma1} into \eqref{eqexp5d} and by invoking properties \eqref{Op}.

\end{proof}

\newpage

\begin{proof}
(Lemma~\ref{lemmaA1} - Supplementary)
We have
\begin{small}
\begin{align}
A_{t+1}   \left[  \dfrac{\sigma_{t,0}^{2h}}{\sigma_{t,1}^{2h}} \right.& \left. \left( \dfrac{n_{0,1} +1 + t\hat{p}_{t,1}   }{n_{0,0}+1+t(1-\hat{p}_{t,1})  }\right)^{2h} -  \dfrac{\sigma_{t,0}^{2h}}{\sigma_{t+1,1}^{2h}} \left( \dfrac{n_{0,1} +1 + t\hat{p}_{t,1} + A_{t+1}   }{n_{0,0}+1+t(1-\hat{p}_{t,1})  }\right)^{2h} \right]  \nonumber \\
&= A_{t+1}  \dfrac{\sigma_{t,0}^{2h}}{( n_{0,0}+1+t(1-\hat{p}_{t,1}))^{2h} } \left[ \dfrac{\left( n_{0,1} +1+t \hat{p}_{t,1} \right)^{2h}}{\sigma_{t,1}^{2h}} - \dfrac{ \left( n_{0,1} +1+t \hat{p}_{t,1}+1 \right)^{2h} }{ \sigma_{t+1,1}^{2h}}    \right]  \nonumber \\
&= A_{t+1}  \dfrac{\sigma_{t,0}^{2h} t^{2h}\hat{p}_{t,1}^{2h}}{ ( n_{0,0}+1+t(1-\hat{p}_{t,1}))^{2h} } \left[ \dfrac{\left( 1+ \frac{n_{0,1}+1}{t\hat{p}_{t,1}}\right)^{2h}}{\sigma_{t,1}^{2h} } -  \dfrac{\left( 1+ \frac{n_{0,1}+2}{t\hat{p}_{t,1}}\right)^{2h}}{\sigma_{t+1,1}^{2h} }     \right]  \label{eqexp2a}\\
& = A_{t+1}  \dfrac{\sigma_{t,0}^{2h} \hat{p}_{t,1}^{2h}}{ (1-\hat{p}_{t,1})^{2h} } \left[ \dfrac{\left( 1+ \frac{n_{0,1}+1}{t\hat{p}_{t,1}}\right)^{2h}}{\sigma_{t,1}^{2h} } -  \dfrac{\left( 1+ \frac{n_{0,1}+2}{t\hat{p}_{t,1}}\right)^{2h}}{\sigma_{t+1,1}^{2h} }     \right] \left(1+\mathcal{O}_P(t^{-1})\right) \label{eqexp2b}\\
&= A_{t+1}  \dfrac{\sigma_{t,0}^{2h} \hat{p}_{t,1}^{2h}}{ (1-\hat{p}_{t,1})^{2h} } \left[   \dfrac{  \sigma_{t+1,1}^{2h}  -\sigma_{t,1}^{2h} + 2h \frac{n_{0,1} +1}{t \hat{p}_{t,1}} \sigma_{t+1,1}^{2h} -2h \frac{n_{0,1} +2}{t \hat{p}_{t,1}} \sigma_{t,1}^{2h}   }{\sigma_{t,1}^{2h}\sigma_{t+1,1}^{2h}} + \mathcal{O}_P(t^{-2}) \right]\left(1+\mathcal{O}_P(t^{-1})\right)  \label{eqexp2c}\\
& = A_{t+1} \dfrac{\sigma_{t,0}^{2h} \hat{p}_{t,1}^{2h}}{ \sigma_{t+1,1}^{2h}\sigma_{t,1}^{2h}(1-\hat{p}_{t,1})^{2h} } \left[  \left( 1+2h \frac{n_{0,1}+1}{t \hat{p}_{t,1}} \right)\left( \sigma_{t+1,1}^{2h} - \sigma_{t,1}^{2h} \right) - \frac{2h}{t\hat{p}_{t,1}}\sigma_{t,1}^{2h} +\mathcal{O}_P(t^{-2}) \right] \left(1+\mathcal{O}_P(t^{-1})\right)  \label{eqexp2d}\\
& = A_{t+1} \dfrac{\sigma_{t,0}^{2h} \hat{p}_{t,1}^{2h}}{ \sigma_{t+1,1}^{2h}\sigma_{t,1}^{2h}(1-\hat{p}_{t,1})^{2h} } \left[ \left( \sigma_{t+1,1}^{2h} - \sigma_{t,1}^{2h} \right) - \frac{2h}{t\hat{p}_{t,1}}\sigma_{t,1}^{2h} \right] + \mathcal{O}_P(t^{-2})\label{eqexp2e}\\
&= A_{t+1} \dfrac{\sigma_{t,0}^{2h} \hat{p}_{t,1}^{2h}}{\sigma_{t+1,1}^{2h}\sigma_{t,1}^{2h}(1-\hat{p}_{t,1})^{2h} } \left[ \left( \sigma_{t,1}^{2} + \Delta \sigma_{t,1}^2 \right)^{h} - \sigma_{t,1}^{2h} - \frac{2h}{t\hat{p}_{t,1}}\sigma_{t,1}^{2h} \right] + \mathcal{O}_P(t^{-2})\label{eqexp2f}\\
& = A_{t+1} \dfrac{\sigma_{t,0}^{2h} \hat{p}_{t,1}^{2h}}{ \sigma_{t+1,1}^{2h}(1-\hat{p}_{t,1})^{2h} } \left[ \left(1+\frac{ \Delta \sigma_{t,1}^2}{\sigma_{t,1}^2} \right)^{h}-1 -\frac{2h}{t\hat{p}_{t,1}}\right] + \mathcal{O}_P(t^{-2})  \label{eqexp2g}\\
& = A_{t+1}h \dfrac{\sigma_{t,0}^{2h} \hat{p}_{t,1}^{2h}}{ \sigma_{t,1}^{2h}(1-\hat{p}_{t,1})^{2h} } \left[ \frac{ \Delta \sigma_{t,1}^2}{\sigma_{t,1}^2} - \frac{2}{t\hat{p}_{t,1}} \right] + \mathcal{O}_P(t^{-2}). 
\end{align}
\end{small}\\
The first equality is obtained leveraging the fact that the left-hand-side  doesn't vanishes only when $A_{t+1}=1$. In \eqref{eqexp2a} we collect the term $t^{2h}\hat{p}_{t,1}^{2h}$ and we retain the dominant part to obtain \eqref{eqexp2b}. The remainder term appears as $\mathcal{O}_P(t^{-1})$ since $\sigma_{t,a}^2$ converges to a finite limit different from 0 almost surely for $a\in\{0,1\}$ and, due to Proposition 1, $\hat{p}_{t,1}$ and $(1- \hat{p}_{t,1})$ converge almost surely to a limit which is different from 0: indeed, we can bound $(1- \hat{p}_{t,1})$ in a compact set which doesn't contain 0 with arbitrarily high probability. 
The terms $\left( 1+ \frac{n_{0,1}+1}{t\hat{p}_{t,1}}\right)^{2h}$ and $\left( 1+ \frac{n_{0,1}+2}{t\hat{p}_{t,1}}\right)^{2h}$  in the left-hand-side of equation  \eqref{eqexp2b} can  be  approximated  by  Tailor   expansion:
{\small $$\left( 1+ \frac{n_{0,1}+1}{t\hat{p}_{t,1}}\right)^{2h}= 1+ 2h\frac{n_{0,1}+1}{t\hat{p}_{t,1}} + \mathcal{O}_P(t^{-2})$$}
and
{\small $$\left( 1+ \frac{n_{0,1}+2}{t\hat{p}_{t,1}}\right)^{2h}= 1+ 2h\frac{n_{0,1}+2}{t\hat{p}_{t,1}} + \mathcal{O}_P(t^{-2}).$$}
Therefore  \eqref{eqexp2b} equals \eqref{eqexp2c}.
The $\mathcal{O}_P(t^{-2})$ in \eqref{eqexp2d} is justified by invoking Lemma \ref{lemma1} and noting that $\Delta \sigma^2_{t,a}= \mathcal{O}_P(t^{-1})$ for $a\in \{0,1\}$. 
The term $2h \frac{n_{0,1}+1}{t \hat{p}_{t,1}} \left( \sigma_{t+1,1}^{2h} - \sigma_{t,1}^{2h} \right)$  in \eqref{eqexp2d} enters the remainder term in  \eqref{eqexp2e}.
In \eqref{eqexp2f} we have rewritten $\sigma_{t+1,1}^{2}$ as $\sigma_{t,1}^{2} + \Delta \sigma_{t,1}^2$.
The equality in \eqref{eqexp2g} follows from a Taylor expansion of $\left(1+\frac{ \Delta \sigma_{t,1}^2}{\sigma_{t,1}^2} \right)^{h}$ . \\
With similar arguments we can prove \eqref{eqexp3}. We have
\begin{small}
\begin{align}
(1-A_{t+1}) &  \left[    \dfrac{\sigma_{t,0}^{2h}}{\sigma_{t,1}^{2h}}  \left( \dfrac{n_{0,1} +1 + t\hat{p}_{t,1}   }{n_{0,0}+1+t(1-\hat{p}_{t,1})  }\right)^{2h} -  \dfrac{\sigma_{t+1,0}^{2h}}{\sigma_{t,1}^{2h}} \left( \dfrac{n_{0,1} +1 + t\hat{p}_{t,1}  }{n_{0,0}+1+t(1-\hat{p}_{t,1}) +1 - A_{t+1}   }\right)^{2h}   \right] \nonumber \\ 
& =(1-A_{t+1})  \dfrac{( n_{0,1}+1+t\hat{p}_{t,1})^{2h} } {\sigma_{t,1}^{2h}}\left[ \dfrac{\sigma_{t,0}^{2h}}{\left( n_{0,0} +1+t(1- \hat{p}_{t,1}) \right)^{2h}} - \dfrac{ \sigma_{t+1,0}^{2h}}{ \left( n_{0,0} +1+t(1- \hat{p}_{t,1}) +1 \right)^{2h} }    \right] \nonumber \\
& = (1-A_{t+1}) \dfrac{( n_{0,1}+1+t\hat{p}_{t,1})^{2h} }{\sigma_{t,1}^{2h}t^{2h}( 1-\hat{p}_{t,1})^{2h} } \left[  \dfrac{\sigma_{t,0}^{2h}}{\left( 1+ \frac{ n_{0,0} +1}{t(1- \hat{p}_{t,1})} \right)^{2h}} -   \dfrac{\sigma_{t+1,0}^{2h}}{\left( 1+ \frac{ n_{0,0} +2}{t(1- \hat{p}_{t,1})} \right)^{2h}}  \right]  \nonumber \\
& = (1-A_{t+1}) \dfrac{\hat{p}_{t,1}^{2h}}{\sigma_{t,1}^{2h}( 1-\hat{p}_{t,1})^{2h} } \left[  \dfrac{\sigma_{t,0}^{2h}}{\left( 1+ \frac{ n_{0,0} +1}{t(1- \hat{p}_{t,1})} \right)^{2h}} -   \dfrac{\sigma_{t+1,0}^{2h}}{\left( 1+ \frac{ n_{0,0} +2}{t(1- \hat{p}_{t,1})} \right)^{2h}}  \right]  \left(1+\mathcal{O}_P(t^{-1})\right)\nonumber \\
& = (1-A_{t+1}) \dfrac{\hat{p}_{t,1}^{2h}}{\sigma_{t,1}^{2h}( 1-\hat{p}_{t,1})^{2h} } \left[  \left( \sigma_{t,0}^{2h} - \sigma_{t+1,0}^{2h} \right)  + \frac{2h (n_{0,0}+2)}{t(1-\hat{p}_{t,1})}\sigma_{t,0}^{2h} -    \nonumber \right. \\ & \left. - \frac{2h (n_{0,0}+1)}{t(1-\hat{p}_{t,1})}\sigma_{t+1,0}^{2h} + \mathcal{O}_P(t^{-2}) \right]\left(1+\mathcal{O}_P(t^{-1})\right) \nonumber \\
&  = (1-A_{t+1}) \dfrac{\hat{p}_{t,1}^{2h}}{\sigma_{t,1}^{2h}( 1-\hat{p}_{t,1})^{2h} } \left[   \left( \sigma_{t,0}^{2h} - \sigma_{t+1,0}^{2h} \right)  +  \frac{2h}{t(1-\hat{p}_{t,1})}\sigma_{t,0}^{2h}    \right] +\mathcal{O}_P(t^{-2})\nonumber \\
& = (1-A_{t+1})  \dfrac{\hat{p}_{t,1}^{2h}\sigma_{t,0}^{2h}}{\sigma_{t,1}^{2h}( 1-\hat{p}_{t,1})^{2h} }h \left[  - \frac{\Delta \sigma_{t,0}^2 }{\sigma_{t,0}^2} +  \frac{2}{t(1-\hat{p}_{t,1})}   \right] + \mathcal{O}_P(t^{-2}). 
\end{align}
\end{small}
This concludes the proof of this auxiliary Lemma.
\end{proof}

\newpage

\begin{proof}
(Proposition~\ref{propb})
In Lemma \ref{lemma2} we have simplified the expression for $p_{t+1,1}-p_{t,1}$, highlighting its principal part. Inspired by this result, we verify that  the updating rule for the randomization probabilities of a BUD can be written as a stochastic approximation of the following form 
  \begin{small}
\begin{align}
p_{t+1,1}&=p_{t,1} +\dfrac{1}{t}({G}_{t+1} + {r}_{t+1})\nonumber \\
&= p_{t,1}  - \dfrac{1}{t} {g}(p_{t,1},\tilde{y}_{t,1},\tilde{y}_{t,0} )+  \dfrac{1}{t}(\Delta {M}_{t+1} + {r}_{t+1}), \label{sa1}
\end{align} 
\end{small}
for a specific process ${G}_{t+1}$, where $g(p_{t,1},\tilde{y}_{t,1},\tilde{y}_{t,0})=-E_{\psi} \left(G_{t+1}\mid \Sigma_t\right)$, ${r}_{t+1}= \mathcal{O}_P(t^{-1}) $ and $\Delta {M}_{t+1}$ is a $\Sigma_t$-martingale difference sequence. 
So, Lemma \ref{lemma2} suggests us to define $ G_{t+1}$, as well as in the main text,  as
\begin{small}
\begin{align}
 G_{t+1}:= & h p_{t,1} (1-p_{t,1}) \left\{ \left[  \dfrac{(v_1+2v_2\tilde{y}_{t,1})(Y_{t+1}-\tilde{y}_{t,1}) }{v(\tilde{y}_{t,1})^2} -2\right] A_{t+1}\left[1+\left(\dfrac{p_{t,1}}{1-p_{t,1}}\right)^{\frac{1}{2h}}  \dfrac{v(\tilde{y}_{t,0})}{v(\tilde{y}_{t,1})}  \right]  \right. \nonumber \\
& \left. + \left[ 2- \dfrac{(v_1+2v_2\tilde{y}_{t,0})(Y_{t+1}-\tilde{y}_{t,0})}{v(\tilde{y}_{t,0})^2} \right] (1-A_{t+1}) \left[1+ \left(\dfrac{1-p_{t,1}}{p_{t,1}}\right)^{\frac{1}{2h}}  \dfrac{v(\tilde{y}_{t,1})}{v(\tilde{y}_{t,0})} \right]   \right\}. 
\label{eqexp5b}
 \end{align}
\end{small}
With this definition of $G_{t+1}$, the randomization probabilities of a BUD meet the above properties of the stochastic approximation \eqref{sa1}.\\
Now, $g(p_{t,1},\tilde{y}_{t,1},\tilde{y}_{t,0})=-E_{\psi} \left(G_{t+1}\mid \Sigma_t\right)$ implies that
\begin{small}
 \begin{align}
g(p_{t,1},\tilde{y}_{t,1},\tilde{y}_{t,0})&=- h p_{t,1} (1-p_{t,1}) \left\{ - 2p_{t,1}\left[1+\left(\dfrac{p_{t,1}}{1-p_{t,1}}\right)^{1/(2h)}  \dfrac{v(\tilde{y}_{t,0})}{v(\tilde{y}_{t,1})}  \right]+ \right. \nonumber\\
& \left. \quad +  2(1-p_{t,1}) \left[ 1+\left(\dfrac{1-p_{t,1}}{p_{t,1}}\right)^{1/(2h)}  \dfrac{v(\tilde{y}_{t,1})}{v(\tilde{y}_{t,0})} \right]  \right\}-\nonumber \\
 &-hp_{t,1}(1-p_{t,1}) \left\{  p_{t,1} \dfrac{   (v_1+2v_2\tilde{y}_{t,1})(b'(\psi_1)-\tilde{y}_{t,1})   }{v(\tilde{y}_{t,1})^2}  \left[1+\left(\dfrac{p_{t,1}}{1-p_{t,1}}\right)^{\frac{1}{2h}}  \dfrac{v(\tilde{y}_{t,0})}{v(\tilde{y}_{t,1})}  \right]  - \right. \\
& - \left.(1-p_{t,1}) \dfrac{ (v_1+2v_2\tilde{y}_{t,0})(b'(\psi_0)-\tilde{y}_{t,0})}{v(\tilde{y}_{t,0})^2} \left[1+\left(\dfrac{1-p_{t,1}}{p_{t,1}}\right)^{\frac{1}{2h}}  \dfrac{v(\tilde{y}_{t,1})}{v(\tilde{y}_{t,0})}  \right] \right\}.
\label{eqexp9} 
\end{align}
\end{small}
Rearranging the right-hand-side of \eqref{eqexp9}, it follows that
\begin{small}
 \begin{align}
g(p_{t,1},\tilde{y}_{t,1},\tilde{y}_{t,0})&=-{2h} \dfrac{v(\tilde{y}_{t,1})}{v(\tilde{y}_{t,0})} (1-p_{t,1})^{\frac{2h+1}{2h}} p_{t,1}^{\frac{2h-1}{2h}} \left[1+\left(\dfrac{p_{t,1}}{1-p_{t,1}}\right)^{\frac{1}{2h}}\frac{v(\tilde{y}_{t,0})}{v(\tilde{y}_{t,1})}\right]^2 \left[ \dfrac{1}{ 1+\left(\dfrac{p_{t,1}}{1-p_{t,1}}\right)^{\frac{1}{2h}}\dfrac{v(\tilde{y}_{t,0})}{v(\tilde{y}_{t,1})}  }- p_{t,1} \right]- \\
&-hp_{t,1}(1-p_{t,1}) \left\{  p_{t,1} \dfrac{   (v_1+2v_2\tilde{y}_{t,1})(b'(\psi_1)-\tilde{y}_{t,1})   }{v(\tilde{y}_{t,1})^2}  \left[1+\left(\dfrac{p_{t,1}}{1-p_{t,1}}\right)^{\frac{1}{2h}}  \dfrac{v(\tilde{y}_{t,0})}{v(\tilde{y}_{t,1})}  \right]  - \right. \\
& - \left.(1-p_{t,1}) \dfrac{ (v_1+2v_2\tilde{y}_{t,0})(b'(\psi_0)-\tilde{y}_{t,0})}{v(\tilde{y}_{t,0})^2} \left[1+\left(\dfrac{1-p_{t,1}}{p_{t,1}}\right)^{\frac{1}{2h}}  \dfrac{v(\tilde{y}_{t,1})}{v(\tilde{y}_{t,0})}  \right] \right\}. \label{eqexp10}
\end{align}
\end{small} 
Nonetheless, $\Delta {M}_{t+1}$, defined as {\small \begin{equation} \Delta {M}_{t+1} :={G}_{t+1} + g(p_{t,1},\tilde{y}_{t,1},\tilde{y}_{t,0}), \label{eqexp11}
\end{equation}} is a $\Sigma_t$- martingale difference sequence, since, by construction, its expectation with respect to $\Sigma_t$ is zero.
Additionally, $t^{-1}r_{t+1}$, defined as $(p_{t+1,1}-p_{t,1})- t^{-1}G_{t+1}$,  determined from \eqref{eqlemma1} and \eqref{eqexp5b}, is $\mathcal{O}_P(t^{-2})$, due to Lemma \ref{lemma2}.
Indeed, $r_{t+1}=\mathcal{O}_P(t^{-1})$ . \\
Analogously, we derive the stochastic approximation for $\tilde{y}_{t+1,a}$ for $a\in \{0,1\}$: equation (ii) of Lemma \ref{lemma1} suggests us to define, as in the main text, 
{\small \begin{equation}
G_{t+1,a}:= 1(A_{t+1}=a) (Y_{t+1}-\tilde{y}_{t,a}) \left[1+  \left(\dfrac{p_{t,a}}{ p_{t,1-a} }\right)^{\frac{1}{2h}} \dfrac{v(\tilde{y}_{t,1-a})}{v(\tilde{y}_{t,a})} \right]
  \label{G2}
\end{equation}}
and
{\small $$
g_a(p_{t,1},\tilde{y}_{t,1},\tilde{y}_{t,0}) := - E_{\psi}(G_{t+1,a}\mid \Sigma_t)= - p_{t,a}(b'(\psi_a)-\tilde{y}_{t,a})  \dfrac{v(\tilde{y}_{t,1-a})}{v(\tilde{y}_{t,a})}, 
$$}
so that $\tilde{y}_{t+1,a}$ satisfies the following recursive rule
\begin{small}
\begin{align}
\tilde{y}_{t+1,a}&=\tilde{y}_{t,a} +\dfrac{1}{t}({G}_{t+1,a} + {r}_{t+1,a})\nonumber \\
&= \tilde{y}_{t,a}  - \dfrac{1}{t} {g}_a(p_{t,1},\tilde{y}_{t,1},\tilde{y}_{t,0} )+  \dfrac{1}{t}(\Delta {M}_{t+1,a} + {r}_{t+1,a}),
\end{align} 
\end{small}
 where  $\Delta {M}_{t+1,a}=G_{t+1,a} +g_a(p_{t,1},\tilde{y}_{t,1},\tilde{y}_{t,0})$ is a $\Sigma_t$-martingale difference sequence and $r_{t+1,a}$, defined as $t (\tilde{y}_{t+1,a}-\tilde{y}_{t,a})-G_{t+1,a}$ from \eqref{G122} and \eqref{G2}, is $\mathcal{O}_P(t^{-1})$. \\
Indeed, joining the above results, we get the stochastic approximation for the vector $[p_{t,1}, \tilde{y}_{t,1}, \tilde{y}_{t,0}]'$ as stated in Proposition \ref{propb}.
 
\end{proof}

\newpage

\begin{proof}
(Theorem~\ref{theo})
The ordinary differential equation associated to the stochastic approximation of Proposition \ref{propb} has the following form
{\small \begin{equation} 
\left\{
\begin{matrix} 
\frac{dp}{dt} =-g(p,\tilde{y}_1,\tilde{y}_0) \\
\frac{d \tilde{y}_1}{dt} =-g_1(p,\tilde{y}_1,\tilde{y}_0) \\
\frac{d \tilde{y}_0}{dt}= -g_0(p,\tilde{y}_1,\tilde{y}_0) \\
\end{matrix}
\right. \label{ODE2}
\end{equation}}
with initial condition 
{\small \begin{equation} 
\left\{
\begin{matrix} 
p(0)= p_{0}\\
\tilde{y}_1(0)=\tilde{y}_{01} \\
\tilde{y}_0(0)=\tilde{y}_{00} \\
\end{matrix}
\right. ,
\end{equation}}
where $[p_0,\tilde{y}_{01},\tilde{y}_{00}]  \in (0,1)\times \mathbb{R}^2$ and $\tilde{g}=[g, g_1,g_0]'$ is defined in the main text. Refer to \cite{benveniste2012adaptive} and \cite{kushner2003stochastic} for a presentation of the mathematical results and theory on stochastic approximation.
We prove that
\begin{itemize}
\item[A1)] the point $[\rho_1, b'(\psi_1), b'(\psi_0)]$ is a stationary point  of the ordinary differential equation \eqref{ODE2};
 \item[A2)] $\tilde{g}$ is differentiable and the minimum eigenvalue of $D\tilde{g}( \rho_1, b'(\psi_1), b'(\psi_0))$ is  $>\dfrac{1}{2}$;
\item[A3)] {\small $
 E_{\psi}(\Delta \tilde{M}_{t+1}\Delta \tilde{M}_{t+1}'\mid\Sigma_t)  $} converges a.s. to a symmetric and definite positive matrix $\tilde{\Gamma}$ and, in particular, {\small $
 \text{Var}_{\psi}({G}_{t+1}\mid\Sigma_t)= E_{\psi}(\Delta M_{t+1}^2\mid\Sigma_t) \underset{{\small t \rightarrow \infty}}{\longrightarrow} \Gamma=\tilde{\Gamma}_{1,1} {\normalsize \text{ a.s.}} $}; 
\item[A4)] for some $\delta>0$, {\small $\underset{t}{\sup }\, E_{\psi}\left(\Vert\Delta \tilde{M}_{t+1}\Vert^{2+\delta}\mid \Sigma_t \right) < \infty $};
\item[A5)] for an $\epsilon>0$, \begin{small}$
(t+1) E_{\psi}\left( \Vert \tilde{r}_{t+1}\Vert^2 \mathbf{1}_{\{\Vert [p_{t,1},\tilde{y}_{t,1 }, \tilde{y}_{t,0}]- [\rho_1, b'(\psi_1), b'(\psi_0)]\Vert <\epsilon\}} \right) \underset{{\small t \rightarrow \infty}}{\longrightarrow}0.
$ \end{small}
\end{itemize}

Thus, from Theorem A.2 on asymptotics of stochastic approximation by Laruelle and Pag\`es in \cite{laruelle2013randomized} (see also Theorem 3 at page 110 in \cite{benveniste2012adaptive}), we can conclude that
\begin{small}
\begin{equation}
t^{1/2}\begin{bmatrix}p_{t,1}-\rho_1 \\ \tilde{y}_{t,1}- b'(\psi_1) \\ \tilde{y}_{t,0}- b'(\psi_0)\end{bmatrix}\underset{{\small t \rightarrow \infty}}{\longrightarrow} \mathcal{N}(0, \tilde{\Sigma}) \label{matrix}
\end{equation}
\end{small}
where \begin{small}
\begin{equation}
\tilde{\Sigma}:= \int_0^{\infty}  \left(e^{ -( D\tilde{g}( \rho_1, b'(\psi_1), b'(\psi_0))  - \frac{I_3}{2})u}\right)' \tilde{\Gamma} e^{ -( D\tilde{g}( \rho_1, b'(\psi_1), b'(\psi_0))  - \frac{I_3}{2})u} \text{d}u . \label{sigmatilde}
\end{equation}
\end{small}
In the following steps we verify that A1)-A5) are satisfied by the stochastic approximation of Proposition \ref{propb}  and we compute the asymptotic variance of $p_{t,1}$. \\
\\
\textit{STEP 1: Assumptions A1)-A2), ODE, stationarity and stability} \\
The unique stationary point of the ODE \eqref{ODE2} is $[\rho_1, b'(\psi_1), b'(\psi_0)]$, since 
\begin{small}
\begin{equation}
\tilde{g}(p_{t,1},\tilde{y}_{t,1 }, \tilde{y}_{t,0})=0 \quad {\normalsize \text{if and only if}} \quad [p_{t,1},\tilde{y}_{t,1 }, \tilde{y}_{t,0}]=[\rho_1,b'(\psi_1), b'(\psi_0)].
\end{equation}
\end{small}
Moreover, standard computations show that the differential of $\tilde{g}$ evaluated at the equilibrium point takes value 
{\small \begin{equation} 
D\tilde{g}( \rho_1, b'(\psi_1), b'(\psi_0))=\begin{bmatrix}1+2h &0 &0\\ 0&1&0  \\ 0&0&1\end{bmatrix}. \label{matrix2}
\end{equation}}
Thus the minimum eigenvalue of $D\tilde{g}( \rho_1, b'(\psi_1), b'(\psi_0))$ is $1> \dfrac{1}{2}$. \\ 
\\
\textit{STEP 2: Assumption A3), finiteness of the limiting variance } \\
In order to prove that the matrix $\tilde{\Gamma}={\small \lim}_{t \to \infty}E_{\psi}(\Delta \tilde{M}_{t+1} \Delta \tilde{M}_{t+1}'\mid \Sigma_t)$ is positive definite  it is sufficient to show that the diagonal elements of the matrix obtained by the triangularization of $\tilde{\Gamma}$ are positive. In fact, by Sylvester's criterion, $\tilde{\Gamma}$ is positive definite if and only if all the $k^{th}$ leading principal minor of the matrix are positive for $k=1,2,3$. Now, by using elementary row operations, the matrix can be reduced to an upper triangular matrix and, since the $k^{th}$ leading principal minor of a triangular matrix is the product of its diagonal elements up to row $k$, Sylvester's criterion is equivalent to checking whether its diagonal elements are all positive.\\
The components of the matrix $\tilde{\Gamma}$ can be determined combining the explicit expression of the conditional expectation of the pairwise products of the components of $\Delta \tilde{M}_{t+1}$, expressed in terms of the explicit expressions of $\tilde{G}_{t+1}$ and $\tilde{g}$, and the following remarks:\\
a) $E_{\psi_1}(A_{t+1}(Y_{t+1}- \tilde{y}_{t,1})\mid \Sigma_t) \underset{{\small t \rightarrow \infty}}{\longrightarrow} 0$ and $E_{\psi_0}((1-A_{t+1})(Y_{t+1}- \tilde{y}_{t,0})\mid \Sigma_t) \underset{{\small t \rightarrow \infty}}{\longrightarrow} 0$ since $\tilde{y}_{t,a}= \dfrac{\sum_{s=1}^t Y_s 1(A_s=a)}{t\hat{p}_{t,a}}+ \mathcal{O}_P(t^{-1})$ for $a \in \{0,1\}$ and the law of large numbers can be applied to the outcomes of the two arms;\\
b) $E_{\psi_1}(b'(\psi_1)- \tilde{y}_{t,1}\mid \Sigma_t) \underset{{\small t \rightarrow \infty}}{\longrightarrow} 0$ and $E_{\psi_0}(b'(\psi_0)- \tilde{y}_{t,0}\mid \Sigma_t) \underset{{\small t \rightarrow \infty}}{\longrightarrow} 0$ due to a similar reasoning as above; \\
c) the conditional expectation of products containing $A_{t+1}$ and $(1-A_{t+1})$ as factors vanishes; \\
d) $p_{t,1},\tilde{y}_{t,1 }, \tilde{y}_{t,0}$ converge.\\
Thus,   
\begin{small}
\begin{align}
\tilde{\Gamma}_{1,1}&=\Gamma = \lim_{t \rightarrow \infty} \text{Var}_{\psi}(G_{t+1}\mid\Sigma_t) \nonumber \\
&=h^2 \rho_1^2(1-\rho_1)^2\left[\dfrac{(v_1+2v_2b'(\psi_1))^2}{\rho_1\sigma_1^2}+  \dfrac{(v_1+2v_2b'(\psi_0))^2}{(1-\rho_1)\sigma_0^2} +\dfrac{4}{\rho_1}+\dfrac{4}{1-\rho_1}\right], \label{varvar} \\
\tilde{\Gamma}_{2,2}& = \lim_{t \rightarrow \infty} \text{Var}_{\psi}(G_{t+1,1}\mid\Sigma_t) \nonumber \\
&=\dfrac{\sigma_1^2}{\rho_1}, \label{varvar3} \\
\tilde{\Gamma}_{3,3}& = \lim_{t \rightarrow \infty} \text{Var}_{\psi}(G_{t+1,0}\mid\Sigma_t) \nonumber \\
&=\dfrac{\sigma_0^2}{1-\rho_1}, \label{varvar2} \\
\tilde{\Gamma}_{1,2}&=\tilde{\Gamma}_{2,1}= \lim_{t \rightarrow \infty} \text{Var}_{\psi}( \Delta M_{t+1}\Delta M_{t+1,1} \mid\Sigma_t) \nonumber \\
&=h(1-\rho_1)(v_1+2v_2b'(\psi_1)), \nonumber \\
\tilde{\Gamma}_{1,3}&=\tilde{\Gamma}_{3,1}= \lim_{t \rightarrow \infty} \text{Var}_{\psi}( \Delta M_{t+1}\Delta M_{t+1,0} \mid\Sigma_t)  \nonumber \\
&= h \rho_1(v_1+2v_2b'(\psi_0)), \nonumber \\
\tilde{\Gamma}_{2,3}&=\tilde{\Gamma}_{3,2}= \lim_{t \rightarrow \infty} \text{Var}_{\psi}( \Delta M_{t+1,1}\Delta M_{t+1,0} \mid\Sigma_t) \nonumber \\
&=0.
\end{align}
\end{small}
To triangularize $\tilde{\Gamma}$, it is sufficient to substitute the first row by a linear combination of the second and third rows, so that the elements (1,2) and (1,3) of the matrix vanish. In particular the entry $(1,1)$ becomes \begin{small}
\begin{align}
\tilde{\Gamma}_{3,3}(\tilde{\Gamma}_{2,2}\tilde{\Gamma}_{1,1}- \tilde{\Gamma}_{1,2}^2)-\tilde{\Gamma}_{1,3}^2\tilde{\Gamma}_{2,2}&=4h^2\sigma_0^2\sigma_1^2 \nonumber \\
&>0. \label{defpos}
\end{align}
\end{small}
Since the above inequality holds and $\tilde{\Gamma}_{2,2}>0, \tilde{\Gamma}_{3,3}>0$, we can conclude that the matrix $\tilde{\Gamma}$ is positive definite.\\
\\
\noindent \textit{ STEP 3: Assumption A4), finiteness of $(2+\delta)^{th}$-moment}\\  
To prove A4) it is sufficient to prove the finiteness of ${\small\underset{t}{\sup}}\;E_{\psi}(\Delta M_{t+1}^{2+\delta}\mid \Sigma_t), {\small\underset{t}{\sup}}\;E_{\psi}(\Delta M_{t+1,1}^{2+\delta}\mid \Sigma_t)$ and ${\small\underset{t}{\sup}}\;E_{\psi}(\Delta M_{t+1,0}^{2+\delta}\mid \Sigma_t)$ separately.
But this follows from  the convergence of $[p_{t,1},\tilde{y}_{t,1 }, \tilde{y}_{t,0}]$ and the finiteness of the moments of distributions in the natural exponential family with quadratic variance function.\\
\\
\textit{STEP 4: Assumption A5), remainder term} \\
Assumption A5) is a consequence of the construction of the remainder term $\tilde{r}_{t+1}$.\\
\noindent Recall that $r_{t+1}, r_{t+1,1}$ and $r_{t+1,0}$ are $\mathcal{O}_P(t^{-1})$ as stated in Proposition 2 and they have been obtained by isolating the dominant terms in the expression for $t(p_{t+1,1}-p_{t,1}),t(\tilde{y}_{t,1 }-b'(\psi_1)),t( \tilde{y}_{t,0}-b'(\psi_0)) $ and subtracting the components of $\tilde{G}_{t+1}$, respectively.\\
Thus, if $\Vert [p_{t,1},\tilde{y}_{t,1 }, \tilde{y}_{t,0}]- [\rho_1, b'(\psi_1), b'(\psi_0)]\Vert <\epsilon$ for some $\epsilon>0$, then also $\mid \sigma_{t,1}^2-\sigma^2_1\mid< \delta_1$, $\mid \sigma_{t,0}^2-\sigma^2_0\mid< \delta_2$ for some $\delta_1, \delta_2$ and $\hat{p}_{t,1} \in K$, where $K$ is a compact subset of $(0,1)$, since it can be computed as in \eqref{eqexphat}. Under this conditions, for $\forall w$, $t^2 \Vert \tilde{r}_{t+1} \Vert ^2$ is, by construction, an algebraic function of random variables that have small variability around their limits, which are different from zero and are finite, and, therefore, it is bounded. 
This implies that 
\begin{small}
\begin{equation}
(t+1) E_{\psi}\left( \Vert \tilde{r}_{t+1}\Vert^2 \mathbf{1}_{\{\Vert [p_{t,1},\tilde{y}_{t,1 }, \tilde{y}_{t,0}]- [\rho_1, b'(\psi_1), b'(\psi_0)]\Vert <\epsilon\}} \right) \underset{{\small t \rightarrow \infty}}{\longrightarrow}0. \label{remaind1}
\end{equation}
\end{small} \\
\textit{STEP 5: SA theorem} \\
From the above steps, we have shown that Theorem A.2 on asymptotics of stochastic approximation by Laruelle and Pag\`es in \cite{laruelle2013randomized} holds: it follows that
 {\small $$ t^{1/2}(p_{t,1} - \rho_1)\underset{{\small t \rightarrow \infty}}{\longrightarrow} \mathcal{N}(0, \Sigma ),$$}
where $\Sigma$ is the entry $(1,1)$ of the matrix $\tilde{\Sigma}$, defined in \eqref{sigmatilde}. Thus, we have
\begin{small}
\begin{equation}
\tilde{\Sigma}=\int_0^{\infty}  \sum_{k=0}^{\infty}\frac{1}{k!}\left(\left(- D\tilde{g}( \rho_1, b'(\psi_1), b'(\psi_0))  + \frac{I_3}{2}\right)^k \right)' u^k  \tilde{\Gamma}   \sum_{j=0}^{\infty}\frac{1}{j!}\left(- D\tilde{g}( \rho_1, b'(\psi_1), b'(\psi_0))  + \frac{I_3}{2}\right)^j u^j \text{d}u .
\end{equation}
\end{small}
and, due to the diagonal structure of the matrix $D\tilde{g}( \rho_1, b'(\psi_1), b'(\psi_0))$ computed in \eqref{matrix2}, we write
\begin{small} 
\begin{align*}
\Sigma &= \int_0^{\infty} \sum_{k,j=0}^{\infty} \frac{1}{k!}\left( -\frac{1}{2}-2h\right)^k \Gamma \frac{1}{j!}\left( -\frac{1}{2}-2h\right)^j u^{k+j} \text{d}u \nonumber \\
& = \int_0^{\infty} {e^{2(-\frac{1}{2}-2h)u}  \Gamma \text{d}u} \nonumber \\
& = \dfrac{\Gamma}{1+4h},
\end{align*}
\end{small}
where $\Gamma$ has been computed in \eqref{varvar}. With similar reasoning as above, we obtain the multivariate Central Limit type result \eqref{matrix} for  $[p_{t,1},\tilde{y}_{t,1 }, \tilde{y}_{t,0}]'$ , where the asymptotic variance-covariance matrix given in \eqref{sigmatilde} becomes 
\begin{small}
\begin{equation}
\tilde{\Sigma}=\begin{bmatrix}\dfrac{\Gamma}{1+4h} &0 &0 \\
0& \dfrac{\sigma_1^2}{\rho_1} &0 \\0&0&\dfrac{\sigma_0^2}{1-\rho_1}\end{bmatrix}. \label{matrixsigma}
\end{equation}
\end{small}
This completes the proof.
 
\end{proof}

\begin{proof}
(Corollary~\ref{prop4})
First, by inverting the definitory equation
\begin{small}
\begin{equation}
p_{t,1}=\dfrac{1}{ 1+ \dfrac{\sigma_{t,0}^{2h}}{\sigma_{t,1}^{2h}} \left( \dfrac{t_{0,1} +1 + t\hat{p}_{t,1}  }{t_{0,0}+1+t(1-\hat{p}_{t,1}) }\right)^{2h} } ,
\end{equation}
\end{small}
and by (i) of Lemma \ref{lemma1}, we have
\begin{small}
\begin{equation}
\hat{p}_{t,1} = \dfrac{1}{1+ \left(\dfrac{p_{t,1}}{1-p_{t,1}}\right)^{\frac{1}{2h}}  \dfrac{v(\tilde{y}_{t,0})}{v(\tilde{y}_{t,1})}  } +  \mathcal{O}_P(t^{-1}).
\label{eqhatpt}
\end{equation}
\end{small}
Second, starting from the asymptotic result \eqref{matrix} with \eqref{matrixsigma}, we can apply a multivariate Delta Method and Slutsky Theorem (see 5.5.17 and 5.5.24 of \cite{casella2002statistical}) to deduce the asymptotic normality of the allocation proportion of a BUD. In particular, we use the principal part of \eqref{eqhatpt}, that is the function $f(p_{t,1},\tilde{y}_{t,1 }, \tilde{y}_{t,0}) =\dfrac{1}{1+ \left(\dfrac{p_{t,1}}{1-p_{t,1}}\right)^{\frac{1}{2h}}  \dfrac{v(\tilde{y}_{t,0})}{v(\tilde{y}_{t,1})}  }$, to conclude  that 
{\small $$ t^{1/2}(f(p_{t,1},\tilde{y}_{t,1 }, \tilde{y}_{t,0})- f(\rho_1, b'(\psi_1), b'(\psi_0))) \underset{{\small t \rightarrow \infty}}{\longrightarrow} \mathcal{N}(0, \nabla f(\rho_1, b'(\psi_1), b'(\psi_0))' \tilde{\Sigma} \nabla f(\rho_1, b'(\psi_1), b'(\psi_0))),$$}where $\tilde{\Sigma}$ is the diagonal matrix given in \eqref{matrixsigma}.  Standard calculations show that
$f(\rho_1, b'(\psi_1), b'(\psi_0)))=\rho_1$ and that
 \begin{small}
$$ \nabla f (\rho_1, b'(\psi_1), b'(\psi_0))'= \left[ \dfrac{1}{2h}, \dfrac{\rho_1(1-\rho_1)}{2\sigma_1^2}(v_1+2v_2b'(\psi_1)), -\dfrac{\rho_1(1-\rho_1)}{2\sigma_0^2}(v_1+2v_2b'(\psi_0)) \right].$$ 
\end{small} Thus, the asymptotic variance of the allocation proportion $\hat{p}_{t,1}$ is equal to {\small $$\dfrac{\Gamma}{4h^2(1+4h)} + \dfrac{\rho_1(1-\rho_1)^2}{4\sigma_1^2}(v_1+2 v_2 b'(\psi_1))^2+ \dfrac{\rho_1^2(1-\rho_1)}{4\sigma_0^2}(v_1+2 v_2 b'(\psi_0))^2, $$} that can be re-written as $$ \dfrac{\rho_1^2(1-\rho_1)^2}{4}\left[ \left(\dfrac{(v_1+2v_2 b'(\psi_1))^2}{\rho_1 \sigma_1^2}+ \dfrac{(v_1+2v_2 b'(\psi_0))^2}{(1-\rho_1) \sigma_0^2} \right)\left(1+ \dfrac{1}{1+4h} \right) + \dfrac{4}{\rho_1(1+4h)}+ \dfrac{4}{(1-\rho_1)(1+4h)}\right]. $$

\end{proof}
\textit{ Behavior of the MLE of $\theta_a$}\\
The maximum-likelihood estimates (MLE) $\widehat \theta_{t,a} $
of the unknown true mean response to treatment 
$a=0,1$ within the NEF of outcome models,
under the sequential BUD design, have the same limiting distribution as the MLE of a study design  with fixed  sample size. Whilst we are considering a response-adaptive procedure, a version of the central limit theorem for the MLE arising in the classical setting of independent and identically distributed random variables is preserved. For the  proof of this result we refer to \cite{hu2006theory} that presents and proves it in a more general framework (see Theorem 3.1). 
\vspace{1cm}

\textit{Regularity conditions (Lemma \ref{lemma4})}\\
In this section, we provide a less stringent set of conditions under which the approximation of the information gain   \eqref{approxus1}    holds. For simplicity, we consider observations $Y_1,\dots,Y_n$ from a single arm. Let us denote by $\mathcal F_n$ the sigma algebra generated by them. \\
Assume that the parameter space $\Theta\subset \mathbb R$ is a bounded open interval, that the true value of the parameter $\theta_0$ is an interior point of $\Theta$ and that the prior is the uniform distribution on $\Theta$.
Let $f(y,\theta)$ and $l(y,\theta)$ denote the density function and the log-likelihood of the observations, respectively.
Denote by \begin{small}
\[
\dot f(y,\theta)=\frac{\partial }{\partial \theta}f(y,\theta),\quad \ddot f(y,\theta)=\frac{\partial^2 }{\partial \theta^2}f(y,\theta)
\]
\end{small}
and by $\hat \theta_n$ the MLE based on a sample of size $n$.
Also, let \begin{small}
\[ f(y, \theta, \rho) = \sup_{|\theta- \theta'|\leq \rho} f(y, \theta ') , \quad \quad Q(y,r)=\sup_{|\theta|>r} f(y, \theta)
\]
\end{small}
for $\rho, r>0$.\\
We require regularity conditions of Johnson in \cite{johnson1970} to hold:\\
a) $f$ is three times continuously differentiable with respect to $\theta$.\\
b) For every $\theta \in \bar{\Theta}$ and $\rho,r>0$, $f(y, \theta, \rho)$ and $Q(y,r)$ are measurable functions of $y$ and that, for sufficiently small $\rho$ and sufficiently large $r$,
\begin{small}
\[E_{\theta_0}\left[\log f(Y,\theta,\rho)\right]^+ <\infty, \quad \quad E_{\theta_0}\left[\log Q(Y,r)\right]^+ <\infty. \]
\end{small}
\noindent c) There exists $G_k$ for $k=1,2$  satisfying
$\mid \frac{\partial^k}{\partial \theta^k}l(y,\theta)\mid\leq G_k(y)$
for $\theta$ in a neighborhood of $\theta_0$ and $E_{\theta_0}G_k(Y)<\infty.$ \\
In addition to assumptions a)-c) we assume that the following conditions are satisfied:\\
d) There exists $G_3$ such that
$\sup_{|\theta-\hat \theta_n|<\delta} |\frac{\partial^3}{\partial \theta^3}l(y,\theta)|\leq G_3(y)$
and
$E_{\theta_0}G_3(Y)<\infty.$\\
e) For some $\delta>0$ 
\begin{small}
\begin{align}
& \int \sup_{|\theta-\hat \theta_n|<\delta} |\ddot f(y,\theta)|dy=O_P(1),\label{bound1}\\
& \int \sup_{|\theta-\hat \theta_n|<\delta}|\ddot f(y,\theta)|\frac{|\dot f(y,\hat \theta_n)|}{f(y,\hat \theta_n)}dy=O_P(1),\label{bound2}\\
& \int \sup_{|\theta-\hat \theta_n|<\delta}\frac{\ddot f(y,\theta)^2}{f(y,\hat \theta_n)}dy=O_P(1),\label{bound3}\\
&\int 
\sup_{|\theta_1-\hat \theta_n|<\delta,|\theta_2-\hat \theta_n|<\delta}
\frac{	|\dot f(y,\theta_1)| }{f(y,\theta_2)}f(y,\hat \theta_n)
dy=O_P(1), \label{bound4}\\
&\int \sup_{|\theta_1-\hat \theta_n|<\delta,|\theta_2-\hat \theta_n|<\delta,|\theta_3-\hat \theta_n|<\delta}
\frac{\dot f(y,\theta_1)^2|\dot f(y,\theta_2)|}{f(y,\theta_3)f(y,\hat \theta_n)}dy=O_P(1). \label{bound5}
\end{align}
\end{small}
\newpage

\textit{Additional Lemma (complementary to the proof of Lemma \ref{lemma4})}\\
\begin{lemma}[Supplementary]\label{lemmaA2}
Let  $Y_1,Y_2, \dots$ be a sequence of random variables satisfying the regularity conditions of the previous section. For every $n$, let $\mathcal F_n$ the sigma algebra generated by $Y_1, \dots, Y_n$. Then,
$\Delta_n:=\text{Var}(\theta\mid \mathcal F_{n})-E(\text{Var}(\theta\mid\mathcal F_{n+1})\mid\mathcal F_n)=I_{\theta_0}^{-1} n^{-2}+o_P(n^{-2})$.
\end{lemma}
Before proving Lemma \ref{lemmaA2} (Supplementary), let us introduce some further notation and preliminary results. \\
First, notice that, for every $\theta$,
\begin{small}
\begin{equation}
\int \dot f(y,\theta)dy=0. \label{intzero}
\end{equation}
\end{small}
\noindent Let {\small $$a_{k,n}(\theta)=\frac{1}{n}\sum_{i=1}^n \frac{\partial^k}{\partial \theta^k}\log f(Y_i,\theta).$$}
By definition $a_{1,n}(\hat{\theta}_n)=0$. Moreover
\begin{small}
\[
a_{2,n}(\hat \theta_n)\rightarrow -I_{\theta_0}\quad a.s.
\]
\end{small}
as $n\rightarrow\infty$,
where 
$I_{\theta}=E_\theta(\frac{\partial}{\partial \theta} \log f(Y,\theta))^2=-E_\theta(\frac{\partial^2}{\partial \theta^2} \log f(Y,\theta))).$
Also,
\begin{small}
\[
\prod_{i=1}^n\frac{f(Y_i,\hat\theta_n+\phi)}{f(Y_i,\hat\theta_n)}=\exp(\frac{1}{2}na_{2,n}(\hat \theta_n)\phi^2+\frac{1}{6}na_{3,n}(\theta^*_n)\phi^3),
\]
\end{small}
\noindent for some $\theta^*_n=\theta^*_n(\phi)$ that satisfies $|\theta^*_n-\hat \theta_n|<\phi$.\\
By the change of variable $u=\sqrt n \phi$, we obtain 
\begin{small}
\[
\prod_{i=1}^n\frac{f(Y_i,\hat\theta_n+u/\sqrt n)}{f(Y_i,\hat\theta_n)}=\exp(a_{2,n}(\hat \theta_n)u^2/2+a_{3,n}(\theta^*_n)u^3/(6\sqrt n)),
\]
\end{small}
\noindent for some $\theta^*_n=\theta^*_n(u)$ that satisfies $|\theta^*_n-\hat \theta_n|<u/\sqrt n$.\\
Let $C_n(u)=a_{2,n}(\hat \theta_n)u^2/2+a_{3,n}(\theta^*_n)u^3/(6\sqrt n)$, then
\begin{small}
\begin{equation}
\prod_{i=1}^n\frac{f(Y_i,\hat\theta_n+u/\sqrt n)}{f(Y_i,\hat\theta_n)}=e^{C_n(u)}. \label{change}
\end{equation}
\end{small}
By Lemma 2.2, Lemma 2.3 and (2.5) in \cite{johnson1970} there exist $\epsilon, \delta$ and $N_0$ such that, $P$-a.s.,
\begin{small}
\begin{equation}
\label{ineq1}
\prod_{i=1}^n\frac{f(Y_i,\hat\theta_n+\phi)}{f(Y_i,\hat\theta_n)}\leq \exp(-\phi^2/12)\quad \mbox{ for }|\phi|\leq \delta \mbox{ and }n\geq N_0
\end{equation}
\end{small}
and
\begin{small}
\begin{equation}
\label{ineq71}
\prod_{i=1}^n\frac{f(Y_i,\hat\theta_n+\phi)}{f(Y_i,\hat\theta_n)}\leq \exp(-n\epsilon)\quad \mbox{ for }|\phi|>\delta \mbox{ and }n\geq N_0.
\end{equation}
\end{small}
By the change of variable $u=\sqrt n \phi$, we obtain
\begin{small}
\begin{equation}
\label{ineq11}
\prod_{i=1}^n\frac{f(Y_i,\hat\theta_n+u/\sqrt n)}{f(Y_i,\hat\theta_n)}\leq \exp(-u^2/(12n))\quad \mbox{ for }|u|\leq \delta\sqrt n \mbox{ and }n\geq N_0 \quad \text{a.s.}
\end{equation}
\end{small}
and
\begin{small}
\begin{equation}
\label{ineq12}
\prod_{i=1}^n\frac{f(Y_i,\hat\theta_n+u/\sqrt n)}{f(Y_i,\hat\theta_n)}\leq \exp(-n\epsilon)\quad \mbox{ for }|u|>\delta\sqrt n \mbox{ and }n\geq N_0\quad \text{a.s.}.
\end{equation}
\end{small}
The first inequality can be rewritten as
\begin{small}
\begin{equation}
\label{ineq111}
e^{C_n(u)}\leq \exp(-u^2/(12n))\quad \mbox{ for }|u|\leq \delta\sqrt n \mbox{ and }n\geq N_0 \quad \text{a.s.}.
\end{equation}
\end{small}

\begin{proof}
(Lemma~\ref{lemmaA2} - Supplementary)
First, in STEP 1, we show that 
\begin{small}
\[\Delta_n
=
\frac{1}{\int  
	\prod_{i=1}^n
	\frac{f(Y_i,\hat \theta_n+\phi)}{f(Y_i,\hat \theta_n)}d\phi}
\left(
\int
\frac{
	\left(
	\int \phi f(y,\hat\theta_n+\phi)
	\prod_{i=1}^n
	\frac{f(Y_i,\hat \theta_n+\phi)}{f(Y_i,\hat \theta_n)}d\phi
	\right)^2
}{
	\int  f(y,\hat\theta_n+\phi)
	\prod_{i=1}^n
	\frac{f(Y_i,\hat \theta_n+\phi)}{f(Y_i,\hat \theta_n)}d\phi
}
dy
-
\frac{
	\left(
	\int \phi 
	\prod_{i=1}^n
	\frac{f(Y_i,\hat \theta_n+\phi)}{f(Y_i,\hat \theta_n)}d\phi
	\right)^2
}{
	\int  
	\prod_{i=1}^n
	\frac{f(Y_i,\hat \theta_n+\phi)}{f(Y_i,\hat \theta_n)}d\phi
}
\right).
\]
\end{small}
\noindent Then, we provide a useful approximation of $\Delta_n$: let $\delta$ be defined such that inequalities \eqref{ineq11} and \eqref{ineq12} hold,  and let 
\begin{small}
\[
\tilde \Delta_n=\frac{1}{\int_{-\delta}^\delta 
	\prod_{i=1}^n
	\frac{f(Y_i,\hat \theta_n+\phi)}{f(Y_i,\hat \theta_n)}d\phi}
\left(
\int
\frac{
	\left(
	\int_{-\delta}^\delta \phi f(y,\hat \theta_n+\phi)
	\prod_{i=1}^n
	\frac{f(Y_i,\hat \theta_n+\phi)}{f(Y_i,\hat \theta_n)}d\phi
	\right)^2
}{
	\int_{-\delta}^\delta  f(y,\hat \theta_n+\phi)
	\prod_{i=1}^n
	\frac{f(Y_i,\hat \theta_n+\phi)}{f(Y_i,\hat \theta_n)}d\phi
}
dy
-
\frac{
	\left(
	\int_{-\delta}^\delta \phi 
	\prod_{i=1}^n
	\frac{f(Y_i,\hat \theta_n+\phi)}{f(Y_i,\hat \theta_n)}d\phi
	\right)^2
}{
	\int_{-\delta}^\delta  
	\prod_{i=1}^n
	\frac{f(Y_i,\hat \theta_n+\phi)}{f(Y_i,\hat \theta_n)}d\phi
}
\right).
\]
\end{small}
\noindent The proof proceeds by proving in STEP 2 that {\small $\tilde \Delta_n=I_{\theta_0}^{-1}n^{-2}+o_P(n^{-2})$} and then in STEP 3 that
\begin{small}
\begin{equation}
\Delta_n=\tilde \Delta_n+o_P(n^{-2}). \label{approxus}
\end{equation}
\end{small}
By extending the above approximation of the information increment to the two arm setting, we conclude the proof of Lemma \ref{lemmaA2} (Supplementary). \\

\noindent \textit{STEP 1: Expression of $\Delta_n$}  \\
\noindent Since 
\begin{small}
\[
\text{Var}(\theta\mid \mathcal F_{n+1})=E(\theta^2\mid \mathcal F_{n+1})-E(\theta\mid\mathcal F_{n+1})^2,
\]
\end{small}
\noindent then
\begin{small}
\[
E(\text{Var}(\theta\mid\mathcal F_{n+1})\mid\mathcal F_n)=E(\theta^2\mid\mathcal F_n)-E(E(\theta\mid\mathcal F_{n+1})^2\mid\mathcal F_n).\]
\end{small}
Thus,
\begin{small}
\[
\begin{aligned}
\Delta_n&=E(E(\theta\mid\mathcal F_{n+1})^2\mid\mathcal F_n)-E(\theta\mid \mathcal F_n)^2\\
&=E(E(\theta-\hat{\theta}_n\mid\mathcal F_{n+1})^2\mid\mathcal F_n)-E(\theta-\hat \theta_n\mid \mathcal F_n)^2.
\end{aligned}
\]
\end{small}
It holds
\begin{small}
\begin{align*}
\Delta_n&=\int
\frac{
	\left(
	\int \phi f(y,\hat\theta_n+\phi)
	\prod_{i=1}^n
	\frac{f(Y_i,\hat \theta_n+\phi)}{f(Y_i,\hat \theta_n)}d\phi
	\right)^2
}{
	\left(
	\int  f(y,\hat\theta_n+\phi)
	\prod_{i=1}^n
	\frac{f(Y_i,\hat \theta_n+\phi)}{f(Y_i,\hat \theta_n)}d\phi
	\right)^2
}
\frac{
	\int f(y,\hat\theta_n+\phi)
	\prod_{i=1}^n
	\frac{f(Y_i,\hat \theta_n+\phi)}{f(Y_i,\hat \theta_n)}d\phi
}{
	\int  
	\prod_{i=1}^n
	\frac{f(Y_i,\hat \theta_n+\phi)}{f(Y_i,\hat \theta_n)}d\phi
}
dy -\\
&
-
\frac{
	\left(
	\int \phi 
	\prod_{i=1}^n
	\frac{f(Y_i,\hat \theta_n+\phi)}{f(Y_i,\hat \theta_n)}d\phi
	\right)^2
}{
	\left(
	\int  
	\prod_{i=1}^n
	\frac{f(Y_i,\hat \theta_n+\phi)}{f(Y_i,\hat \theta_n)}d\phi
	\right)^2
}
\\ &=\int
\frac{
	\left(
	\int \phi f(y,\hat\theta_n+\phi)
	\prod_{i=1}^n
	\frac{f(Y_i,\hat \theta_n+\phi)}{f(Y_i,\hat \theta_n)}d\phi
	\right)^2
}{
	\int  f(y,\hat\theta_n+\phi)
	\prod_{i=1}^n
	\frac{f(Y_i,\hat \theta_n+\phi)}{f(Y_i,\hat \theta_n)}d\phi
	\int  
	\prod_{i=1}^n
	\frac{f(Y_i,\hat \theta_n+\phi)}{f(Y_i,\hat \theta_n)}d\phi
}
dy
-
\frac{
	\left(
	\int \phi 
	\prod_{i=1}^n
	\frac{f(Y_i,\hat \theta_n+\phi)}{f(Y_i,\hat \theta_n)}d\phi
	\right)^2
}{
	\left(
	\int  
	\prod_{i=1}^n
	\frac{f(Y_i,\hat \theta_n+\phi)}{f(Y_i,\hat \theta_n)}d\phi
	\right)^2
}
\\ &
=
\frac{1}{\int  
	\prod_{i=1}^n
	\frac{f(Y_i,\hat \theta_n+\phi)}{f(Y_i,\hat \theta_n)}d\phi}
\left(
\int
\frac{
	\left(
	\int \phi f(y,\hat\theta_n+\phi)
	\prod_{i=1}^n
	\frac{f(Y_i,\hat \theta_n+\phi)}{f(Y_i,\hat \theta_n)}d\phi
	\right)^2
}{
	\int  f(y,\hat\theta_n+\phi)
	\prod_{i=1}^n
	\frac{f(Y_i,\hat \theta_n+\phi)}{f(Y_i,\hat \theta_n)}d\phi
}
dy
-
\frac{
	\left(
	\int \phi 
	\prod_{i=1}^n
	\frac{f(Y_i,\hat \theta_n+\phi)}{f(Y_i,\hat \theta_n)}d\phi
	\right)^2
}{
	\int  
	\prod_{i=1}^n
	\frac{f(Y_i,\hat \theta_n+\phi)}{f(Y_i,\hat \theta_n)}d\phi
}
\right)
\end{align*}
\end{small}
\\

\noindent \textit{STEP 2: $\tilde \Delta_n=I_{\theta_0}^{-1}n^{-2}+o_P(n^{-2})$} \\
\noindent We have 
\begin{small}
\begin{align}
\tilde \Delta_n &=\frac{1}{\int_{-\delta}^\delta 
	\prod_{i=1}^n
	\frac{f(Y_i,\hat \theta_n+\phi)}{f(Y_i,\hat \theta_n)}d\phi}
\left(
\int
\frac{
	\left(
	\int_{-\delta}^\delta \phi f(y,\hat \theta_n+\phi)
	\prod_{i=1}^n
	\frac{f(Y_i,\hat \theta_n+\phi)}{f(Y_i,\hat \theta_n)}d\phi
	\right)^2
}{
	\int_{-\delta}^\delta  f(y,\hat \theta_n+\phi)
	\prod_{i=1}^n
	\frac{f(Y_i,\hat \theta_n+\phi)}{f(Y_i,\hat \theta_n)}d\phi
}
dy
-
\frac{
	\left(
	\int_{-\delta}^\delta \phi 
	\prod_{i=1}^n
	\frac{f(Y_i,\hat \theta_n+\phi)}{f(Y_i,\hat \theta_n)}d\phi
	\right)^2
}{
	\int_{-\delta}^\delta  
	\prod_{i=1}^n
	\frac{f(Y_i,\hat \theta_n+\phi)}{f(Y_i,\hat \theta_n)}d\phi
}
\right)
 \label{deftilde}\\
&=\frac{1}{n}
\frac{1}{\int_{-\delta\sqrt n}^{\delta \sqrt n} 
	e^{C_n(u)}du}
\left(
\int
\frac{
	\left(
	\int_{-\sqrt n\delta}^{\sqrt n\delta} u f(y,\hat \theta_n+u/\sqrt n)e^{C_n(u)}
	du
	\right)^2
}{
	\int_{-\sqrt n\delta}^{\sqrt n\delta}  f(y,\hat \theta_n+u/\sqrt n)e^{C_n(u)}
	du
}
dy
-
\frac{
	\left(
	\int_{-\sqrt n\delta}^{\sqrt n\delta}u e^{C_n(u)} 
	du
	\right)^2
}{
	\int_{-\sqrt n \delta}^{\sqrt n\delta} e^{C_n(u)}  
	du
}
\right), \label{changeofvar}
\end{align}
\end{small}
\noindent where \eqref{deftilde} is the definition of $\tilde \Delta_n$ and \eqref{changeofvar} is a consequence of the change of variable $u=\sqrt n \phi$ and \eqref{change}.
Thus,
\begin{small}
\begin{align}
\tilde \Delta_n&=\frac{1}{n}
\frac{1}{\int_{-\delta\sqrt n}^{\delta \sqrt n} 
	e^{C_n(u)}du}
\left(
\int
\frac{
	\left(
	\int_{-\sqrt n\delta}^{\sqrt n\delta} u f(y,\hat \theta_n+u/\sqrt n)e^{C_n(u)}
	du
	\right)^2
}{
	\int_{-\sqrt n\delta}^{\sqrt n\delta}  f(y,\hat \theta_n+u/\sqrt n)e^{C_n(u)}
	du
}
dy
-
\frac{
	\left(
	\int_{-\sqrt n\delta}^{\sqrt n\delta}u f(y,\hat{\theta}_n)e^{C_n(u)} 
	du
	\right)^2
}{
	\int_{-\sqrt n \delta}^{\sqrt n\delta} f(y,\hat{\theta}_n) e^{C_n(u)}  
	du
}
\right) \nonumber \\
&=\frac{1}{n}\frac{1}{\int_{-\delta\sqrt n}^{\delta \sqrt n}e^{C_n(u)}du}
\int\left(\frac{{A_n'(y)}^2}{B_n'(y)}-\frac{A_n(y)^2}{B_n(y)}\right)dy. \label{AB}
\end{align}

\end{small}
where
\begin{small}
\[
\begin{aligned}
&A_n'(y)=	\int_{-\sqrt n\delta}^{\sqrt n\delta} u f(y,\hat \theta_n+u/\sqrt n)e^{C_n(u)}
	du\\	
	&B_n'(y)=
	\int_{-\sqrt n\delta}^{\sqrt n\delta}  f(y,\hat \theta_n+u/\sqrt n)e^{C_n(u)}
	du
\\
&	A_n(y)=
	\int_{-\sqrt n\delta}^{\sqrt n\delta}u f(y,\hat{\theta}_n)e^{C_n(u)} 
	du
	\\
	&B_n(y)=	\int_{-\sqrt n \delta}^{\sqrt n\delta} f(y,\hat{\theta}_n) e^{C_n(u)}  
	du
	\end{aligned}
	\]
\end{small}
\noindent Therefore, we can rewrite \eqref{AB} as follows 
\begin{small}
\begin{align*}
\tilde \Delta_n &=\frac{1}{n}\left(\frac{1}{\int_{-\delta\sqrt n}^{\delta \sqrt n}e^{C_n(u)}du}\right)^2
\left[ 2\int \frac{A_n(y)(A_n'(y)-A_n(y))}{f(y,\hat{\theta}_n)}dy+\int \frac{(A_n'(y)-A_n(y))^2}{f(y,\hat \theta_n)}dy- \right. \nonumber \\
& \left.-\int \frac{{A_n'(y)}^2}{B_n'(y)}\frac{B_n'(y)-B_n(y)}{f(y,\hat{\theta}_n)}dy
	\right]
\end{align*}
\end{small}
\noindent We will show that 
{\small $\tilde \Delta_n=I_{\theta_0}^{-1}n^{-2}+o_P(n^{-2}),$} by showing that:
\begin{description}
	\item[1.] {\small $\int \frac{A_n(y)(A_n'(y)-A_n(y))}{f(y,\hat\theta_n)}dy=o_P(\frac{1}{n})$}
	\item[2.] {\small $\int\frac{(A'_n(y)-A_n(y))^2}{f(y,\hat \theta_n)}dy = \frac{1}{n}2\pi I_{\theta_0}^{-2.}+ o_P(\frac{1}{n})$}
	\item[3.] {\small $\int \frac{{A'(y)}^2}{B'(y)}\frac{B(y)-B'(y)}{f(y,\hat \theta_n)}dy=o_P(\frac{1}{n})$.}
	\item[4.] {\small $\int_{-\sqrt n \delta}^{\sqrt n\delta}e^{C_n(u)}du= (2\pi)^{1/2}I_{\theta_0}^{-1/2}+o_P(1)$}
	\end{description}

\noindent {\bf 1.} Let us show that $\int \frac{A_n(y)(A_n'(y)-A_n(y))}{f(y,\hat\theta_n)}dy=o_P(\frac{1}{n})$.\\
There exists $\theta^{(1)}_n$ such that $|\theta_n^{(1)}-\hat{\theta}_n|<\delta$ and

\begin{small}
\begin{align}
\int \frac{A_n(y)(A_n'(y)-A_n(y))}{f(y,\hat\theta_n)}dy&=\int \left(
\int_{-\sqrt n\delta}^{\sqrt n\delta}ue^{C_n(u)}du\int_{-\sqrt n\delta}^{\sqrt n\delta}u\left(
\dot f(y,\hat{\theta}_n)\frac{u}{\sqrt n}+\ddot f(y,\theta^{(1)}_n)\frac{u^2}{n}
\right)e^{C_n(u)}du
\right)dy \nonumber \\
&=\int_{-\sqrt n\delta}^{\sqrt n\delta}ue^{C_n(u)}du\int \int_{-\sqrt n\delta}^{\sqrt n\delta}\ddot f(y,{\theta}^{(1)}_n)\frac{u^3}{n}e^{C_n(u)}du dy \label{inequ2ab} \\
&\leq \frac{1}{n} \sup_\theta \int |\ddot f(y,{\theta})|dy \;\int_{-\sqrt n\delta}^{\sqrt n\delta}ue^{C_n(u)}du \int_{-\sqrt n\delta}^{\sqrt n\delta}u^3e^{C_n(u)}du ,\label{inequ2a}
\end{align}
\end{small}
where in \eqref{inequ2ab} we have used \eqref{intzero} and the fact that $\int u^2 e^{C_n(u)}du < \infty$.
Furthermore,
\begin{small}
\begin{align}
\int_{-\sqrt n\delta}^{\sqrt n\delta}ue^{C_n(u)}du&=\int_{-\sqrt n\delta}^{\sqrt n\delta}ue^{a_{2,n}(\hat \theta_n)u^2/2+a_{3,n}(\theta^*_n)u^3/(6\sqrt n)}du \nonumber \\
&=\int_{-\sqrt n\delta}^{\sqrt n\delta}ue^{a_{2,n}(\hat \theta_n)u^2/2}\left(1+(e^{a_{3,n}(\theta^*_n)u^3/(6\sqrt n)}-1)\right)du\nonumber \\
&=\int_{-\sqrt n\delta}^{\sqrt n\delta}ue^{a_{2,n}(\hat \theta_n)u^2/2}(e^{a_{3,n}(\theta^*_n)u^3/(6\sqrt n)}-1)du \nonumber \\
&\leq \int_{-\sqrt n\delta}^{\sqrt n\delta}\frac{u^4}{6\sqrt n}|a_{3,n}(\theta^*_n)|e^{a_{2,n}(\hat \theta_n)u^2/2+a_{3,n}(\theta^*_n)u^3/(6\sqrt n)}du \label{inequ2bb} \\
&\leq \frac{1}{\sqrt n}\int_{-\sqrt n\delta}^{\sqrt n\delta}\frac{u^4}{6}|a_{3,n}(\theta^*_n)|e^{C_n(u)}du \nonumber \\
&\leq \left(\frac{1}{n}\sum_{i=1}^n G_3(Y_i)\right)\frac{1}{\sqrt n}\int_{-\sqrt n\delta}^{\sqrt n\delta}\frac{u^4}{6}e^{C_n(u)}du \nonumber \\
&=O_P(\frac{1}{\sqrt n}), \label{inequ2b}
\end{align}
\end{small}
\noindent where \eqref{inequ2bb} follows from $e^x -1 \leq x e^x$ for any $x \in \mathbb{R}$ and equation \eqref{inequ2b} is a consequence of assumption d), equation \eqref{ineq111} and dominated convergence theorem.
On the other hand, by (\ref{ineq111}),
\begin{small}
\begin{equation}
\int_{-\sqrt n\delta}^{\sqrt n \delta}|u|^3e^{C_n(u)}du<\infty. \label{inequ2}
\end{equation}
\end{small}
Thus, combining \eqref{inequ2a} and \eqref{inequ2} with \eqref{bound1}, we get
\begin{small}
\[
\int \frac{A_n(y)(A_n'(y)-A_n(y))}{f(y,\hat\theta_n)}dy=O_P(\frac{1}{n\sqrt n})
\]
\end{small}

\noindent{\bf 2.} Let us show that $\int\frac{(A'_n(y)-A_n(y))^2}{f(y,\hat \theta_n)}dy = \frac{1}{n}2\pi I_{\theta_0}^{-2} + o_P(\frac{1}{n})$.\\
There exists $\theta^{(2)}_n$ such that $|\theta^{(2)}_n-\hat \theta_n|<\delta$ and
\begin{small}
\[\begin{aligned}
\int\frac{(A'_n(y)-A_n(y))^2}{f(y,\hat \theta_n)}dx
&=\int
\left(
\int_{-\sqrt n\delta}^{\sqrt n \delta}
u
\left(
\dot f(y,\hat \theta_n)\frac{u}{\sqrt n}+\ddot f(y,\theta^{(2)}_n)\frac{u^2}{2n}
\right)
e^{C_n(u)}du
\right)^2
\frac{1}{f(y,\hat \theta_n)}dy\\
&=  \frac{1}{n}\int \frac{\dot f(y,\hat \theta_n)^2}{f(y,\hat \theta_n)^2}f(y,\hat \theta_n)dy
\left(
\int_{-\sqrt n\delta}^{\sqrt n\delta}u^2e^{C_n(u)}du
\right)^2\\
&
+\frac{1}{4n^2}\int 
\left(
\int_{-\sqrt n\delta}^{\sqrt n\delta}
\ddot f(\theta^{(2)}_n)u^3 e^{C_n(u)}du
\right)^2\frac{1}{f(y,\hat \theta_n)}du+R_n
\end{aligned}
\]
\end{small}
with
\begin{small}
\[\begin{aligned}
|R_n|&
\leq  \frac{1}{n\sqrt n}
\int \sup_{|\theta-\hat \theta_n|<\delta}|\ddot f(y,\theta)|\frac{|\dot f(y,\hat \theta_n)|}{f(y,\hat \theta_n)}dy \left( \int_{-\sqrt n\delta}^{\sqrt n\delta}u^4 e^{C_n(u)}
du \right)^2\\
&=O_P(\frac{1}{n\sqrt n}),
\end{aligned}
\]
\end{small}
by \eqref{bound2} and \eqref{ineq111}.\\
On the other hand,
\begin{small}
\[
\frac{1}{n}\int \frac{\dot f(y,\hat \theta_n)^2}{f(y,\hat \theta_n)^2}f(y,\hat \theta_n)dy
\left(
\int_{-\sqrt n\delta}^{\sqrt n\delta}u^2e^{C_n(u)du}
\right)^2\sim \frac{1}{n}E_{\hat \theta_n}\left(\frac{\dot f(Y,\hat \theta_n)^2}{f(Y,\hat \theta_n)^2}\right)I_{\theta_0}^{-2}\frac{2\pi}{I_{\theta_0}}\sim \frac{1}{n}2\pi I_{\theta_0}^{-2}.
\]
\end{small}
Furthermore,
\begin{small}
\[\begin{aligned}
\frac{1}{n^2}
\int
\left(
\int_{-\sqrt n\delta}^{\sqrt n\delta}
\ddot f(\theta^{(2)}_n)u^3e^{C_n(u)}du
\right)^2\frac{1}{f(y,\hat{\theta}_n)}dy
&\leq \frac{1}{n^2}
\left(
\int_{-\sqrt n\delta}^{\sqrt n\delta}
u^3e^{C_n(u)}du
\right)^2 
\int \sup_{|\theta-\hat \theta_n|<\delta}\ddot f(x,\theta)^2\frac{1}{f(y,\hat \theta_n)}dy\\
&=o_P(\frac{1}{n}),
\end{aligned}
\]
\end{small}
where last equality follows from \eqref{bound3}.\\

\noindent{\bf 3.} Let us show that $\int \frac{{A'(y)}^2}{B'(y)}\frac{B(y)-B'(y)}{f(y,\hat \theta_n)}dy=o_P(\frac{1}{n})$.\\
There exist $\theta^{(3)}_n$ and $\theta^{(4)}_n$ such that $|\theta^{(3)}_n-\hat \theta_n|<\delta$, 
$|\theta^{(4)}_n-\hat \theta_n|<\delta$ and 
\begin{small}
\begin{align}
\int &\frac{{A'(y)}^2}{B'(y)}\frac{B'(y)-B(y)}{f(y,\hat \theta_n)}dy
=\int
\frac{
	\left(
	\int_{-\sqrt n\delta}^{\sqrt n\delta}uf(y,\hat \theta_n+\frac{u}{\sqrt n})e^{C_n(u)}du
	\right)^2
}{\int_{-\sqrt n\delta}^{\sqrt n\delta}f(y,\hat \theta_n+\frac{u}{\sqrt n})e^{C_n(u)}du
}
\frac{
\left(
\int_{-\sqrt n\delta}^{\sqrt n\delta}
f(y,\hat \theta_n+\frac{u}{\sqrt n})-f(y,\hat \theta_n)\right)e^{C_n(u)}du
}{
f(y,\hat \theta_n)
}
dy \nonumber \\
&=\int
\frac{
	\left(
	f(y,\hat \theta_n) \int_{-\sqrt n\delta}^{\sqrt n\delta}ue^{C_n(u)}du
	+\int_{-\sqrt n\delta}^{\sqrt n\delta}\dot f(y,\theta^{(3)}_n)\frac{u^2}{\sqrt n}e^{C_n(u)}du
	\right)^2
}{\int_{-\sqrt n\delta}^{\sqrt n\delta}e^{C_n(u)}f(y,\hat \theta_n+\frac{u}{\sqrt n})du
}
\frac{
	\int_{-\sqrt n\delta}^{\sqrt n\delta}
	\dot f(y,\theta^{(4)}_n)\frac{u}{\sqrt n}e^{C_n(u)}du
}{
	f(y,\hat \theta_n)
}
dy \nonumber \\
&\leq 4
\int
\frac{
	f(y,\hat \theta_n)^2 \left(\int_{-\sqrt n\delta}^{\sqrt n\delta}ue^{C_n(u)}du\right)^2
	+\left(\int_{-\sqrt n\delta}^{\sqrt n\delta}\dot f(y,\theta^{(3)}_n)\frac{u^2}{\sqrt n}e^{C_n(u)}du
	\right)^2
}{\int_{-\sqrt n\delta}^{\sqrt n\delta}e^{C_n(u)}f(y,\hat \theta_n+\frac{u}{\sqrt n})du
}
\frac{
	\int_{-\sqrt n\delta}^{\sqrt n\delta}
	\dot f(y,\theta^{(4)}_n)\frac{u}{\sqrt n}e^{C_n(u)}du
}{
	f(y,\hat \theta_n)
}
dy \label{4a}\\
&\leq 4
\left(\int_{-\sqrt n\delta}^{\sqrt n\delta}ue^{C_n(u)}du\right)^2
\int \frac{ \int_{-\sqrt n\delta}^{\sqrt n\delta}
	\dot f(y,\theta^{(4)}_n)\frac{u}{\sqrt n}e^{C_n(u)}du
}{\int_{-\sqrt n\delta}^{\sqrt n\delta}e^{C_n(u)}f(y,\hat \theta_n+\frac{u}{\sqrt n})du
}f(y,\hat \theta_n)
dy \nonumber \\
&+ 4
\int
\frac{
\left(\int_{-\sqrt n\delta}^{\sqrt n\delta}\dot f(y,\theta^{(3)}_n)\frac{u^2}{\sqrt n}e^{C_n(u)}du
	\right)^2	\int_{-\sqrt n\delta}^{\sqrt n\delta}
	\dot f(y,\theta^{(4)}_n)\frac{u}{\sqrt n}e^{C_n(u)}du
	}{f(y,\hat \theta_n)\int_{-\sqrt n\delta}^{\sqrt n\delta}e^{C_n(u)}f(y,\hat \theta_n+\frac{u}{\sqrt n})du}
dy \nonumber
\\
&\leq 4
\frac{1}{\sqrt n}\frac{\left(\int_{-\sqrt n\delta}^{\sqrt n\delta}ue^{C_n(u)}du\right)^3}{
\int_{-\sqrt n\delta}^{\sqrt n\delta}
e^{C_n(u)}du
}
\int 
\sup_{|\theta_1-\hat \theta_n|<\delta,|\theta_2-\hat \theta_n|<\delta}
\frac{	|\dot f(y,\theta_1)| }{f(y,\theta_2)}f(y,\hat \theta_n)
dy \nonumber \\
&+ 4\frac{1}{n\sqrt n}
\frac{
	\left(\int_{-\sqrt n\delta}^{\sqrt n\delta}u^2e^{C_n(u)}du
	\right)^2	\int_{-\sqrt n\delta}^{\sqrt n\delta}
	ue^{C_n(u)}du
}{\int_{-\sqrt n\delta}^{\sqrt n\delta}e^{C_n(u)}du}
\int \sup_{|\theta_1-\hat \theta_n|<\delta,|\theta_2-\hat \theta_n|<\delta,|\theta_3-\hat \theta_n|<\delta}
\frac{\dot f(y,\theta_1)^2|\dot f(y,\theta_2)|}{f(y,\theta_3)f(y,\hat \theta_n)}
dy \nonumber \\
&=\frac{1}{n^2}O_P(1)\left(
\int 
\sup_{|\theta_1-\hat \theta_n|<\delta,|\theta_2-\hat \theta_n|<\delta}
\frac{	|\dot f(y,\theta_1)| }{f(y,\theta_2)}f(y,\hat \theta_n)
dy\right. \nonumber \\
&\left.\quad \quad\quad+
\int \sup_{|\theta_1-\hat \theta_n|<\delta,|\theta_2-\hat \theta_n|<\delta,|\theta_3-\hat \theta_n|<\delta}
\frac{\dot f(y,\theta_1)^2|\dot f(y,\theta_2)|}{f(y,\theta_3)f(y,\hat \theta_n)}
dy
\right) \label{4c} \\
&= o_P(\frac{1}{n}) \label{4b},
\end{align}
\end{small}
where in \eqref{4a} we have used $(a+b)^2 \leq 4(a^2+b^2)$ and in \eqref{4c} we have invoked \eqref{ineq111} and \eqref{inequ2b}. Equation \eqref{4b} follows from \eqref{bound4} and \eqref{bound5}.\\

\noindent {\bf 4.} Let us show that   {\small $\int_{-\sqrt n \delta}^{\sqrt n\delta}e^{C_n(u)}du\rightarrow (2\pi)^{1/2}I_{\theta_0}^{-1/2}$}. \\
We have, for $\mid \theta_n^*- \hat{\theta}_n \mid < \delta$,
\begin{small}
\begin{align*}
\int_{-\sqrt n \delta}^{\sqrt n\delta}e^{C_n(u)}du&= \int_{-\sqrt n \delta}^{\sqrt n\delta}e^{a_{2,n}(\hat \theta_n)u^2/2+a_{3,n}(\theta_n^*)u^3/(6\sqrt n)} du \\
&= \int_{-\sqrt n \delta}^{\sqrt n\delta}e^{a_{2,n}(\hat \theta_n)}du + \tilde{R}_n
\end{align*}
\end{small}
\noindent with
\begin{small}
\begin{align*}
\mid \tilde{R}_n \mid & \leq \int_{-\sqrt n \delta}^{\sqrt n\delta}e^{a_{2,n}(\hat \theta_n)} \left( e^{a_{3,n}(\theta_n^*)u^3/(6\sqrt n)} -1  \right)du\\
& \leq \int_{-\sqrt n \delta}^{\sqrt n\delta}e^{a_{2,n}(\hat \theta_n)}a_{3,n}(\theta_n^*)\frac{u^3}{6\sqrt n} e^{a_{3,n}(\theta_n^*)u^3/(6\sqrt n)} du \\
& \leq \frac{1}{\sqrt n} \mid \frac{1}{n}\sum_{i=1}^n G_3(Y_i) \mid \int_{-\sqrt n \delta}^{\sqrt n\delta} u^3 e^{C_n(u)}du \\
& = o_P(1).
\end{align*}
\end{small}
\noindent Since 
 \begin{small}
\[
e^{a_{2,n}(\hat \theta_n)u^2/2}  \rightarrow e^{-I_{\theta_0}u^2/2} \quad \text{a.s.}
\]
\end{small}
\noindent then
\begin{small}
\[
\int_{-\sqrt n \delta}^{\sqrt n\delta}e^{C_n(u)}du = (2\pi)^{1/2}I_{\theta_0}^{-1/2}+ o_P(1).
\]
\end{small}

\noindent \textit{STEP 3: Asymptotic behavior of $\Delta_n-\tilde \Delta_n$} \\
Let us show that {\small $\Delta_n-\tilde \Delta_n=o_P(\frac{1}{n^2})$}.\\
It holds that
\begin{small}
\begin{align*}
\Delta_n &-\tilde \Delta_n=\frac{1}{\int 
	\prod_{i=1}^n
	\frac{f(Y_i,\hat \theta_n+\phi)}{f(Y_i,\hat \theta_n)}d\phi}
\left[
\int
\frac{
	\left(
	\int \phi f(y,\hat \theta_n+\phi)
	\prod_{i=1}^n
	\frac{f(Y_i,\hat \theta_n+\phi)}{f(Y_i,\hat \theta_n)}d\phi
	\right)^2
}{
	\int  f(y,\hat \theta_n+\phi)
	\prod_{i=1}^n
	\frac{f(Y_i,\hat \theta_n+\phi)}{f(Y_i,\hat \theta_n)}d\phi
}
dy
-
\frac{
	\left(
	\int \phi 
	\prod_{i=1}^n
	\frac{f(Y_i,\hat \theta_n+\phi)}{f(Y_i,\hat \theta_n)}d\phi
	\right)^2
}{
	\int  
	\prod_{i=1}^n
	\frac{f(Y_i,\hat \theta_n+\phi)}{f(Y_i,\hat \theta_n)}d\phi
}
\right] \\ 
&-\frac{1}{\int_{-\delta}^\delta 
	\prod_{i=1}^n
	\frac{f(Y_i,\hat \theta_n+\phi)}{f(Y_i,\hat \theta_n)}d\phi}
\left[
\int
\frac{
	\left(
	\int_{-\delta}^\delta \phi f(y,\hat \theta_n+\phi)
	\prod_{i=1}^n
	\frac{f(Y_i,\hat \theta_n+\phi)}{f(Y_i,\hat \theta_n)}d\phi
	\right)^2
}{
	\int_{-\delta}^\delta  f(y,\hat \theta_n+\phi)
	\prod_{i=1}^n
	\frac{f(Y_i,\hat \theta_n+\phi)}{f(Y_i,\hat \theta_n)}d\phi
}
dy
-
\frac{
	\left(
	\int_{-\delta}^\delta \phi 
	\prod_{i=1}^n
	\frac{f(Y_i,\hat \theta_n+\phi)}{f(Y_i,\hat \theta_n)}d\phi
	\right)^2
}{
	\int_{-\delta}^\delta  
	\prod_{i=1}^n
	\frac{f(Y_i,\hat \theta_n+\phi)}{f(Y_i,\hat \theta_n)}d\phi
}
\right]\\
& \leq \frac{\int_{[-\delta,\delta]^c} \prod_{i=1}^n	\frac{f(Y_i,\hat \theta_n+\phi)}{f(Y_i,\hat \theta_n)}d\phi}{
		\int \prod_{i=1}^n	\frac{f(Y_i,\hat \theta_n+\phi)}{f(Y_i,\hat \theta_n)}d\phi
}|\tilde \Delta_n| 
+
\frac{1}{\int 
	\prod_{i=1}^n
	\frac{f(Y_i,\hat \theta_n+\phi)}{f(Y_i,\hat \theta_n)}d\phi}
\left[\left|
\int
\frac{
	\left(
	\int \phi f(y,\hat \theta_n+\phi)
	\prod_{i=1}^n
	\frac{f(Y_i,\hat \theta_n+\phi)}{f(Y_i,\hat \theta_n)}d\phi
	\right)^2
}{
	\int  f(y,\hat \theta_n+\phi)
	\prod_{i=1}^n
	\frac{f(Y_i,\hat \theta_n+\phi)}{f(Y_i,\hat \theta_n)}d\phi
}
dy
- \right. \right.\\
& -
\left. \left. \int
\frac{
	\left(
	\int_{-\delta}^\delta \phi f(y,\hat \theta_n+\phi)
	\prod_{i=1}^n
	\frac{f(Y_i,\hat \theta_n+\phi)}{f(Y_i,\hat \theta_n)}d\phi
	\right)^2
}{
	\int_{-\delta}^\delta  f(y,\hat \theta_n+\phi)
	\prod_{i=1}^n
	\frac{f(Y_i,\hat \theta_n+\phi)}{f(Y_i,\hat \theta_n)}d\phi
}
dy
\right|
+\left|
\frac{
	\left(
	\int \phi 
	\prod_{i=1}^n
	\frac{f(Y_i,\hat \theta_n+\phi)}{f(Y_i,\hat \theta_n)}d\phi
	\right)^2
}{
	\int  
	\prod_{i=1}^n
	\frac{f(Y_i,\hat \theta_n+\phi)}{f(Y_i,\hat \theta_n)}d\phi
} \right. \right. -\\
& \left. \left.
-\frac{
	\left(
	\int_{-\delta}^\delta \phi 
	\prod_{i=1}^n
	\frac{f(Y_i,\hat \theta_n+\phi)}{f(Y_i,\hat \theta_n)}d\phi
	\right)^2
}{
	\int_{-\delta}^\delta  
	\prod_{i=1}^n
	\frac{f(Y_i,\hat \theta_n+\phi)}{f(Y_i,\hat \theta_n)}d\phi
}
\right|
\right].
\end{align*}
\end{small}
\noindent By STEP 2 \begin{small}
\begin{align*}
\int_{[-\delta,\delta]} \prod_{i=1}^n \frac{f(Y_i,\hat \theta_n+\phi)}{f(Y_i,\hat \theta_n)}d\phi = \frac{1}{\sqrt n} \int_{-\sqrt n \delta}^{\sqrt n\delta}e^{C_n(u)}du =\frac{1}{\sqrt n}
(2\pi)^{1/2}I_{\theta_0}^{-1/2} +o_P(\frac{1}{\sqrt n}).
\end{align*}
\end{small}
\noindent On the other hand, by \eqref{ineq71},
\begin{small}
\[
\int_{[-\delta,\delta]^c} \prod_{i=1}^n \frac{f(Y_i,\hat \theta_n+\phi)}{f(Y_i,\hat \theta_n)}d\phi\leq C' e^{-n\epsilon}.
\]
\end{small}
Thus,
\begin{small}
\[
\int \prod_{i=1}^n \frac{f(Y_i,\hat \theta_n+\phi)}{f(Y_i,\hat \theta_n)}d\phi = C\frac{1}{\sqrt n}+o_P(\frac{1}{\sqrt n})
\]
\end{small}
and
\begin{small}
\[
\frac{\int_{[-\delta,\delta]^c} \prod_{i=1}^n	\frac{f(Y_i,\hat \theta_n+\phi)}{f(Y_i,\hat \theta_n)}d\phi}{
	\int \prod_{i=1}^n	\frac{f(Y_i,\hat \theta_n+\phi)}{f(Y_i,\hat \theta_n)}d\phi
}|\tilde \Delta_n|=o_P(\frac{1}{n^2}).
\]
\end{small}
Moreover,
\begin{small}
\begin{align*}
&\left|
\int
\frac{
	\left(
	\int \phi f(y,\hat \theta_n+\phi)
	\prod_{i=1}^n
	\frac{f(Y_i,\hat \theta_n+\phi)}{f(Y_i,\hat \theta_n)}d\phi
	\right)^2
}{
	\int  f(y,\hat \theta_n+\phi)
	\prod_{i=1}^n
	\frac{f(Y_i,\hat \theta_n+\phi)}{f(Y_i,\hat \theta_n)}d\phi
}
dy
-
\int
\frac{
	\left(
	\int_{-\delta}^\delta \phi f(y,\hat \theta_n+\phi)
	\prod_{i=1}^n
	\frac{f(Y_i,\hat \theta_n+\phi)}{f(Y_i,\hat \theta_n)}d\phi
	\right)^2
}{
	\int_{-\delta}^\delta  f(y,\hat \theta_n+\phi)
	\prod_{i=1}^n
	\frac{f(Y_i,\hat \theta_n+\phi)}{f(Y_i,\hat \theta_n)}d\phi
}
dy
\right|\\
 &\leq
\left|
\int
\frac{
	\left(
	\int \phi f(y,\hat \theta_n+\phi)
	\prod_{i=1}^n
	\frac{f(Y_i,\hat \theta_n+\phi)}{f(Y_i,\hat \theta_n)}d\phi
	\right)^2-\left(
	\int_{-\delta}^\delta \phi f(y,\hat \theta_n+\phi)
	\prod_{i=1}^n
	\frac{f(Y_i,\hat \theta_n+\phi)}{f(Y_i,\hat \theta_n)}d\phi
	\right)^2
}{
	\int  f(y,\hat \theta_n+\phi)
	\prod_{i=1}^n
	\frac{f(Y_i,\hat \theta_n+\phi)}{f(Y_i,\hat \theta_n)}d\phi
}dy\right| + \\
&+\int \left(\frac{
\int_{-\delta}^\delta \phi f(y,\hat \theta_n+\phi)
\prod_{i=1}^n
\frac{f(Y_i,\hat \theta_n+\phi)}{f(Y_i,\hat \theta_n)}d\phi
}{
\int_{-\delta}^\delta  f(y,\hat \theta_n+\phi)
\prod_{i=1}^n
\frac{f(Y_i,\hat \theta_n+\phi)}{f(Y_i,\hat \theta_n)}d\phi
}\right)^2
\int_{[-\delta,\delta]^c}f(y,\hat \theta_n +\phi)\prod_{i=1}^n
\frac{f(Y_i,\hat \theta_n+\phi)}{f(Y_i,\hat \theta_n)}d\phi dy\\
&\leq
2
\int
\frac{
	\int |\phi| f(y,\hat \theta_n+\phi)
	\prod_{i=1}^n
	\frac{f(Y_i,\hat \theta_n+\phi)}{f(Y_i,\hat \theta_n)}d\phi
}
{
\int  f(y,\hat \theta_n+\phi)
\prod_{i=1}^n
\frac{f(Y_i,\hat \theta_n+\phi)}{f(Y_i,\hat \theta_n)}d\phi
}
	\int_{[-\delta,\delta]^c} \phi f(y,\hat \theta_n+\phi)
	\prod_{i=1}^n
	\frac{f(Y_i,\hat \theta_n+\phi)}{f(Y_i,\hat \theta_n)}d\phi dy+ \\
&
+\int \left(\frac{
	\int_{-\delta}^\delta \phi f(y,\hat \theta_n+\phi)
	\prod_{i=1}^n
	\frac{f(Y_i,\hat \theta_n+\phi)}{f(Y_i,\hat \theta_n)}d\phi
}{
	\int_{-\delta}^\delta  f(y,\hat \theta_n+\phi)
	\prod_{i=1}^n
	\frac{f(Y_i,\hat \theta_n+\phi)}{f(Y_i,\hat \theta_n)}d\phi
}\right)^2
\int_{[-\delta,\delta]^c}f(y,\hat \theta_n+\phi)\prod_{i=1}^n
\frac{f(Y_i,\hat \theta_n+\phi)}{f(Y_i,\hat \theta_n)}d\phi dy\\
&
\leq C''e^{-n\epsilon},
\end{align*}
\end{small}
where last inequality comes from \eqref{ineq71}.
Analogously,
\begin{small}
\begin{align*}
&\left|
\frac{
	\left(
	\int \phi 
	\prod_{i=1}^n
	\frac{f(Y_i,\hat \theta_n+\phi)}{f(Y_i,\hat \theta_n)}d\phi
	\right)^2
}{
	\int  	\prod_{i=1}^n
	\frac{f(Y_i,\hat \theta_n+\phi)}{f(Y_i,\hat \theta_n)}d\phi
}
-
\frac{
	\left(
	\int_{-\delta}^\delta \phi 
	\prod_{i=1}^n
	\frac{f(Y_i,\hat \theta_n+\phi)}{f(Y_i,\hat \theta_n)}d\phi
	\right)^2
}{
	\int_{-\delta}^\delta  
	\prod_{i=1}^n
	\frac{f(Y_i,\hat \theta_n+\phi)}{f(Y_i,\hat \theta_n)}d\phi
}
\right|\\
&
\leq
\left|
\frac{
	\left(
	\int \phi 
	\prod_{i=1}^n
	\frac{f(Y_i,\hat \theta_n+\phi)}{f(Y_i,\hat \theta_n)}d\phi
	\right)^2-\left(
	\int_{-\delta}^\delta \phi 
	\prod_{i=1}^n
	\frac{f(Y_i,\hat \theta_n+\phi)}{f(Y_i,\hat \theta_n)}d\phi
	\right)^2
}{
	\int  
	\prod_{i=1}^n
	\frac{f(Y_i,\hat \theta_n+\phi)}{f(Y_i,\hat \theta_n)}d\phi
}\right|+ \\
&
+\left(\frac{
	\int_{-\delta}^\delta \phi 
	\prod_{i=1}^n
	\frac{f(Y_i,\hat \theta_n+\phi)}{f(Y_i,\hat \theta_n)}d\phi
}{
	\int_{-\delta}^\delta  
	\prod_{i=1}^n
	\frac{f(Y_i,\hat \theta_n+\phi)}{f(Y_i,\hat \theta_n)}d\phi
}\right)^2
\int_{[-\delta,\delta]^c}\prod_{i=1}^n
\frac{f(Y_i,\hat \theta_n+\phi)}{f(Y_i,\hat \theta_n)}d\phi \\
&
\leq
2
\frac{
	\int |\phi| 
	\prod_{i=1}^n
	\frac{f(Y_i,\hat \theta_n+\phi)}{f(Y_i,\hat \theta_n)}d\phi
}
{
	\int  
	\prod_{i=1}^n
	\frac{f(Y_i,\hat \theta_n+\phi)}{f(Y_i,\hat \theta_n)}d\phi
}
\int_{[-\delta,\delta]^c} \phi 
\prod_{i=1}^n
\frac{f(Y_i,\hat \theta_n+\phi)}{f(Y_i,\hat \theta_n)}d\phi +\\
&
+\int \left(\frac{
	\int_{-\delta}^\delta \phi 
	\prod_{i=1}^n
	\frac{f(Y_i,\hat \theta_n+\phi)}{f(Y_i,\hat \theta_n)}d\phi
}{
	\int_{-\delta}^\delta  
	\prod_{i=1}^n
	\frac{f(Y_i,\hat \theta_n+\phi)}{f(Y_i,\hat \theta_n)}d\phi
}\right)^2
\int_{[-\delta,\delta]^c}\prod_{i=1}^n
\frac{f(Y_i,\hat \theta_n+\phi)}{f(Y_i,\hat \theta_n)}d\phi \\
&\leq C'''e^{-n\epsilon}.
\end{align*}
\end{small}
This concludes the proof of STEP 3 and, thus, of Lemma \ref{lemmaA2} (Supplementary).

\end{proof}

\begin{proof}
(Lemma~\ref{lemma4})
Assumptions of Lemma \ref{lemma4} imply regularity conditions (a)-(e) given above. The result follows from Lemma \ref{lemmaA2} (Supplementary).
\end{proof}

\begin{proof}
(Proposition~\ref{propc})
It is enough to prove the results for $a=1$. 
For two-arm BUDs we have
\begin{small}
\[
\Delta_t(a)= \text{Var}(\theta_a\mid \Sigma_t)
- E(\text{Var}(\theta_a\mid \Sigma_{t+1})\mid A_{t+1}=a,\Sigma_t)
\]
\end{small}
for $a \in\{0,1\}$.
Define 
\begin{small}
\begin{small}
\[
F_t=-\hat p_{t,1}+
\frac{ 
	\Delta_t(1)  ^h
}{
  \Delta_t(1) ^h+  \Delta_t(0) ^h
}
\]
\end{small}
\end{small}
and
\begin{small}
\begin{small}
\[
\tilde F_t=-\hat p_{t,1}+\frac{I_{\theta_{0,1}}^{-h}\left(t \hat{p}_{t,1}\right)^{-2h} }{ I_{\theta_{0,1}}^{-h}\left(t \hat{p}_{t,1}\right)^{-2h}+ I_{\theta_{0,0}}^{-h}\left(t (1-\hat{p}_{t,1})\right)^{-2h}}.
\]
\end{small}
\end{small}
\noindent As a function of $\hat p_{t,1}$, $\tilde F_t$ is strictly decreasing. 
The unique root of $\tilde F_t=0$ is
\begin{small}
\begin{equation}
\rho_1:=\frac{
	I_{\theta_{0,1}}^{-\frac{h}{2h+1}}
}{I_{\theta_{0,0}}^{-\frac{h}{2h+1}}+I_{\theta_{0,1}}^{-\frac{h}{2h+1}}}.
\end{equation}
\end{small}
\noindent Now, we show that $F_t-\tilde F_t$ converges to zero a.s. as $t\rightarrow\infty$.
Recall that if $a_n$, $b_n$, $a_n'$ and $b_n'$ are sequences of positive numbers, then
\begin{small}
\[
\left |
\frac{a_n}{a_n+b_n}
-
\frac{a_n'}{a_n'+b_n'}
\right |\leq 
\min\left(
\left |
\frac{a_n}{b_n}-\frac{a_n'}{b_n'}\right |,
\left | \frac{b_n}{a_n}-\frac{b_n'}{a_n'}\right |
\right).
\]
\end{small}
Thus, we get
\begin{small}
\[
\mid F_t-\tilde F_t\mid \leq \min\left(
\left | \frac{\Delta_t(1)^h}{\Delta_t(0)^h}- \frac{I_{\theta_{0,1}}^{-h}\left(t \hat{p}_{t,1}\right)^{-2h} }{ I_{\theta_{0,0}}^{-h}\left(t (1-\hat{p}_{t,1})\right)^{-2h}}
\right |,
\left | \frac{\Delta_t(0)^h}{\Delta_t(1)^h}- \frac{ I_{\theta_{0,0}}^{-h}\left(t (1-\hat{p}_{t,1})\right)^{-2h}}{I_{\theta_{0,1}}^{-h}\left(t \hat{p}_{t,1}\right)^{-2h} }
\right |
\right).
\]
\end{small}
Hence
\begin{small}
\begin{align}
\mid F_t-\tilde F_t\mid
\leq &\min \left ( \left | \frac{\Delta_t(1)^h I_{\theta_{0,0}}^{-h}\left(t (1-\hat{p}_{t,1})\right)^{-2h}- \Delta_t(0)^h I_{\theta_{0,1}}^{-h}\left(t \hat{p}_{t,1}\right)^{-2h} }{\Delta_t(0)^h I_{\theta_{0,0}}^{-h}\left(t (1-\hat{p}_{t,1})\right)^{-2h}}\right | ,\right. \nonumber \\ & \quad \quad \left. \left | \frac{ \Delta_t(0)^h I_{\theta_{0,1}}^{-h}\left(t \hat{p}_{t,1}\right)^{-2h}- \Delta_t(1)^h I_{\theta_{0,0}}^{-h}\left(t (1-\hat{p}_{t,1})\right)^{-2h}}{  \Delta_t(1)^h I_{\theta_{0,1}}^{-h}\left(t \hat{p}_{t,1}\right)^{-2h}     } \right| 
\right). \label{inequ}
\end{align}
\end{small}
Additionally, by Lemma \ref{lemma4}, for every $a=0,1$, we have
\begin{small}
$\Delta_t(a)=I_{\theta_{0,a}}^{-1}(t\hat p_{t,a})^{-2}+o_P((t \hat p_{t,a})^{-2})$ 

\end{small} \noindent and, indeed, by properties of $o_P(\cdot)$,
\begin{small}
\begin{equation}
\Delta_t(a)^h = I_{\theta_{0,a}}^{-h}(t \hat p_{t,a})^{-2h}+ o_P((t \hat p_{t,a})^{-2h}). \label{op}
\end{equation}
\end{small}
\noindent Thus, plugging \eqref{op} into the right-hand side of \eqref{inequ}, we conclude that $F_t-\tilde F_t\rightarrow 0$.
Now, let $c$ be such that $\tilde F_t<-2c$ if
 $\hat p_{t,1}>\rho_1+\epsilon$ and 
 $\tilde F_t>2c$ if $\hat p_{t,1}<\rho_1-\epsilon$. 
 Since $F_t-\tilde F_t\rightarrow 0$, 
there exists a random time $T$ such that $\mid F_t-\tilde F_t\mid<c$ for all $t\geq T$. 
For every $t\geq T$, $F_t<-c$ if $\hat p_{t,1}>\rho_1+\epsilon$ and $F_t>c$ if $\hat p_{t,1}<\rho_1-\epsilon$.  Based on basics of stochastic approximation theory, it follows that $\hat p_{t,1}  \rightarrow \rho_1 $ almost surely.
 Additionally, by definition of $p_{t,1}$, equation \eqref{op} and properties of $o_P(\cdot)$, we have\\
\begin{small}
\begin{equation}
p_{t,1} = \frac{I_{\theta_{0,1}}^{-h}\left(t \hat{p}_{t,1}\right)^{-2h}   }{I_{\theta_{0,1}}^{-h}\left(t \hat{p}_{t,1}\right)^{-2h}  + I_{\theta_{0,0}}^{-h}\left(t (1-\hat{p}_{t,1})\right)^{-2h}}+o_P(1).
\end{equation}
\end{small}
Therefore, applying continuous mapping theorem, 
we have \begin{small}
\begin{equation*}
p_{t,1}  \underset{{\small t \rightarrow \infty}}{\longrightarrow} \rho_1  \, \,{\normalsize  \text{a.s. } } .
\end{equation*}
\end{small}

\end{proof}

\end{document}